\documentclass[11pt]{article}
\usepackage{amsmath}\begin{normalsize}
\end{normalsize}
\usepackage{algorithm}
\usepackage[noend]{algpseudocode}
\usepackage{color}
\usepackage{bm,verbatim}
\usepackage[authoryear]{natbib}
\usepackage[colorlinks,citecolor=blue,linkcolor=blue,urlcolor=blue]{hyperref}
\usepackage{latexsym, epsfig, amssymb, amsmath, amsthm, graphicx, mathrsfs, booktabs}
\usepackage{rotating, tabularx, setspace,paralist}
\usepackage{graphicx}
\usepackage[OT1]{fontenc}
\usepackage{lscape}
\usepackage{multirow,multicol,mathtools}
\usepackage{anysize}
\usepackage{graphicx}
\usepackage{subcaption}
\renewcommand{\baselinestretch}{1.2}
\marginsize{1.1in}{1.1in}{0.55in}{0.8in} %
\usepackage{ifthen,amsmath,amssymb}

\makeatletter
\def\singlespace{\def\baselinestretch{1}\@normalsize}

\newtheorem{remark}{Remark}%[section]
\newtheorem{example}{Example}%[section]
\renewcommand{\theequation}{%\thesection.
\arabic{equation}%
}

\makeatother

\newcommand{\ba}{\mbox{\bf a}}

\newcommand{\be}{\mbox{\bf e}}

\newcommand{\bv}{\mbox{\bf v}}

\newcommand{\bx}{\mbox{\bf x}}
\newcommand{\by}{\mbox{\bf y}}

\newcommand{\bA}{\mbox{\bf A}}
\newcommand{\bB}{\mbox{\bf B}}

\newcommand{\bD}{\mbox{\bf D}}

\newcommand{\bH}{\mbox{\bf H}}
\newcommand{\bI}{\mbox{\bf I}}

\newcommand{\bQ}{\mbox{\bf Q}}

\newcommand{\bS}{\mbox{\bf S}}
\newcommand{\bP}{\mbox{\bf P}}

\newcommand{\bM}{\mbox{\bf M}}
\newcommand{\bV}{\mbox{\bf V}}
\newcommand{\bW}{\mbox{\bf W}}
\newcommand{\bX}{\mbox{\bf X}}

\newcommand{\bpi}{\mbox{\boldmath $\pi$}}

\newcommand{\bTheta}{\mbox{\boldmath $\Theta$}}

\newcommand{\bleta}{\mbox{\boldmath $\eta$}}

\newcommand{\bSig}{\mbox{\boldmath $\Sigma$}}

\newcommand{\bPi}{\mbox{\boldmath $\Pi$}}

\newcommand{\sbx}{\scriptsize \bx}
\newcommand{\sby}{\scriptsize \by}

\newcommand{\var}{\mathrm{var}}

\newcommand{\cov}{\mathrm{cov}}

\newcommand{\Sig}{\mathbf{\Sigma}}

\newcommand{\tr}{\mathrm{tr}}
\newcommand{\diag}{\mathrm{diag}}

\def\toD{\overset{\mathscr{D}}{\longrightarrow}}

\def\independenT#1#2{\mathrel{\setbox0\hbox{$#1#2$}%
\copy0\kern-\wd0\mkern4mu\box0}}

\def\beginn{\begin{eqnarray*}}
\def\endn{\end{eqnarray*}}
\def\beginy{\begin{eqnarray}}
\def\endy{\end{eqnarray}}
\def\begine{\begin{enumerate}}
\def\ende{\end{enumerate}}
\definecolor{bittersweet}{rgb}{0.8, 0.5, 0.2}

\theoremstyle{plain}
\newtheorem{thm}{Theorem}
\newtheorem{lem}{Lemma}
\newtheorem{coro}{Corollary}

\newtheorem{deff}{Definition}

\newtheorem{cond}{Condition}
\newcommand{\non}{\nonumber \\}
\newcommand{\bbA}{{\bf A}}

\newcommand{\bbB}{{\bf B}}

\newcommand{\bbc}{{\bf c}}
\newcommand{\bbD}{{\bf D}}

\newcommand{\bbe}{{\bf e}}

\newcommand{\bbf}{{\bf f}}

\newcommand{\bbP}{{\bf P}}

\newcommand{\bbH}{{\bf H}}

\newcommand{\bbw}{{\bf w}}
\newcommand{\bbI}{{\bf I}}
\newcommand{\bbi}{{\bf i}}

\newcommand{\bbM}{{\bf M}}

\newcommand{\bbQ}{{\bf Q}}

\newcommand{\bbs}{{\bf s}}

\newcommand{\bbS}{{\bf S}}

\newcommand{\bbu}{{\bf u}}
\newcommand{\bbV}{{\bf V}}
\newcommand{\bbv}{{\bf v}}

\newcommand{\bbW}{{\bf W}}

\newcommand{\bbX}{{\bf X}}

\newcommand{\bbx}{{\bf x}}

\newcommand{\bbY}{{\bf Y}}
\newcommand{\bby}{{\bf y}}

\newcommand{\bbz}{{\bf z}}
\newcommand{\bbb}{{\bf b}}

\newcommand{\ep}{\ensuremath{\epsilon}}

\renewcommand{\baselinestretch}{1.3}

\usepackage[draft,inline,nomargin,index]{fixme}
\fxsetup{theme=color,mode=multiuser}
\FXRegisterAuthor{yf}{ayf}{\color{magenta} YF}

\begin{document}
	
\title{SIMPLE: Statistical Inference on Membership Profiles in Large Networks
	\thanks{Jianqing Fan is Frederick L. Moore '18 Professor of Finance, Department of Operations Research and Financial Engineering, Princeton University, Princeton, NJ 08544, USA (E-mail: jqfan@princeton.edu). %
		Yingying Fan is Centennial Chair in Business Administration and Professor, Data Sciences and Operations Department, Marshall School of Business, University of Southern California, Los Angeles, CA 90089 (E-mail: \textit{fanyingy@marshall.usc.edu}). %
		Xiao Han is Professor, International Institute of Finance, Department of Statistics and Finance, University of Science and Technology of China, Hefei, China 230026 (E-mail: \textit{xhan011@ustc.edu.cn}). %
		Jinchi Lv is Kenneth King Stonier Chair in Business Administration and Professor, Data Sciences and Operations Department, Marshall School of Business, University of Southern California, Los Angeles, CA 90089 (E-mail: \textit{jinchilv@marshall.usc.edu}). %
		This work was supported by NIH grant R01-GM072611-16, NSF grants DMS-1712591,  DMS-1953356, DMS-2052926, and DMS-2052964, NSF CAREER Award DMS-1150318, a grant from the Simons Foundation, Adobe Data Science Research Award, and a grant from NSF of China (No.12001518). Corresponding authors:  Jianqing Fan and Xiao Han.}
	%\date{\today}
	\date{May 3, 2021}
	\author{Jianqing Fan$^1$, Yingying Fan$^2$, Xiao Han$^{3}$ and Jinchi Lv$^2$
		\medskip\\
		Princeton University$^1$, University of Southern California$^2$\\
		and University of Science and Technology of China$^3$
		\\
	} %
}

\maketitle

\begin{abstract}
Network data is prevalent in many contemporary big data applications in which a common interest is to unveil important latent links between different pairs of nodes. Yet a simple fundamental question of how to precisely quantify the statistical uncertainty associated with the identification of latent links still remains largely unexplored. In this paper, we propose the method of statistical inference on membership profiles in large networks (SIMPLE) in the setting of degree-corrected mixed membership model, where the null hypothesis assumes that the pair of nodes share the same profile of community memberships. In the simpler case of no degree heterogeneity, the model reduces to the mixed membership model for which an alternative more robust test is also proposed. Both tests are of the Hotelling-type statistics based on the rows of empirical eigenvectors or their ratios, whose asymptotic covariance matrices are very challenging to derive and estimate. Nevertheless, their analytical expressions are unveiled and the unknown covariance matrices are consistently estimated. Under some mild regularity conditions, we establish the exact limiting distributions of the two forms of SIMPLE test statistics under the null hypothesis and contiguous alternative hypothesis. They are the chi-square distributions and the noncentral chi-square distributions, respectively, with degrees of freedom depending on whether the degrees are corrected or not. We also address the important issue of estimating the unknown number of communities and establish the asymptotic properties of the associated test statistics. The advantages and practical utility of our new procedures in terms of both size and power are demonstrated through several simulation examples and real network applications.
\end{abstract}
	
\textit{Running title}: SIMPLE
	

\section{Introduction} \label{Sec1}
	
%The nodes can be  broadly defined such as individuals, economic entities, documents, or medical disorders in social, economic, text, or health networks.
	
Large-scale network data that describes the pairwise relational information among  objects is commonly encountered in many applications such as the studies of citation networks, protein-protein interaction networks, health networks, financial networks, trade networks, and social networks. The popularity of such applications has motivated a spectrum of research with network data. Popularly used methods include algorithmic ones and model-based ones, where the former uses algorithms to optimize some carefully designed criteria (e.g., \cite{Newman_2013, PhysRevE.88.042822, Zhang18144}), and the latter relies on specific structures of some probabilistic models (see, e.g., \cite{Goldenberg:2010} for a review). This paper belongs to the latter group.   In the literature, a number of probabilistic models have been proposed for modeling network data. As arguably the simplest model with planted community identity, the stochastic block model (SBM) \citep{SBM1983, WangWong1987, Abbe2017CommunityDA} has received a tremendous amount of attention in the last decade. To overcome the limitation and increase the flexibility  in the basic stochastic block model, various variants have been proposed. To name a few, the degree-corrected SBM \citep{DCSBM2011} introduces a degree parameter for each node to make the expected degrees match the observed ones. The overlapping SBM, such as the mixed membership model \citep{Airoldi:2008}, allows the communities to overlap by assigning  each node  a profile of community memberships. See also \cite{Newman2015} for a review of network models.
	
%In the statistical literature, a most frequently studied problem with network data is community detection, aiming at grouping the nodes into a number of communities within which nodes  are more densely connected.
%As one of the most scalable tool for community detection, spectral methods have received tremendous amount of attention.   Most popularly used methods include spectral based methods \cite{rohe2011} \cite{lei2015} \cite{jin2015}, likelihood based methods \cite{ab13} \cite{yan2016}, and modularity based methods \citep{Newman_2013, PhysRevE.88.042822, Zhang18144, bbi09}.  There has been a huge literature on community detection in the past decades under the previously discussed  probabilistic models.  See, for example, , , , , , \cite{zhang2016}, \cite{wf16} for community detection methods under SBM;   %for community detection methods with the assistance of node features under SBM;
% \cite{zhao12} for community detection methods under DCSBM,   {\color{red} need to classify them into the three groups}
%; \cite{mossel2012stochastic}. {\color{red}add more references} See also \cite{Abbe2017CommunityDA} for a comprehensive review of recent developments in community detection under SBMs.
An important problem in network analysis  is to unveil the true latent links between different pairs of nodes, where nodes can be  broadly defined such as individuals, economic entities, documents, or medical disorders in social, economic, text, or health networks. There is a growing literature on network analysis with various methods available for clustering the nodes into different communities within which nodes  are more densely connected, based on the observed adjacency matrices or the similarity matrices constructed using the node information.
	These methods focus mainly on the clustering aspect of the problem, outputting subgroups with predicted membership identities. Yet the statistical inference aspect such as quantifying the statistical uncertainty associated with the identification of latent links has been largely overlooked.   %Due to the nature of these methods, there are no associated confidence levels given for the clustering results. In fact, almost community detection problems so far have focused on estimating rather than inference.
	This paper aims at filling this crucial gap by proposing new statistical tests for testing whether any given pair of nodes share the same membership profiles, and providing the associated $p$-values. %{\color{red} we need to be careful about the p-value claim because it needs the knowledge of $K$}

	%\yfnote
	{Knowing the statistical significance of membership profiles can bring more confidence to practitioners in decision making. Taking the stock market for example, investors often want to form diversified portfolios by including stocks with little or no correlation in their returns. The correlation matrix of stock returns can then be used to construct an affinity matrix, and stocks with relatively highly correlated returns can be regarded as in the same community. Obtaining the pairwise p-values of stocks can help investors form diversified portfolios with statistical confidence. For instance, if one is interested in the Apple stock, then the pairwise p-values of Apple and all other candidate stocks can be calculated, and stocks with the smallest p-values can be included to form portfolios. Another important application is in legislation. For example, illegal logging greatly conflicts with indigenous and local populations,  contributing to violence, human rights abuses, and corruption. The DNA sequencing technology has been used to identify the region of logs. In such application, an affinity matrix can be calculated according to the similarity of DNA sequences. Then applying our method, p-values can be calculated and used in court as statistical evidence in convicting illegal logging.}

	To make the problem concrete, we consider the family of degree-corrected mixed membership models, which includes the mixed membership model and the stochastic block model as special cases. In the degree-corrected mixed membership model, node $i$ is assumed to have a membership profile characterized by a community membership probability vector $\bpi_i\in \mathbb{R}^K$, where $K$ is the number of communities and the $k$th entry of $\bpi_i$ specifies the mixture proportion of node $i$ in community $k$ \citep{Airoldi:2008,zhao12}. For example, a book can be 30\% liberal and 70\% conservative.  In addition, each node is allowed to have its own degree. For any given pair of nodes $i$ and $j$, we investigate whether they have the same membership profile or not by testing  the hypothesis  $H_0: \bpi_i=\bpi_j$ vs. $H_a: \bpi_i\neq \bpi_j$.  Two forms of statistical inference on membership profiles in large networks (SIMPLE) test are proposed.  Under the mixed membership model where all nodes  have the same degree, we construct the first form of SIMPLE test by resorting to the $i$th and $j$th rows of the spiked eigenvector matrix of the observed adjacency matrix. We establish the asymptotic null and alternative distributions of the test statistic, where under the null hypothesis the asymptotic distribution is chi-square  with $K$ degrees of freedom and under the alternative hypothesis, the asymptotic distribution is noncentral chi square with a location parameter determined by how distinct the membership profiles of nodes $i$ and $j$ are.

In the more general degree-corrected mixed membership model, where nodes are allowed to have heterogeneous degrees, we build the second form of SIMPLE test based on the ratio statistic proposed in \cite{jin2015}. %, and investigate the asymptotic properties of the test statistic under both the null and alternative hypotheses.
	We show that the asymptotic null distribution is chi-square with $K-1$ degrees of freedom, and under the alternative hypothesis and some mild regularity conditions, the test statistic diverges to infinity with asymptotic probability one. We prove that these asymptotic properties continue to hold even with estimated population parameters (including the number of communities $K$) provided that these parameters can be estimated reasonably well. We then suggest specific estimators of these unknown parameters and show that they achieve the desired estimation precision.  %remains to hold with consistently estimated model parameters such as the number of communities $K$.
	These new theoretical results enable us to construct rejection regions that are pivotal to the unknown parameters for each of these two forms of
	the SIMPLE test, and to calculate $p$-values explicitly. Our method is more applicable than most existing ones in the community detection literature where $K$ is required to be known. Although the second form of SIMPLE test can be applied to both cases with and without degree heterogeneity, we would like to point out that the first test is empirically more stable since it does not involve any ratio calculations. %When $K$ is known apriori, p-values for each tests can be calculated explicitly.
	To the best of our knowledge, this paper is the first in the literature to provide quantified uncertainty levels in community membership estimation and inference. % estimation without the need to know the true number of communities.
	
		%\yfnote
		{Our test is most useful when one cares about local information of the network. For instance, if  the interest is whether two (or several) nodes belong to the same community with quantified significance level, then SIMPLE can be used. Indeed, our statistics do not rely on any pre-determined membership information. Compared to community detection methods, our work has at least three advantages: 1) we do not need to assign memberships to nodes that are not of interests; 2) our method can provide the level of significance, which can be very important in scientific discoveries; and 3) if partial membership information is known in a network, then the nodes with missing membership information can be recovered with statistical confidence by applying our tests. }	
	
	Both forms of SIMPLE test are constructed using the spectral information of the observed adjacency matrix.  In this sense, our work is related to the class of spectral clustering methods, which is one of the most scalable tools for community detection and has been popularly used in the literature.  See, e.g.,  \cite{von2007} for a tutorial of spectral clustering methods. %and the references mentioned previous for their applications to network data.
	See also  \cite{rohe2011,lei2015, jin2015} among many others for the specifics on the implementation of spectral methods for community detection. In addition, the optimality for the case of two communities has been established by \cite{abbe2019entrywise}.  Our work is related to but substantially different from the link prediction problem \citep{Liben-Nowell:2007, WuLevinaZhu2018}, which can be thought of as predicting pairs of nodes as linked or non-linked. The major difference is that in link prediction, only part of the adjacency matrix is observed and one tries to predict the latent links among the nodes which are unobserved. Moreover, link prediction methods usually do not provide statistical confidence levels.

	Our work falls into the category of hypothesis testing with network data. In the literature, hypothesis testing has been used for different purposes.  %such as estimating and testing the number of communities $K$.
For example,
\cite{arias-castro2014} and \cite{verzelen2015} formalized the problem of community detection in a given random graph as a hypothesis testing problem in dense and sparse random networks, respectively.
Under the stochastic block model assumption, \cite{Bickel2016} proposed a recursive bipartitioning algorithm to automatically estimate the number of communities using hypothesis test constructed from the largest principal eigenvalue of the suitably centered and scaled adjacency matrix. The null hypothesis of their test is that the network has only $K=1$ community. %, that is, it is generated from an Erd\"{o}s-R\'{e}nyi random graph.
	\cite{L16} generalized their idea and proposed a test allowing for $K\geq 1$ communities in the stochastic block model under the null hypothesis. The number of communities can then be estimated by sequential testing. \cite{wang2017} proposed a likelihood ratio test for selecting the correct $K$ under the setting of SBM.

The rest of the paper is organized as follows. Section \ref{Sec2} introduces the model setting and technical preparation. We present the SIMPLE method and its asymptotic theory as well as the implementation details
%and applications
of SIMPLE in Section \ref{Sec3}. Sections \ref{Sec4} and \ref{Sec5} provide several simulation and real data examples illustrating the finite-sample performance and utility of our newly suggested method. We discuss some implications and extensions of our work in Section \ref{Sec6}. All the proofs and technical details are provided in the Supplementary Material.

\section{Statistical inference in large networks} \label{Sec2}

\subsection{Model setting} \label{Sec2.1}
Consider an undirected graph $\mathcal{N} = (V, E)$ with $n$ nodes, where $V = \{1,\cdots, n\}$ is the set of nodes and $E$ is the set of links. Throughout the paper, we use the notation $[n] = \{1,\cdots, n\}$. Let $\bbX= (x_{ij})\in \mathbb{R}^{n\times n}$ be the symmetric adjacency matrix representing the connectivity structure of graph $\mathcal N$, where $x_{ij}=1$ if there is a link connecting nodes $i$ and $j$, and $x_{ij}=0$ otherwise. We consider the general case when graph $\mathcal N$ may or may not admit self loops, where in the latter scenario $x_{ii}=0$ for all $i\in[n]$.  Under a probabilistic model, we will assume that $x_{ij}$ is an independent realization from a Bernoulli random variable for all upper triangular entries of random matrix $\bbX$.
%Throughout we assume the number of communities $K$ is known. For the case where $K$ is unknown for Section ??? for more discussions.

To model the connectivity pattern of graph $\mathcal N$, consider a symmetric binary random matrix $\bX^*$ with the following  latent structure
\begin{equation} \label{eq: model}
	\bbX^*=\bbH+\bbW^*,
\end{equation}
where $\bbH=(h_{ij})\in \mathbb{R}^{n\times n}$ is the deterministic mean matrix (or probability matrix) of low rank $K \geq 1$ (see \eqref{new.eq006} later for a specification) and $\bbW^*=(w_{ij}^*)\in \mathbb{R}^{n\times n}$ is a symmetric random matrix with mean zero and independent entries on and above the diagonal.  Assume that the observed adjacency matrix $\bbX$ is either $\bbX^*$ or $\bX^* - \diag(\bbX^*)$, corresponding to the cases with or without self loops, respectively. In either case, we have the following decomposition
\begin{equation} \label{eq: model.general}
	\bX = \bH + \bbW,	
\end{equation}
where $\bbW = \bbW^*$ in the presence of self loops and $\bbW = \bbW^* - \diag(\bbX^*)$ in the absence of self loops. We can see that in either case, $\bW$ in (\ref{eq: model.general}) is symmetric with independent entries on and above the diagonal. Our study will cover both cases. Hereafter to simplify the presentation, we will slightly abuse the notation by referring to $\bbH$ as the mean matrix and $\bbW$ as the noise matrix.

Assume that there is an underlying latent community structure that the network $\mathcal N$ can be decomposed into $K$ latent disjoint communities
\begin{equation*}
\mathcal{C}_1, \cdots,  \mathcal{C}_K,
\end{equation*}
where each node $i$ is associated with the community membership probability vector \\
$\bpi_i = (\bpi_i(1),\cdots, \bpi_i(K))^T\in \mathbb{R}^K$  such that%within each community $V^{(i)}$ share some common characteristics.
\begin{equation} \label{new.eq004}
P(\text{node } i \text{ belongs to community } \mathcal{C}_k) = \bpi_i(k), \quad k=1,\cdots, K.
\end{equation}
%This follows automatically that
%\begin{equation}\label{eq: mixed}
%\bpi_i(k)\in [0,1], \ \sum_{k=1}^K\bpi_i(k)=1.
%\end{equation}
Throughout the paper, we assume that the number of communities $K$ is  \textit{unknown} but bounded away from infinity.

For any given pair of nodes $i,j \in V$ with $i\neq j$, our goal is to infer whether they share the same community identity or not with quantified uncertainty level from the observed adjacency matrix $\bbX$ in the general model (\ref{eq: model.general}).  In other words, for each pair of nodes $i,j\in V$ with $i\neq j$, we are interested in testing the hypothesis
\begin{align}\label{eq: hypothesis}
H_0: \bpi_i=\bpi_j \quad \text{ versus } \quad H_a:\bpi_i\neq \bpi_j.
\end{align}
Throughout the paper, we consider the preselected pair $(i,j)$ and thus nodes $i$ and $j$ are fixed.

To make the problem more explicit, we consider the degree-corrected mixed membership (DCMM) model. Using the same formulation as in \cite{JKL17},
%Under the DCMM model, node $i$ from community $\mathcal{C}_k$ connects  with node $j$ from community $\mathcal{C}_l$ for $j \neq i$ with probability
%\begin{equation} \label{new.eq005}
%P(x_{ij}=1| i \in \mathcal{C}_k, j\in \mathcal{C}_l) = \theta_i\theta_jp_{kl},
%\end{equation}
%where $\theta_i>0$, $i \in [n]$, stands for the degree heterogeneity.
%Consequently,
the probability of a link between nodes $i$ and $j$ with $i\neq j$ under the DCMM model can be written as
\begin{equation} \label{new.eq006}
P(x_{ij}=1) = \theta_i\theta_j\sum_{k=1}^K\sum_{l=1}^K\bpi_i(k)\bpi_j(l)p_{kl}.
\end{equation}
Here, $\theta_i>0$, $i \in [n]$, measures the degree heterogeneity,
and $p_{kl}$ can be interpreted as the probability of a typical member ($\theta_i=1$, say) in community $\mathcal{C}_k$ connects with a typical member ($\theta_j=1$, say) in community $\mathcal{C}_l$, as in the stochastic block model.
Writing \eqref{new.eq006} in the matrix form, we have

\begin{equation} \label{DCMM}
\bbH = \bTheta\bPi\bbP\bPi^T\bTheta,
\end{equation}
where $\bTheta=\diag(\theta_1,\cdots,\theta_n)$ stands for the degree heterogeneity matrix, $\bPi = (\bpi_1,\cdots, \bpi_n)^T \in \mathbb{R}^{n\times K}$ is the matrix of community membership probability vectors, and $\bbP = (p_{kl})\in \mathbb{R}^{K\times K}$ is a nonsingular matrix with $p_{kl}\in[0,1]$, $1\le k,l\le K$.  %Note that deterministic matrix $\bbH$ is assumed to be of rank $K$ as in the general model (\ref{eq: model.general}).

The family of DCMM models in (\ref{DCMM}) contains several popularly used network models for community detection as special cases. For example, when $\bTheta= \sqrt{\theta} \bI_n$  and $\bpi_i \in \{\be_1,\cdots, \be_K\}$ with $\be_k$ a unit vector whose $k$th component is one and all other components are zero,  the model reduces to the stochastic block model with non-overlapping communities. When $\bTheta =\sqrt{ \theta} \bI_n$ and $\bpi_i$'s are general community membership probability vectors, the model becomes the mixed membership model.   Each of these models has been studied extensively in the literature. Yet almost all these existing works have focused on the community detection perspective, which is a statistical estimation problem. In this paper, however we will concentrate on the statistical inference problem  \eqref{eq: hypothesis}.

\subsection{Technical preparation} \label{Sec2.2}

When $\bTheta = \sqrt{\theta} \bI_n$, we have $\bH = \theta \bPi\bbP\bPi^T$.  Thus the column space spanned by $\bPi$ is the same as the eigenspace spanned by the top $K$ eigenvectors of matrix $\bH$. In other words,  the membership profiles of the network are encoded in the eigen-structure of the mean matrix $\bbH$.
Denote by $\bH=\bbV\bbD\bbV^T$ the eigen-decomposition of the mean matrix,
where  $\bbD=\diag(d_1,\cdots,d_K)$ with {$|d_1|\geq |d_2| \geq \cdots \geq |d_K|>0$} is the matrix of all $K$ nonzero eigenvalues and {\color{black}$\bbV=(\bbv_1,\cdots,\bbv_K) \in \mathbb{R}^{n\times K}$} is the corresponding orthonormal matrix of eigenvectors. %Our model includes the popularly used network models such as the stochastic block model, the mixed membership model, and the degree corrected mixed membership  model. Throughout the paper, the number of communities $K$ is assumed to be known. We are interested in inferring the community structure from the adjacency matrix $\bbX$.
%It is well known that the matrices $\bV$ and $\bD$ carry all the information on the community structure in the network model.
In practice, one replaces the matrices $\bD$ and $\bV$ by those of the observed adjacency  matrix $\bX$.    Denote by $\widehat d_1, \cdots , \widehat d_n$ the eigenvalues of matrix $\bX$ and  $\widehat\bbv_1, \cdots, \widehat{\bbv}_n$ the   corresponding eigenvectors. Without loss of generality, assume that $|\widehat d_1| \geq |\widehat d_2| \geq \cdots \geq |\widehat d_n|$ and let $\widehat\bbV=(\widehat \bbv_1,\cdots,\widehat \bbv_K)\in \mathbb{R}^{n\times K}$. Denote by $\bbW = (w_{ij})$ and define  $\alpha_n=\{\max_{1\leq j\leq n}\sum_{i=1}^n \var(w_{ij})\}^{1/2}$, which is simply the maximum standard deviation of the {\color{black}column} sums (node degrees).

The {\color{black}asymptotic} mean of the empirical eigenvalue $\widehat{d}_k$ for $k\in [K]$ has been derived in \cite{FFHL19}, which is a population quantity $t_k$ and will be used frequently in our paper.  Its definition is somewhat complicated which we now describe as follows.
Let $a_k$ and $b_k$ be defined as
$$a_k=\begin{cases}
\frac{d_k}{1+c_0/2} & \text{ if } d_k>0\cr (1+c_0/2)d_k &\text{ if } d_k<0
\end{cases},  \quad b_k=\begin{cases}
(1+c_0/2)d_k & \text{ if } d_k>0\cr \frac{d_k}{1+c_0/2} &\text{ if } d_k<0
\end{cases},$$
where the eigen-ratio gap constant $c_0 > 0$ is given in Condition \ref{as3} in Section \ref{Sec3.1}.
For any deterministic real-valued matrices $\bbM_1$ and $\bbM_2$ of appropriate dimensions and complex number $z\neq 0$, define  %{\bf Q: readers will have no idea what are doing?  Adding a few sentences?}
\begin{equation}\label{1119.1}
\mathcal{R}(\bbM_1,\bbM_2,z)=-\frac{1}{z}\bbM_1^T\bbM_2 -\sum_{l=2}^L\frac{1}{z^{l+1}}\bbM_1^T\mathbb{E}\bbW^l\bbM_2
\end{equation}
with $L$ the smallest positive integer such that uniformly over $k \in [K]$,
\begin{equation}\label{csl}
\left(\frac{\alpha_n}{|z|}\right)^{L}\le \min\left\{\frac{1}{n^4}, \frac{1}{|z|^4}\right\}, \quad z\in [a_k,b_k],
\end{equation}
where $|z|$ denotes the modulus of complex number $z$.
We can see that as long as $\frac{|d_K|}{\alpha_n}\ge n^{\ep}$ with some positive constant $\ep$, which is guaranteed by Condition \ref{as3} and Condition \ref{cond1} (or \ref{cond5}) in Section \ref{Sec3.1} (or Section \ref{Sec3.2}), the existence of the desired positive integer $L$ can be ensured.

We are now ready to define the asymptotic mean $t_k$ of the sample eigenvalue $\widehat d_k$. For each $k\in [K]$,  define $t_k$ as the solution to equation
\begin{equation}\label{0515.3.1}
1+d_k\left\{\mathcal{R}(\bbv_k,\bbv_k,z)-\mathcal{R}(\bbv_k,\bbV_{-k},z)\left[\bbD_{-k}^{-1}+\mathcal{R}(\bbV_{-k},\bbV_{-k},z)\right]^{-1}\mathcal{R}(\bbV_{-k},\bbv_k,z)\right\}=0
\end{equation}
when restricted to the interval $z\in [a_k, b_k ]$,
where $\bbV_{-k}$ is the submatrix of $\bV$ formed by removing the $k$th column and $\bD_{-k}$ is formed by removing the $k$th diagonal entry of $\bD$.  Then as shown in \cite{FFHL19}, for each $k \in [K]$, $t_k$ is the asymptotic mean of the sample eigenvalue $\widehat d_k$ and $t_k/d_k\rightarrow 1$ as $n\rightarrow\infty$.  %These theoretical results are formally summarized in \cite{FFHL19} and
See also
Lemma \ref{lem: tk} in Section \ref{SecC.6} of Supplementary Material, where the existence, uniqueness, and asymptotic property of $t_k$'s are stated.

To facilitate the technical presentation, we further introduce some notation that will be used throughout the paper.   We use $a\ll b$ to represent $a/b\rightarrow0$. %If there exists some positive constant $C$ such that $\frac{a}{b}\in(C^{-1},C)$, we say $a\sim b$.
For a matrix $\bA = (\bbA_{ij})$, denote by $\lambda_j(\bbA)$ the $j$th largest eigenvalue, and $\|\bbA\|_F=\sqrt{\tr(\bbA\bbA^T)}$,  $\|\bA\|_2$, and $\|\bbA\|_{\infty}=\max_{i,j}|\bbA_{ij}|$ the Frobenius norm, the spectral norm, and the entrywise maximum norm, respectively. In addition, we use $\bbA(k)$ to denote the $k$th row of a matrix $\bbA$, and $\ba(k)$ to denote the $k$th component of a vector $\ba$.  For a unit vector $\bx = (x_1,\cdots, x_n)^T$, let $d_{\sbx}=\max_{1\leq i \leq n}|x_i|$. Also define $\theta_{\max}=\max_{1\leq i\leq n}\theta_i$ and $\theta_{\min}=\min_{1\leq i\leq n}\theta_i$ as the maximum and minimum node degrees, respectively. For each $1 \leq k \leq K$, denote by $\mathcal{N}_k=\{i: 1 \leq i \leq n,\,\bpi_i(k)=1\}$ the set of pure nodes in community $k$, where each pure node belongs to only a single community. Some additional definitions and notation are given at the beginning of Section \ref{SecA}.

\section{SIMPLE and its asymptotic theory} \label{Sec3}

\subsection{SIMPLE for mixed membership models} \label{Sec3.1}
We first consider the hypothesis testing problem (\ref{eq: hypothesis}) in the mixed membership model without degree heterogeneity whose mean matrix takes the form \eqref{DCMM} with $\bTheta= \sqrt{\theta} \bI_n$, that is,
\begin{equation}\label{0115.1}
\mathbb{E}\bbX = \bH =\theta\bPi\bbP\bPi^T.  \qquad  %=\bbV\bbD\bbV^T,
\end{equation}
Here $\theta$ is allowed to converge to zero as $n\rightarrow \infty$. This model is a simple version of the mixed membership stochastic block (MMSB) model considered in  \cite{Airoldi:2008}.
As mentioned before, this model includes the stochastic block model with non-overlapping communities as a special case.

Under model \eqref{0115.1}, if $\bpi_i = \bpi_j$ then nodes $i$ and $j$ are exchangeable and it holds that $\bbV(i) = \bbV(j)$ by a simple permutation argument (see the beginning of the proof of Theorem~\ref{0126-1} in Section \ref{SecA.1}). Motivated by this observation, we consider the following test statistic for assessing the membership information of the $i$th and $j$th nodes
{%\color{blue}
	\begin{align}\label{eq: T-test}
T_{ij} = \left[\widehat\bbV(i)-\widehat\bbV(j)\right]^T\bSig_1^{-1}\left[\widehat\bbV(i)-\widehat\bbV(j)\right], %\widehat{\lambda}_k(\widehat\bbv_k(i)-\widehat\bbv_k(j)).
\end{align}
where $\bSig_1$ is the asymptotic variance of $\widehat\bbV(i)-\widehat\bbV(j)$ that is challenging to derive and estimate.  Nevertheless, we will show that $\bSig_1 =  \cov[(\be_i-\be_j)^T\bW\bV\bbD^{-1}]$}  whose expression is given in  \eqref{eq: cov1-expan} later, and provide an estimator with required accuracy.

%It is seen from the above definition that $T_{ij}$ is not computable because of the unknown covariance matrix $\cov\big((\be_i-\be_j)^T\bW\bV\bD^{-1}\big)$.

We need the following regularity conditions in establishing the asymptotic null and alternative distributions of test statistic $T_{ij}$.

\begin{cond} \label{as3}
	There exists some positive constant $c_0$ such that $$\min\{\frac{|d_i|}{|d_{j}|}\ :1\le i<j\le K, d_i\neq -d_j\}\ge 1+c_0.$$ In addition, $\alpha_n\rightarrow \infty$ as $n\rightarrow \infty$. % {\bf Q: the gap is big}{\color{blue} I modified the gap.}
\end{cond}

\begin{cond} \label{cond1}
	There exist some constants $0<c_0<1$, $0\le c_2<1/2$, $0<c_1<1-2c_2$ such that
	$\lambda_K(\bPi^T\bPi)\ge c_0n$, $\lambda_K(\bbP)\ge n^{-c_2}$, and $\theta\ge n^{-c_1}$.
\end{cond}

\begin{cond} \label{cond2}
	As $n\rightarrow \infty$,	all the eigenvalues of
	$\theta^{-1} \bbD\bSig_1\bbD$ are bounded away from 0 and $\infty$.
\end{cond}

%\yfnote
{The constant $c_0$ in  Condition \ref{as3} can be replaced with some $o(1)$ term that vanishes as $n$ grows at the cost of significantly more tedious calculations in our technical analysis.  This condition is imposed to exclude the complicated case of multiplicity, which can lead to the singularity of $\bSig_1$, making our test ill-defined. A potential remedy is to use the Moore--Penrose generalized inverse of matrix $\bSig_1$ in defining $T_{ij}$, which we will leave to the future study due to the extra technical challenge.  We acknowledge that in some special models, results on community detection have been established allowing multiplicity (e.g., \cite{gao2018}).}  Conditions \ref{cond1} is a standard regularity assumption imposed for the case of mixed membership models. In particular, $\theta$ measures the degree density and is allowed to converge to zero at the polynomial rate $n^{-c_1}$ with constant $c_1$ arbitrarily close to one.  Condition \ref{cond2} is a technical condition for establishing the asymptotic properties of $T_{ij}$. We provide sufficient conditions for ensuring Condition \ref{cond2} in Section \ref{SecCCond2} of Supplementary file. As shown in the proof of Theorem \ref{0126-1}, under Conditions \ref{as3} and \ref{cond1}, we have $\var[(\bbe_i-\bbe_j)^T\bbW\bbv_k] \sim \theta$ for all $k=1,\cdots, K$, which explains the normalization factor $\theta^{-1}$ in Condition \ref{cond2}. Our conditions accommodate the case where the magnitudes of spiked eigenvalues $|d_1|, \cdots, |d_K|$ are of different orders.

%\yfnote{The discussion here need to be updated because we no longer need eigenvalues of the same orders. }
%We would like to point out that the constraints on $\theta$ imposed by Conditions \ref{cond1} and \ref{cond2} do not contradict with each other. To see this, note that by the proof of Theorem \ref{0126-1} we have $d_k\sim n\theta$, $k=1,\cdots, K$, under Condition \ref{cond1}. For the special case of $d_1=\cdots = d_K$, Condition \ref{cond2} reduces to that $\theta^{-1}\cov[(\be_i-\be_j)^T\bW\bV]$ has bounded eigenvalues. On the other hand, as shown in the proof of Theorem \ref{0126-1}, under Conditions \ref{as3} and \ref{cond1}, we have $\var[(\bbe_i-\bbe_j)^T\bbW\bbv_k] \sim \theta$ for all $k=1,\cdots, K$, which is compatible with the reduced form of Condition \ref{cond2}.

{%\color{magenta}
\begin{example}
Consider SBM with $K=2$ communities of equal sizes $n_1=n_2 = n/2$ and $n^{-c_1}\leq \theta <1$. Further assume that $\bbP$ has diagonal entries equal to $a$ and off-diagonal entries equal to $b$, with $a$ and $b$ some positive constants satisfying $a>b$. Then we have $d_1 = n(a+b)\theta $ and $d_2 = n(a-b)\theta$. Some direct calculations show that Conditions \ref{as3}--\ref{cond2} all hold.
	\end{example}
}

The following theorem summarizes the asymptotic distribution of test statistic $T_{ij}$ under the null and alternative hypotheses.

\begin{thm}\label{0126-1}
Assume that Conditions \ref{as3}--\ref{cond1} hold under the mixed membership model \eqref{0115.1}.
	\begin{enumerate}
		\item[i)]  Under the null hypothesis $H_0: \bpi_i=\bpi_j$, if in addition Condition \ref{cond2} holds, 	then we have
			\begin{eqnarray}\label{0126.1h}
			T_{ij} \toD \chi_{K}^2
			\end{eqnarray}
			as $n\rightarrow \infty$, where $\chi_{K}^2$ is the chi-square distribution with $K$ degrees of freedom.	
		
		%then the same result \eqref{0126.1}  in Theorem \ref{0104-1h}i) holds.
		%\end{thm}
		%\begin{thm}
		
		\item[ii)] Under the contiguous alternative hypothesis $H_a: \bpi_i\neq\bpi_j$ but $n^{1/2-c_2}\sqrt{\theta}\|\bpi_i-\bpi_j\|\rightarrow \infty$, then  for arbitrarily large constant $C>0$, we have
			\begin{eqnarray}\label{0120.1}
			P(T_{ij} >C)\rightarrow 1
			\end{eqnarray}
		as $n\rightarrow \infty$.
		Moreover, if Condition \ref{cond2} holds, $c_2=0$, $\|\bpi_i-\bpi_j\|\sim \frac{1}{\sqrt{n\theta}}$, and
		{%\color{blue}
			$[\bbV(i)-\bbV(j)]^T\bSig_1^{-1}[\bbV(i)-\bbV(j)]\rightarrow \mu$} with $\mu$ some constant,
		%	\textbf{is this a vector or scalar?}
		%	and
		%	$$n^2\theta Cov(\frac{(\bbe_i-\bbe_j)^T\bbW\bbv_1}{t_1},\cdots,\frac{(\bbe_i-\bbe_j)^T\bbW\bbv_K}{t_K})\rightarrow \Sigma>0,$$
		then it holds that
		\begin{eqnarray}\label{0126.1}
		T_{ij} \toD \chi_{K}^2(\mu)
		\end{eqnarray}
		as $n\rightarrow \infty$, where $\chi_{K}^2(\mu)$ is a noncentral chi-square distribution with mean $\mu$ and $K$ degrees of freedom.
	\end{enumerate}
\end{thm}

%\textcolor{red}{Xiao: can you provide some intuition on why larger $\theta$ means easier testing problem?}

\begin{remark} Under the joint null hypotheses $H_{0,ij}: \bpi_i = \bpi_j$ for all $1\leq i\neq j\leq n$, we have in fact proved a uniform version of the result in \eqref{0126.1h}:
\begin{equation}\label{eq:uniform}
\lim_{n\rightarrow \infty}\sup_{1\leq i\neq j\leq n}|P(T_{ij}\leq x) - P(X \leq x)| = 0 \ \text{ for all } x\in \mathbb R,
\end{equation}
where $X\sim \chi_K^2$.  See Section \ref{sec: uniform} of Supplementary Material for more details.
\end{remark}

In the special case of stochastic block model with non-overlapping communities, we can see that $\|\bpi_i - \bpi_j\| = 0$ under the null hypothesis $H_0$, and $\|\bpi_i - \bpi_j\| = \sqrt{2}$ under the alternative hypothesis $H_a$. Thus under the null hypothesis $H_0$ and Conditions \ref{as3}--\ref{cond2}, the test statistic $T_{ij}$ has asymptotic distribution \eqref{0126.1h}.  Under the alternative hypotheses $H_a$ and Conditions \ref{as3}--\ref{cond1}, we have $\sqrt{n\theta}\|\bpi_i-\bpi_j\| \rightarrow \infty$ and thus the limiting result \eqref{0120.1} holds. %{\bf Q: nice for consistency $K$ communities}

The test statistic $T_{ij}$ is, however, not directly applicable because of the unknown population parameters $K$ and $\bSig_1$. We next show that for consistent estimators satisfying the following conditions
%{%\color{blue}
	\begin{align}
&P(\widehat K = K) = 1-o(1), \label{eq: khat}\\
&\theta^{-1}\|\bbD(\widehat{\bS}_1 - \bSig_1)\bbD\|_2 = o_p(1),\label{0425.1}
\end{align}
%}
the asymptotic results in Theorem \ref{0126-1} continue to hold.

%\begin{definition}
%	An estimator $\widehat{K}$ of $K$ is acceptable is $P(\widehat K = K) = 1-o(1)$. An estimator $\widehat{\bS}$ of  $\bSig_1$ is acceptable if $n^2\theta\|\widehat{\bS} - \bSig_1\|_2 = o_p(1)$.
%\end{definition}

\begin{thm}\label{0519-1}
	Assume that estimators $\widehat K$ and $\widehat\bS_1$ satisfy \eqref{eq: khat} and \eqref{0425.1}, respectively. Let $\widehat{T}_{ij}$ be the test statistic constructed by replacing $K$ and $\bSig_1$ in \eqref{eq: T-test} with $\widehat K$ and $\widehat \bS_1$, respectively. Then %if $\max_{1\leq i\leq K}\|\bv_i\|_\infty^2 = O(n^{-1})$,
	 Theorem \ref{0126-1} holds with $T_{ij}$ replaced by $\widehat T_{ij}$ under the same conditions.
\end{thm}

Theorem \ref{0519-1} suggests that at significance level $\alpha$, to test the null hypothesis $H_0$ in \eqref{eq: hypothesis}, we can construct the following rejection region
\begin{equation} \label{new.eq007}
\{\widehat T_{ij}> \chi_{\widehat K, 1-\alpha}^2\},
\end{equation}
where $\chi_{\widehat K, 1-\alpha}^2$ is the $100(1-\alpha)$th percentile of the chi-square distribution with $\widehat K$ degrees of freedom.  The following corollary justifies the asymptotic size and power of our test. %By Theorem \ref{0519-1}, we can see that the resulting size will be asymptotically $\alpha$ and the power will be asymptotically one.

\begin{coro}\label{coro1}
Assume that  $\widehat K$ and $\widehat\bS_1$ satisfy \eqref{eq: khat} and \eqref{0425.1}, respectively.	Under the same conditions for ensuring \eqref{0126.1h},  event \eqref{new.eq007} holds with asymptotic probability $\alpha$. Under the same conditions for ensuring \eqref{0120.1}, event \eqref{new.eq007} holds with asymptotic probability one.
\end{coro}

\subsection{SIMPLE for degree-corrected mixed membership models} \label{Sec3.2}

In this section, we further consider the hypothesis testing problem \eqref{eq: hypothesis} in the more general DCMM model \eqref{DCMM}.
%\yfnote
{Degree heterogeneity in network models has been explored in the statistics literature. To name a few, \cite{JKL17} considered the estimation of node membership assuming the average degree of the nodes to be much larger than $\log n$.  \cite{jin2017sharp} established a sharp lower bound for the estimated node membership allowing the average node degree to diverge with the order $\log^2 n$ or faster. %The refined algorithm  to \cite{JKL17} without theoretical guarantee can be found in \cite{jin2018score+}.
	\cite{19M127} proposed a spectral-based detection algorithm to recover the node membership assuming that $\theta_{\max}/\theta_{\min}$ is bounded by some positive constant.  Our assumption on the degree heterogeneity is similar to that in 	\cite{19M127} and will be presented in Condition \ref{cond5} below.}

The test statistic $T_{ij}$ defined in Section \ref{Sec3.1} is no longer applicable due to the degree heterogeneity. A simple algebra shows that degree heterogeneity can be eliminated by the ratios of eigenvectors (columnwise division).  Thus, following \cite{jin2015}, to correct the degree heterogeneity  we define the following componentwise ratio
\begin{align}\label{def-Y}
Y(i,k)=\frac{\widehat\bv_{k}(i)}{\widehat\bv_{1}(i)}, \quad 1\le i\le n,\, 2\le k\le K,
\end{align}
where $0/0$ is defined as 1 by convention. Note that the division here is to get rid of the degree heterogeneity and the equality
\begin{equation}\label{fan1}
\frac{\bv_{k}(i)}{\bv_{1}(i)} = \frac{\bv_{k}(j)}{\bv_{1}(j)}, \qquad 2\le k\le K
\end{equation}
holds under the null hypothesis, which is due to the exchangeability of nodes $i$ and $j$ under the mixed membership model; see \eqref{1231.6h} at the beginning of the proof of Theorem~\ref{0104-1} in Section \ref{SecA.3}.
Denote by $\bbY_i=(Y(i,2),\cdots,Y(i,K))^T$. Our new test statistic will be built upon $\bbY_i$. %\textcolor{red}{Xiao: you need to change the notation in the proofs}{\color{blue} Dear Prof. Fan, in the proof of theorem 3 I used $\bbY_i$ to replace $\bbY(i)$. }

To test the null hypothesis $H_0: \bpi_i = \bpi_j$, using
\eqref{def-Y} and \eqref{fan1}, we propose to use the following test statistic
\begin{equation} \label{new.eq008}
G_{ij} = (\bbY_i-\bbY_j)^T\bSig_2^{-1}(\bbY_i-\bbY_j)
\end{equation}
for assessing the null hypothesis $H_0$ in \eqref{eq: hypothesis}, where $\bSig_2$ is the asymptotic variance of $\bbY_i-\bbY_j$.   This is even much harder to derive and estimate.  Nevertheless, we will show $\bSig_2=\cov(\bbf)$ with
$\bbf=(f_2,\cdots,f_K)^T$ and
\begin{equation} \label{new.eq002}
f_k=\frac{\be_i^T\bW\bbv_k}{ t_k\bbv_1(i)}-\frac{\be_j^T\bW\bbv_k}{ t_k\bbv_1(j)}-\frac{\bbv_k(i)\be_i^T\bW\bbv_1}{ t_1\bbv^2_1(i)}+\frac{\bbv_k(j)\be_j^T\bW\bbv_1}{t_1\bbv^2_1(j)}.
\end{equation}
The entries of $\bSig_2$ are given by \eqref{eq: cov2-expan} later that also involves the asymptotic mean of $\widehat d_k$.

%Throughout this paper, we consider the mixed membership models and the DCMM models that satisfy the following two sets of regularity conditions, respectively.
The following conditions are needed for investigating the asymptotic properties of test statistic $G_{ij}$.

\begin{cond} \label{cond5}
	There exist some constants $c_2\in[0,1/2)$, $c_3\in (0,1-2c_2)$, $c_5\in (0,1)$   and $c_4>0$ such that
	$\lambda_K(\bbP)\ge n^{-c_2}$, $\min_{1\le k\le K}|\mathcal{N}_k|\ge c_5n$, $\theta_{\max}\le c_4\theta_{\min}$, and $\theta^2_{\min}\ge n^{-c_3}$.
\end{cond}

\begin{cond} \label{cond6}
	Matrix	$\bP = (p_{kl})$ is positive definite, irreducible, and has unit diagonal entries.  Moreover $n \min_{1\le k\le K,\, t=i,j}\var(\be_t^T\bbW\bbv_k)\sim n\theta_{\max}^2\rightarrow \infty$.
\end{cond}

\begin{cond} \label{cond7}
	It holds that all the eigenvalues of
	$(n\theta_{\max}^2)^{-1}\bbD\cov(\bbf)\bbD$
	are bounded away from 0 and $\infty$. %, where matrix $\bbf$ is given in (\ref{new.eq002}). %\textcolor{red}{same here; do we really need this assumption?}{\color{blue} Similar to the previous revision, I use $\theta_{\min}$.}
\end{cond}
\begin{cond}\label{cond8}
 Let $\bleta_1$ be the first right singular vector of  $\bbP\bPi^\top\bTheta^2\bPi$. It holds that
 $$\min_{1\le k\le K}\bleta_1(k)>0,\quad \text{and}\quad \frac{\max_{1\le k\le K}\bleta_1(k)}{\min_{1\le k\le K}\bleta_1(k)}\le C,$$
 for some positive constant $C$, where $\bleta_1(k)$ is the $k$-th entry of $\bleta_1$.
\end{cond}

Conditions \ref{cond5}--\ref{cond8} are  similar to those in \cite{JKL17}.  In particular, Conditions \ref{cond5}, \ref{cond6} and \ref{cond8} are special cases of (2.13), (2.14) and (2.16) therein. Same as in the previous section, the degree density is measured by $\theta_{\min}^2$ and is allowed to converge to zero at rate $n^{-c_3}$, and our conditions accommodate the case where $|d_1|, \cdots, |d_K|$ are of different orders.

%\yfnote{An anonymous referee provided great intuition that the quantities measuring how hard the problem is, are a) the degree correction, b) the difference in the edge probabilities of nodes in the same community and nodes in different communities and c) the difference in community membership profiles of nodes $i$ and $j$. We next discuss that our conditions are indeed related to these measures. Condition \ref{cond5} requires that the degree heterogeneity cannot be too large ($\theta_{\max}/\theta_{\min} < c_1$), and the model cannot be too sparse ($\theta_{\min}^2>n^{-c_3}$).  Conditions \ref{cond5} and \ref{cond6} put constraints on the smallest eigenvalue of $\mathbf P$, which is related to b). To see the connection, consider DCSBM with $K=2$. Assume $\mathbf P$ has diagonals equal to 1, and off-diagonals equal to $a$. Then how much smaller $a$ is compared to 1 measures the  the difference in the edge probabilities of nodes in the same community and nodes in differerent
%	communities. Condition 4 on $\lambda_K(\mathbf P)$ requires that $1-a\gg n^{-c_2}$.  Finally, results in Theorem \ref{0126-1}ii) indicate that if $\|\pi_i - \pi_j\|$ is small, the power of our test may be low; and if  $\|\pi_i - \pi_j\|$ is large, the power can asymptotically tend to 1. These are consistent with c) above.  {\bf Q: do we need this?}
%}

\begin{thm}\label{0104-1}
	Assume that Conditions \ref{as3} and \ref{cond5}--\ref{cond8} hold under the degree-corrected mixed membership model \eqref{DCMM}.
	\begin{enumerate}
		\item[i)]  	Under the null hypothesis $H_0: \bpi_i = \bpi_j$, we have as $n\rightarrow \infty$,
		\begin{equation}\label{Gij-null}
			G_{ij}
		\toD \chi_{K-1}^2.
		\end{equation}

		\item[ii)] 	Under the contiguous alternative hypothesis with  $\lambda_2(\bpi_i\bpi_i^T+\bpi_j\bpi_j^T)\gg \frac{1}{n^{1-2c_2}\theta^2_{\min}}$, we have for any arbitrarily large constant $C>0$,
	\begin{equation}\label{Gij-alt}
		P(G_{ij}>C) \rightarrow 1 \text{ as } n\rightarrow \infty.
	\end{equation}  %probability tending to one,
		%$G_{ij} %= (\bbY(i)-\bbY(j))^T(\cov(\bbf))^{-1}(\bbY(i)-\bbY(j))
		%\rightarrow \infty$ as $n\rightarrow \infty$.

	\end{enumerate}
	%\end{thm}
%\begin{thm}\label{0315-1}
\end{thm}
% The proof of Theorem \ref{0104-1}ii) is almost the same as (\ref{0120.1}) and thus we omit it.

A uniform result similar to \eqref{eq:uniform} has also been proved in Section \ref{sec: uniform} of Supplementary Material under the DCMM. The test statistic $G_{ij}$ is not directly applicable in practice due to the presence of the unknown population parameters $K$ and $\bSig_2$. Nevertheless, certain consistent estimators can be constructed and the results in Theorem \ref{0104-1} remain valid. In particular, for the estimator $\widehat K$ of $K$, we require condition \eqref{eq: khat} and for the estimator $\widehat{\bbS}_2$ of $\bSig_2$, we need the following property
%{%\color{blue}
	\begin{equation}\label{0425.3}
(n\theta_{\max}^2)^{-1}\|\bbD(\widehat\bS_2 - \bSig_2)\bbD\|_2 = o_p(1).
\end{equation}
%}

\begin{thm}\label{0524-1}
	Assume that the estimators $\widehat K$ and $\widehat \bS_2$ of parameters $K$ and $\bSig_2$ satisfy \eqref{eq: khat} and \eqref{0425.3}, respectively. Let  $\widehat G_{ij}$ be the test statistic constructed by replacing $K$ and $\bSig_2$ with $\widehat K$ and $\widehat \bS_2$, respectively. Then Theorem \ref{0104-1} holds with $G_{ij}$ replaced by $\widehat G_{ij}$ under the same conditions.
\end{thm}

Theorem \ref{0524-1} suggests that with significance level $\alpha$, the rejection region can be constructed as
\begin{equation} \label{new.eq009}
\{\widehat{G}_{ij} > \chi_{\widehat K-1,1-\alpha}^2\}.
\end{equation}
We have similar results to Corollary \ref{coro1} regarding the type I and type II errors of the above rejection region. %   asymptotic size and power will be $\alpha$ and one,  respectively.

\begin{coro}\label{coro2}
	Assume that  $\widehat K$ and $\widehat\bS_2$  satisfy \eqref{eq: khat} and \eqref{0425.3}, respectively.	Under the same conditions for ensuring \eqref{Gij-null},  event \eqref{new.eq009} holds with asymptotic probability $\alpha$. Under the same conditions for ensuring \eqref{Gij-alt},  event \eqref{new.eq009} holds with asymptotic probability one.
\end{coro}

It is worth mentioning that since the DCMM model \eqref{DCMM} is more general than the mixed membership model  \eqref{0115.1}, the test statistic $\widehat G_{ij}$ can be applied even under model \eqref{0115.1}. However, as will be shown in our simulation studies in Section \ref{Sec4}, the finite-sample performance of $\widehat T_{ij}$ can be better than that of $\widehat G_{ij}$ in such a model setting, which is not surprising since the latter involves ratios (see \eqref{def-Y}) in its definition and has two sources of variations from both numerators and denominators. This is also reflected in losing one degree of freedom in \eqref{new.eq009}

\subsection{Estimation of unknown parameters} \label{Sec3.3}

%\subsection{An estimation procedure} \label{Sec7.new.1}

%Suppose that we are going to do hypothesis testing:
%$$H_0:\text{The population eigenvectors are $\bbv_1$,\cdots,$\bbv_K$}, \ H_1: \text{The population eigenvectors are not $\bbv_1$,\cdots,$\bbv_K$}.$$
%In order to use our theoretical results to solve this problem, we need to provide a good estimation of the mean and variance of $\bbu^T\widehat\bbv_k$. Since under $H_0$, the population eigenvectors are given, we only need to estimate $\bbD$ and $\bbW$.

%The test statistics $T_{ij}$  and $G_{ij}$ discussed in the previous section are non-pivotal in the sense that they depend on unknown population parameters $\cov((\be_i-\be_j)^T\bW\bV\bD^{-1})$ and $\cov(\bff)$, respectively.   We discuss how to estimate these unknown parameters in this section.

%In practice, when implementing the test $T_{ij}$, we need to estimate the unknown matrix $\cov\big((\be_i-\be_j)^T\bW\bV\bD^{-1}\big)$. The following proposition provides an expansion of this covariance matrix.

We now discuss some consistent estimators of $K$, $\bSig_1$, and $\bSig_2$ that satisfy conditions \eqref{eq: khat}, \eqref{0425.1}, and \eqref{0425.3}, respectively. There are some existing works concerning the  estimation of parameter $K$. For example, \cite{L16,  ChenLei2018, Daudin2008, Latouche-etal2012, saldana2017many, wang2017}, among others.  Most of these works consider specific network models such as the stochastic block model or degree-corrected stochastic block model. %The basic idea comes from the fact that the largest eigenvalue of a Wigner matrix asymptotically follows the Tracy--Widom distribution. Under the null hypothesis $H_0: K=K_0$,  the adjacency matrix after appropriate  rescaling can be close to the Wigner matrix and  thus the largest eigenvalue of the rescaled matrix asymptotically follows the Tracy--Widom distribution. Motivated by this, \cite{L16} introduced a goodness-of-fit test for testing $H_0: K=K_0$ by resorting to the distribution of the largest eigenvalue (in magnitude) of the rescaled affinity matrix. To estimate the number of communities $K$, they proposed to sequentially perform the goodness-of-fit test with $ K_0=1, 2, \cdots$ until failing to reject the null hypothesis. They proved that the resulting $K_0$ can consistently estimate the underlying true number of communities.

In our paper, since we consider the general  DCMM model \eqref{DCMM} which allows for mixed memberships, the existing methods are no longer applicable. To overcome the difficulty, we suggest a simple thresholding estimator defined as
\begin{equation}\label{eq: k-threshold}
\widehat K=\Big|\Big\{\widehat d_i: \ \widehat{d}_i^2>2.01(\log n) \check d_n ,  i \in [n] \Big\} \Big|,
\end{equation}
where $|\cdot|$ stands for the cardinality of a set, the constant 2.01 can be replaced with any other constant that is slightly larger than 2, and $\check d_n = \max_{1 \leq l \leq n}\sum_{j=1}^nX_{lj}$ is the maximum degree of the network. That is, we count the number of eigenvalues of matrix $\bX$ whose magnitudes exceed a certain threshold.
The following lemma justifies the consistency of $\widehat K$ defined in \eqref{eq: k-threshold} as an estimator of the true number of communities $K$.

\begin{lem} \label{prop: k-threshold}
Assume that Condition \ref{as3} holds, $|d_K|\gg \sqrt{\log(n)} \alpha_n$ %, $\max_{1\leq i,j \leq n} \mathbb{E}X_{ij}<1$,
 and $\alpha_n\ge n^{c_5}$ for some positive constant $c_5$. Then $\widehat K$ defined in \eqref{eq: k-threshold} is consistent, that is, it satisfies condition \eqref{eq: khat}.
\end{lem}

Observe in Theorems \ref{0126-1}--\ref{0524-1} that we need the condition of $K\geq 1$ for test statistic $\widehat T_{ij}$  and the condition of $K\geq 2$ for test statistic $\widehat {G}_{ij}$. Motivated by such an observation, we propose to use $ \max\{\widehat K, 1\}$  and $ \max\{\widehat K, 2\}$   as the estimated number of communities  in implementing test statistics $\widehat T_{ij}$ and $\widehat G_{ij}$, respectively.

We next discuss the estimation of $\bSig_1$ and $\bSig_2$.
%\yfnote{the definitions of $\Sig_1$ and $\Sig_2$ were changed. Have you updated the theoretical results below?}{\color{blue}The definition of $\Sigma_2$ is not changed, we modified Condition \ref{cond7}. The proof of theorem 5 is modified in Section \ref{SecA.5}.}
The following two lemmas provide the expansions of these two matrices which serve as the foundation for our proposed estimators.

\begin{lem} \label{0516-2}
The $(a,b)$th entry of matrix $\bSig_1$ is given by
	\begin{equation}\label{eq: cov1-expan}
	%{%\color{blue}
		\frac{1}{d_ad_b}\left\{\sum_{t\in \{i,j\}}\sum_{l=1}^n\sigma^2_{tl}\bbv_a(l)\bbv_b(l)-\sigma^2_{ij}\left[\bbv_a(j)\bbv_b(i)+\bbv_a(i)\bbv_b(j)\right]\right\},
	%}
	\end{equation}
	where $\sigma_{ab}^2=\var(w_{ab})$ for $1\leq a,b\leq n$.
\end{lem}

\begin{lem} \label{0516-3}
The $(a,b)$th entry of matrix $\bSig_2$ is given by
\begin{align}
\nonumber &\frac{1}{t_1^2}\Big\{\sum_{l=1,\,l\neq j}^n\sigma^2_{il}\left[\frac{t_1\bbv_{a+1}(l)}{t_{a+1}\bbv_1(i)}-\frac{\bbv_{a+1}(i)\bbv_1(l)}{\bbv_1(i)^2}\right]\left[\frac{t_1\bbv_{b+1}(l)}{t_{b+1}\bbv_1(i)}-\frac{\bbv_{b+1}(i)\bbv_1(l)}{\bbv_1(i)^2}\right]\\
\nonumber &\quad+\sum_{l=1,\,l\neq i}^n \sigma^2_{jl}\left[\frac{t_1\bbv_{a+1}(l)}{t_{a+1}\bbv_1(j)}-\frac{\bbv_{a+1}(j)\bbv_1(l)}{\bbv_1(j)^2}\right]\left[\frac{t_1\bbv_{b+1}(l)}{t_{b+1}\bbv_1(j)}-\frac{\bbv_{b+1}(j)\bbv_1(l)}{\bbv_1(j)^2}\right] \\
\nonumber &\quad+\sigma^2_{ij}\left[\frac{t_1\bbv_{a+1}(j)}{t_{a+1}\bbv_1(i)}-\frac{\bbv_{a+1}(i)\bbv_1(j)}{\bbv_1(i)^2}-\frac{t_1\bbv_{a+1}(i)}{t_{a+1}\bbv_1(j)}+\frac{\bbv_{a+1}(j)\bbv_1(i)}{\bbv_1(j)^2}\right]\\
& \quad \times\left[\frac{t_1\bbv_{b+1}(j)}{t_{b+1}\bbv_1(i)}-\frac{\bbv_{b+1}(i)\bbv_1(j)}{\bbv_1(i)^2}-\frac{t_1\bbv_{b+1}(i)}{t_{b+1} \bbv_1(j)}+\frac{\bbv_{b+1}(j)\bbv_1(i)}{\bbv_1(j)^2}\right]\Big\}. \label{eq: cov2-expan}
\end{align}
\end{lem}

The above expansions in Lemmas \ref{0516-2}--\ref{0516-3} suggest that the covariance matrices $\bSig_1$ and $\bSig_2$ can be estimated by plugging in the sample estimates to replace the unknown population parameters. In particular,  $\bbv_a$ and $d_a$ can be estimated by $\widehat\bbv_a$ and $\widehat d_a$, respectively, and the last result in Lemma \ref{lem: tk} suggests that $t_k$ can be estimated by $\widehat d_k$ very well.  The estimation of $\sigma_{ab}^2$ is more complicated and we will discuss it in more details below.

Recall that $\sigma_{ab}^2 = \var(w_{ab})$.   With estimated $\widehat K$, a naive estimator of $\sigma_{ab}^2$ is $\widehat w_{0, ab}^2$ with $\widehat\bbW_0  = (\widehat w_{0,ab}) =\bbX-\sum_{k=1}^{\widehat K}\widehat d_k\widehat\bbv_k\widehat\bbv_k^T$.  The good news is that it appears in \eqref{eq: cov1-expan} and \eqref{eq: cov2-expan} in the form of the average and hence the variance will be averaged out.
However, this estimator is not good enough to make \eqref{0425.1} and \eqref{0425.3} hold due to the well-known fact that $\hat{d}_k$ is biased up. Thus we propose the following one-step refinement procedure to estimate $\sigma_{ab}^2$, which is motivated from the higher-order asymptotic expansion of empirical eigenvalue $\widehat d_k$  in our theoretical analysis and shrinks $\widehat d_k$ to make the bias at a more reasonable level.

\begin{enumerate}
	\item[1).] Calculate the initial estimator $\widehat\bbW_0  =\bbX-\sum_{k=1}^{\widehat K}\widehat d_k\widehat\bbv_k\widehat\bbv_k^T$.

\item[2).] With the initial estimator $\widehat\bW_0 $,
update the estimator of eigenvalue $d_k$ as
	$$\widetilde d_k=\Big[\frac{1}{\widehat d_k}+\frac{\widehat\bbv_k^T \diag(\widehat\bbW_0^2)\widehat\bbv_k}{\widehat d_k^3}\Big]^{-1}.$$

\item[3).] Then update the  estimator of $\bbW$ as $\widehat\bbW\equiv  (\widehat w_{ij}) =\bbX-\sum_{k=1}^{\widehat K}\widetilde d_k\widehat\bbv_k\widehat\bbv_k^T$ and estimate $\sigma_{ab}^2$ as $\widehat{\sigma}_{ab}^2 = \widehat w_{ab}^2$.

\end{enumerate}

To summarize, we propose to estimate matrix $\bSig_1$ by replacing $d_k$, $\bbv_k$, and $\sigma_{ab}^2$ with $\widehat d_k$, $\widehat \bbv_k$, and $\widehat{\sigma}_{ab}^2$, respectively, in \eqref{eq: cov1-expan}. The covariance matrix $\bSig_2$ can be estimated in a similar way by replacing $t_k$, $\bbv_k$, and $\sigma_{ab}^2$ with $\widehat d_k$, $\widehat\bbv_k$, and $\widehat\sigma_{ab}^2$, respectively, in \eqref{eq: cov2-expan}. Denote by $\widehat{\bS}_1$ and $\widehat{\bS}_2$ the resulting estimators, respectively. The following lemma justifies the effectiveness of these two estimators.

\begin{thm}\label{0425-1}
Under Conditions \ref{as3}--\ref{cond2}, estimator $\widehat{\bS}_1$ satisfies condition \eqref{0425.1}. Under Conditions \ref{as3}  and \ref{cond5}--\ref{cond8}, estimator  $\widehat\bS_2$ satisfies condition \eqref{0425.3}.
%\begin{equation}\label{0425.1}
%n\theta^2\|\breve\Sigma-\Sigma\|=o_p(1).
%\end{equation}
%\begin{equation}\label{0425.1}
%n^2\theta\|\widehat\bS_1-\Sigma_1\|=o_p(1).
%\end{equation}
%\begin{equation}\label{0425.3}
%n\theta^2_{\min}\|\widehat\bS_2-\Sigma_2\|=o_p(1),
%\end{equation}
%\textcolor{red}{Xiao: we need to be more specific about the conditions above.}
\end{thm}

%\begin{thm}\label{0425-1}
%Under the conditions given in the corresponding applications, we have
%%\begin{equation}\label{0425.1}
%%n\theta^2\|\breve\Sigma-\Sigma\|=o_p(1).
%%\end{equation}
%\begin{equation}\label{0425.1}
%\|\widehat\bS_1-\Sigma_1\|=o_p(1).
%\end{equation}
%\begin{equation}\label{0425.3}
%\|\widehat\bS_2-\Sigma_2\|=o_p(1),
%\end{equation}
%\end{thm}

%\subsection{Applications of SIMPLE} \label{Sec3.4}
%Some potential applications of SIMPLE.

\section{Simulation studies} \label{Sec4}
We use simulation examples to examine the finite-sample performance of our new SIMPLE test statistics $\widehat T_{ij}$ and $\widehat G_{ij}$ with true and estimated numbers of communities $K$, respectively. In particular, we consider the following two model settings.

\textit{Model 1: the mixed membership model \eqref{0115.1}}. We consider $K=3$ communities, where there are $n_0$ pure nodes within each community. Thus for the $k$th community, the community membership probability vector for each pure node is $\bpi = \be_k \in \mathbb{R}^K$.  The remaining $n-3n_0$ nodes are divided equally into 4 groups, where within the $l$th group  all nodes have mixed memberships with community membership probability vector $\ba_l$, $l=1,\cdots, 4$. We set $\ba_1 = (0.2,0.6,0.2)^T$, $\ba_2 = (0.6,0.2,0.2)^T$, $\ba_3 = (0.2, 0.2, 0.6)^T$, and $\ba_4 = (\frac{1}{3},\frac{1}{3},\frac{1}{3})^T$. Matrix $\bbP$ has diagonal entries one and  $(i,j)$th entry equal to $\frac{\rho}{|i-j|}$ for $i\neq j$. We experiment with two sets of parameters $(\rho,n,n_0)=(0.2,3000,500)$ and $(0.2,1500,300)$, and vary the value of  $\theta$ from $ 0.2$ to $0.9$ with step size 0.1. It is clear that parameter $\theta$ has direct impact on the average degree and hence measures the signal strength.
	
\textit{Model 2: the DCMM model \eqref{DCMM}}. Both matrices $\bPi$ and $\bbP$ are the same as in Model 1. For the degree heterogeneity matrix $\bTheta=\diag(\theta_1,\cdots,\theta_n)$, we simulate $\frac{1}{\theta_i}$ as independent and identically distributed (i.i.d.) random variables from the uniform distribution on $[\frac{1}{r},\frac{2}{r}]$ with $r\in (0,1]$. We consider different choices of $r$ with $r^2\in \{0.2,0.3,0.4,0.5,0.6,0.7,0.8,0.9\}$. We can see that as parameter $r^2$ increases, the signal becomes stronger.

%\subsection{Justify Theorem \ref{0126-1}}\label{simu1}
%Recalling the model (\ref{0115.1}), for $\bold\Pi$, we choose 4 different mixed memberships $(x,1-2x,x)$, $(1-2x,x,x)$, $(x,x,1-2x)$ and $(\frac{1}{3},\frac{1}{3},\frac{1}{3})$. For each membership we have $\frac{n-3n_0}{4}$ nodes. The remaining nodes are pure memberships $(1,0,0)$, $(0,1,0)$ and $(0,0,1)$, each with $n_0$ nodes. We choose $(K,x,\rho,n,n_0)=(3,0.2,0.2,2000,500)$. The matrix $\bbP$ has diagonal $1$ and  $(i,j)$-th entry being $\frac{\rho}{|i-j|}$, $i\neq j$.

\subsection{Hypothesis testing with $K$ known} \label{Sec4.1}
Recall that our test statistics are designed to test the membership information for each preselected pair of nodes $(i,j)$ with $1\leq i \neq j\leq n$. To examine the empirical size of our tests, we preselect $(i,j)$ as two nodes with community membership probability vector $(0.2,0.6,0.2)^T$. To examine the empirical power of our tests, we preselect $i$ as a node with community membership probability vector $(0.2,0.6,0.2)^T$ and $j$ as a node with community membership probability vector $(0,1,0)^T$. The nominal significance level is set to be 0.05 when calculating the critical points and the number of repetitions is chosen as 500.

\begin{figure}[h]
	\begin{center}
	\begin{subfigure}{\textwidth}
\includegraphics[scale=0.55, trim=00 40 10 30, clip]{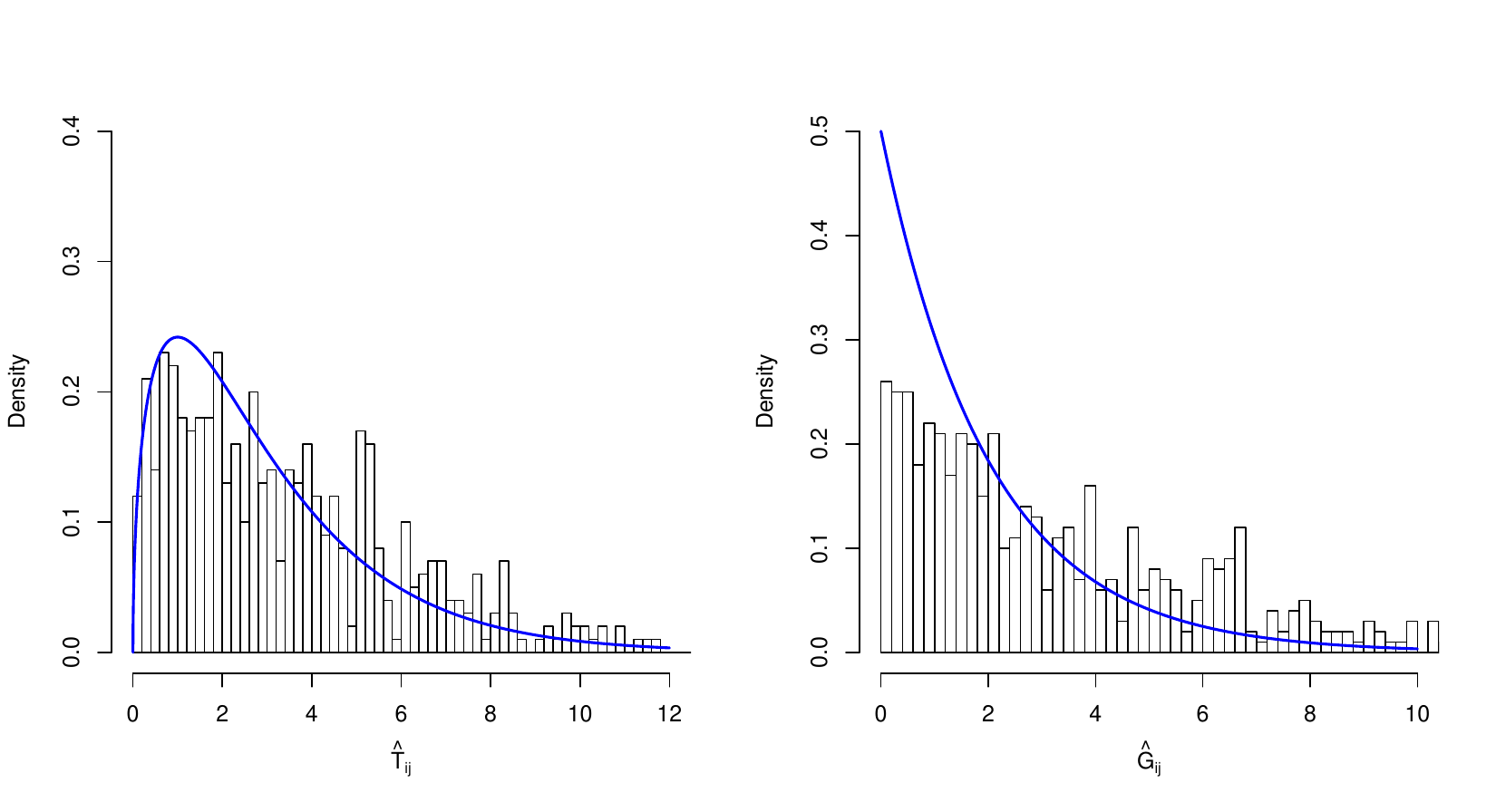}
\end{subfigure}
\begin{subfigure}{\textwidth}
	\includegraphics[scale=0.55, trim=00 40 10 30, clip]{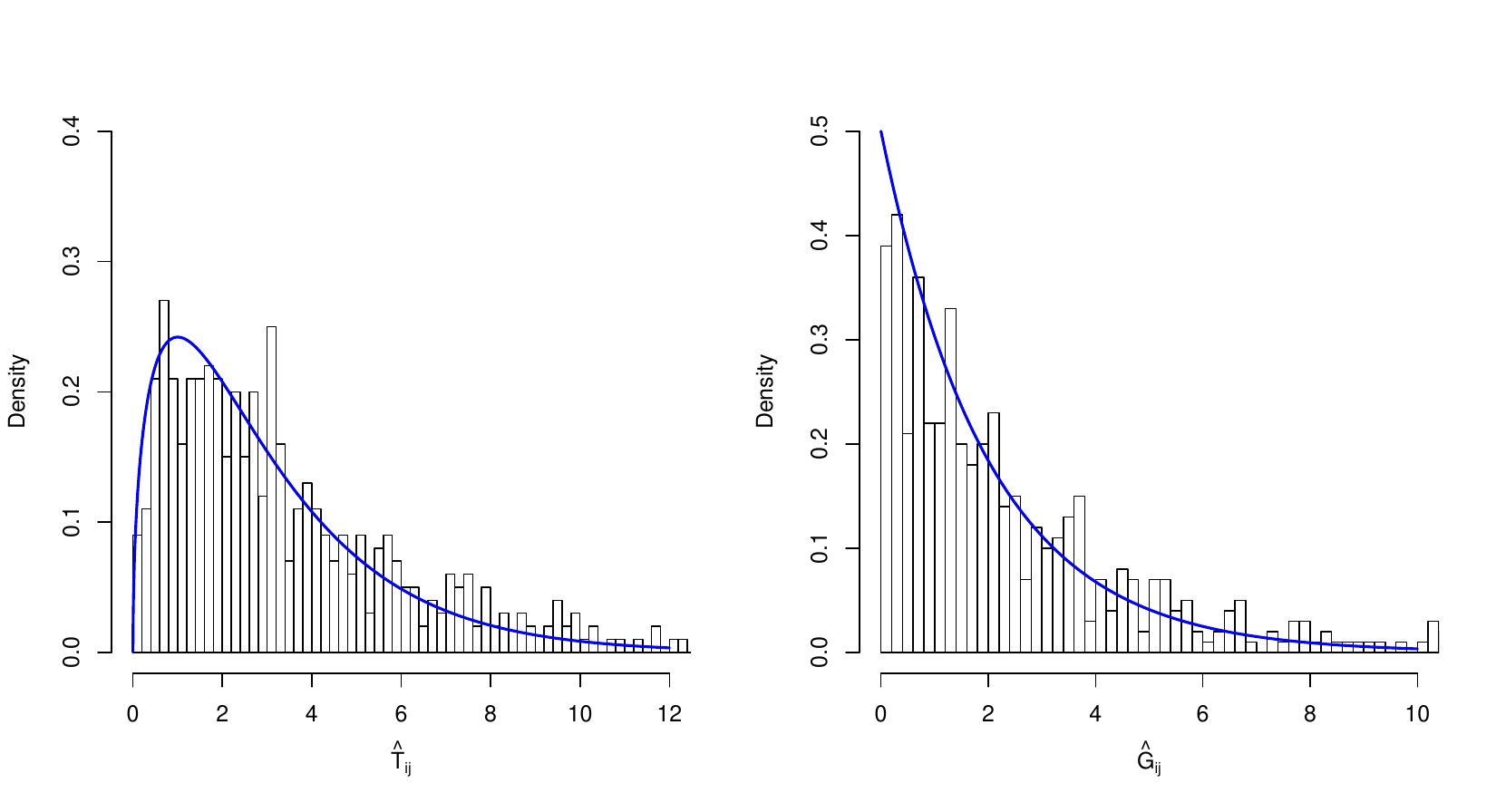}
\end{subfigure}
	\end{center}
	\caption{Left: the histogram of test statistic $\widehat T_{ij}$ under null hypothesis with known $K$ when $\theta=\frac{1}{2\log n}$. Blue curve is the density function of $\chi_{3}^2$. Right: the histogram of test statistic $\widehat G_{ij}$ under null hypothesis with known $K$ when $r^2=\frac{1}{2\log n}$. Blue curve is the density function of $\chi_{2}^2$. Top panel is for sample size $n=1500$ and bottom panel is for sample size $n=3000$. Here $n_0=\frac{n}{5}$.}\label{fig1}
\end{figure}

We first generate simulated data from Model 1 introduced above and examine the empirical size and power of test statistic $\widehat T_{ij}$ with estimated  $\bSig_1$, but with the true value of $K$.
Then we consider Model 2 and examine the empirical size and power of test statistic $\widehat G_{ij}$ with estimated $\bSig_2$ and the true value of $K$. The empirical size and power at different signal levels are reported in Tables \ref{tab1h} and \ref{tab1}, corresponding to sample sizes $n=1500$ and 3000, respectively.  %We also plot the histogram of test statistic $\widehat{T}_{ij}$ for the case of $\theta=0.9$ and the histogram of test statistic $\widehat G_{ij}$ for the case of $r^2=0.9$ in Figure \ref{fig1} under the null hypothesis when $n=3000$.
As shown in Tables \ref{tab1h} and \ref{tab1}, the size and power of our tests converge quickly to the nominal significance level $0.05$ and the value of one, respectively, as the signal strength $\theta$ (related to effective sample size) increases. As demonstrated in Figure \ref{fig1}, the empirical null distributions are well described by our theoretical results. These results provide stark empirical evidence supporting our theoretical findings, albeit complicated formulas \eqref{eq: cov1-expan} and \eqref{eq: cov2-expan}.

\begin{table}[h]
	\centering
	\caption{The size and power of test statistics $\widehat T_{ij}$ and $\widehat G_{ij}$ when the true value of $K$ is used. The nominal level is 0.05 and sample size is $n=1500$. }\label{tab1h}
	\begin{tabular}{l|lllllllll}
		\toprule
		\multirow{3}{*}{Model 1} & $\theta$ &  0.2  & 0.3   & 0.4   & 0.5   & 0.6   & 0.7   & 0.8& 0.9  \\ \cline{2-10}
		& Size     & 0.058  & 0.046  & 0.06  & 0.05 & 0.05 &0.058& 0.036  & 0.05 \\
		& Power    &  0.734 & 0.936    & 0.986     & 0.998     & 1     & 1     & 1 &1   \\ \toprule
		\multirow{3}{*}{Model 2} & $r^2$    &  0.2  & 0.3   & 0.4   & 0.5   & 0.6   & 0.7   & 0.8& 0.9  \\ \cline{2-10}
		& Size     &  0.076  & 0.062  & 0.072  & 0.062 & 0.074 &0.046& 0.044  & 0.056 \\
		& Power    &  0.426 & 0.562    & 0.696    & 0.77    & 0.89     & 0.93     & 0.952 &0.976   \\ \toprule
	\end{tabular}
\end{table}

% Please add the following required packages to your document preamble:
% \usepackage{multirow}
\begin{table}[h]
	\centering
	\caption{The size and power of test statistics $\widehat T_{ij}$ and $\widehat G_{ij}$ when the true value of $K$ is used. The nominal level is 0.05 and sample size is $n=3000$. }\label{tab1}
	\begin{tabular}{l|lllllllll}
		\toprule
		\multirow{3}{*}{Model 1} & $\theta$ &  0.2  & 0.3   & 0.4   & 0.5   & 0.6   & 0.7   & 0.8& 0.9  \\ \cline{2-10}
		& Size     &  0.082  & 0.066  & 0.052  & 0.052 & 0.044 &0.042& 0.038  & 0.062 \\
		& Power    &  0.936 & 0.994     & 1     & 1     & 1     & 1     & 1 &1   \\ \toprule
		\multirow{3}{*}{Model 2} & $r^2$    &  0.2  & 0.3   & 0.4   & 0.5   & 0.6   & 0.7   & 0.8& 0.9  \\ \cline{2-10}
		& Size     &  0.082  & 0.06  & 0.062  & 0.058 & 0.062 &0.066& 0.064  & 0.06 \\
		& Power    &  0.67 & 0.842     & 0.918     & 0.972     & 0.99     & 1     & 1 &1    \\ \toprule
	\end{tabular}
\end{table}

Figure \ref{fig1} presents how the asymptotic null distributions change with sample size $n$ when $\theta = 1/(2\log n)$ and $r^2 = 1/(2\log n)$, respectively, for Model 1 and Model 2. It is seen that the network become sparser as its size increases. The top panel shows the histogram plots when $n=1500$ and the bottom panel corresponds to $n=3000$. One can observe that as sample size increases, the $\chi^2$-distribution fits the empirical null distribution better, which is consistent with our theoretical results.

\subsection{Hypothesis testing with estimated $K$} \label{Sec4.2}

We now examine the finite-sample performance of our test statistics $\widehat{T}_{ij}$ and estimated $\widehat{G}_{ij}$ with estimated $K$. The simulation settings are identical to those in Section \ref{Sec4.1} except that we explore only the setting with sample size $n=3000$.

 In Table \ref{tab5}, we report the proportion of correctly estimated $K$ using the thresholding rule \eqref{eq: k-threshold} in both simulation settings of Models 1 and 2. It is seen that as the signal becomes stronger (i.e., as $\theta$ or $r^2$ increases), the estimation accuracy becomes higher. We also observe that for relatively weak signals, the thresholding rule in \eqref{eq: k-threshold} tends to underestimate $K$, resulting in low estimation accuracy.  We can see from the same table that over all repetitions, $K$ is either correctly estimated or underestimated.  %Recall that our goal in this paper is to test the community membership information using test statistics $\widehat T_{ij}$ and $\widehat G_{ij}$.
  The critical values are constructed based on these estimated values of $K$.

\begin{table}[htbp]
	\centering
	\caption{Estimation accuracy of $K$ using the thresholding rule \eqref{eq: k-threshold}}
	\begin{tabular}{l|lllllllll}
		\toprule
		& $\theta$ or $r^2$       & 0.2  & 0.3   & 0.4   & 0.5   & 0.6   & 0.7   & 0.8&0.9 \\
		\hline
		\multicolumn{1}{l|}{Model 1} & $P(\widehat{K}=K)$   & 1 & 1  & 1  & 1  & 1  & 1  & 1 &1 \\
	&$P(\widehat{K}\leq K)$ 	 & 1 & 1  & 1  & 1  & 1  & 1  & 1 &1\\
	\toprule
		\multicolumn{1}{l|}{Model 2} &$P(\widehat{K}=K)$ &  0  & 0  & 0  & 1 & 1  & 1 & 1  &1\\
			&$P(\widehat{K}\leq K)$ 	 & 1 & 1  & 1  & 1  & 1  & 1  & 1 &1\\
		\toprule
	\end{tabular}\label{tab5}
\end{table}%
%\begin{table}[htbp]
%	\centering
%	\caption{Ratio of $\widehat K \le K $}
%	\begin{tabular}{rrrrrrrrrr}
%		\toprule
%		$\theta$( $r^2$)     & 0.1  & 0.2  & 0.3   & 0.4   & 0.5   & 0.6   & 0.7   & 0.8&0.9 \\
%		\midrule
%		\multicolumn{1}{l}{Model 1} &1  & 1 & 1  & 1  & 1  & 1  & 1  & 1 &1 \\
%		\multicolumn{1}{l}{Model 2} & 1  & 1 & 1  & 1  & 1  & 1  & 1  & 1 &1\\
%		
%		\bottomrule
%	\end{tabular}\label{tab7}
%\end{table}%

\begin{table}[h]
	\centering
	\caption{The size and power of test statistics $\widehat T_{ij}$ and $\widehat G_{ij}$ when the estimated value of $K$ is used. The nominal level is 0.05 and sample size is $n=3000$.   }\label{tab2}
	\begin{tabular}{l|lllllllll}
		\toprule
		\multirow{3}{*}{Model 1} & $\theta$   & 0.2  & 0.3   & 0.4   & 0.5   & 0.6   & 0.7   &0.8& 0.9  \\ \cline{2-10}
		& Size       & 0.082& 0.066& 0.052 & 0.052 & 0.044  & 0.042 & 0.038&0.062 \\
		& Power     &0.936 &0.994 &1    & 1     & 1     & 1     & 1 &1    \\ \toprule
		\multirow{3}{*}{Model 2} & $r^2$      & 0.2  & 0.3   & 0.4   & 0.5   & 0.6   & 0.7   &0.8& 0.9  \\ \cline{2-10}
		& Size      & 0.054& 0.058& 0.062 & 0.058& 0.062 & 0.066   & 0.064&0.06 \\
		& Power     &0.074 &0.042 &0.918    & 0.972     & 0.99     & 1     & 1&1    \\ \toprule
	\end{tabular}
\end{table}

Same as in Section \ref{Sec4.1}, we also examine the empirical size and power of our tests at different levels of signal strength. The results are presented in Table \ref{tab2}. It is seen that the performance of $\widehat T_{ij}$  is identical to that in Table \ref{tab1}, and the performance of $\widehat G_{ij}$ is the same as in Table \ref{tab1} for all $r^2>0.3$. This is expected because of the nearly perfect estimation of $K$ as shown in Table \ref{tab5} in these scenarios and/or the relatively strong signal strength.  When $r^2\leq 0.3$, $\widehat G_{ij}$ has poor power because of the underestimated $K$ (see Table \ref{tab5}).  %  which is expected because of the additional estimation error in estimating $K$.
Nevertheless,  we observe the same trend as the signal strength increases, which provides support for our theoretical results. We have also applied our tests to nodes with more distinct membership probability vectors $(0.2,0.6,0.2)^T$ and $(0,0,1)^T$, and the impact of estimated $K$ is much smaller. These additional simulation results are available upon request.

%We then  substitute $\widehat K$ into $T_{ij}$ and $G_{ij}$ to justify Theorems  \ref{0126-1} and \ref{0104-1}. i.e. For each generated adjacency matrix $\bbX$, we estimate $\widehat K$ first and calculate $T_{ij}$ and $G_{ij}$ based on the first $\widehat K$ eigenvectors. Noticing that for $\theta=0.01$ in Table \ref{tab6}, the ratio of rejection is $0.0700$ corresponding to the null hypothesis although $\widehat K$ is not equal to $K$. The reason why we observe this phenomenon is that the estimated $\widehat K$ is smaller than $K$ and therefore $T_{ij}(\widehat K)$ is smaller than $T_{ij}(K)$. It is  harder for us to reject the null hypothesis for  $\widehat K<K$.

%\begin{table}[h]
%	\centering
%		\caption{Justifying Theorems  \ref{0126-1} and \ref{0104-1} by $\max\{\widehat K,2\}$, n=3000.   }\label{tab2}
%	\begin{tabular}{l|llllllllll}
%		\hline\hline
%		\multirow{3}{*}{Model 1} & $\theta$ & 0.1  & 0.2  & 0.3   & 0.4   & 0.5   & 0.6   & 0.7   &0.8& 0.9  \\ \cline{2-11}
%		& Size     & 0.026  & 0.054& 0.074& 0.064 & 0.056 & 0.044  & 0.052 & 0.058&0.036 \\
%		& Power    & 1 &1 &1 &1    & 1     & 1     & 1     & 1 &1    \\ \hline\hline
%		\multirow{3}{*}{Model 2} & $r^2$    & 0.1  & 0.2  & 0.3   & 0.4   & 0.5   & 0.6   & 0.7   &0.8& 0.9  \\ \cline{2-11}
%		& Size     & 0.038  & 0.002& 0.016& 0.058 & 0.056 & 0.044  & 0.064 & 0.052&0.06 \\
%		& Power    & 0.944 &1 &1 &1    & 1     & 1     & 1     & 1&1    \\ \hline\hline
%	\end{tabular}
%\end{table}

\section{Real data applications} \label{Sec5}

\subsection{U.S. political data} \label{Sec5.2}
The U.S. political data set consists of 105 political books sold by an online bookseller in the year of 2004. Each book is represented by a node and links between nodes represent the frequency of co-purchasing of books by the same buyers. The network was compiled by V. Krebs (source: \url{http://www.orgnet.com}). The books have been assigned manually three labels (conservative, liberal, and neutral) by M. E. J. Newman based on the reviews and descriptions of the books. Note that such labels may not be very accurate. In fact, as argued in multiple papers (e.g., \cite{KF08}), the mixed membership model may better suit this data set.

Since our SIMPLE tests $\widehat T_{ij}$ and $\widehat G_{ij}$ do not differentiate network models with or without mixed memberships,  we will view the network as having $K=2$ communities (conservative and liberal) and treat the neutral nodes as having mixed memberships. To connect our results with the literature, we consider the same 9 books reported in \cite{JKL17}. Another reason of considering the same 9 books as in \cite{JKL17} is that our test statistic $\widehat G_{ij}$ is constructed using the SCORE statistic which is closely related to \cite{JKL17}.
The book names as well as labels (provided by Newman) are reported in Table \ref{tab10}.
 The p-values based on test statistics $\widehat T_{ij}$ and $\widehat{G}_{ij}$ for  testing the pairwise membership profiles of these 9 nodes are summarized in Tables \ref{tab8} and \ref{tab9}, respectively.
 %\textcolor{red}{Maybe we should remove Table 8 because Table 9 is easier to justify if we compare to the results in Jin et al.}

 From Table \ref{tab9}, we see that our results based on test statistic $\widehat G_{ij}$ are mostly consistent with the labels provided by Newman and also very consistent with those in Table 5 of \cite{jin2015}. For example, books 59 and 50 are both labeled as ``conservative" by Newman and our tests return large p-values between them. These two books generally have much smaller p-values with books labeled as ``neutral." Book 78, which was labeled as ``conservative" by Newman, seems to be more similar to some neutral books. This phenomenon was also observed in \cite{JKL17}, who interpreted this as a result of having a liberal author.   Among the nodes labeled by Newman as ``neutral,"  ``All the Shah's Men," or book 29, has relatively larger p-values with conservative books. However, this book has even larger p-values with some other neutral books such as book 104, ``The Future of Freedom," which is consistent with the results in \cite{JKL17} who reported that these two books have very close membership probability vectors. In summary, our SIMPLE method provides statistical significance for the membership probability vectors estimated in \cite{JKL17}.

\begin{figure}[ht]
	\centering
				\begin{subfigure}[b]{0.46\textwidth}
		\includegraphics[scale=0.4, trim=85 190 0 120,clip]{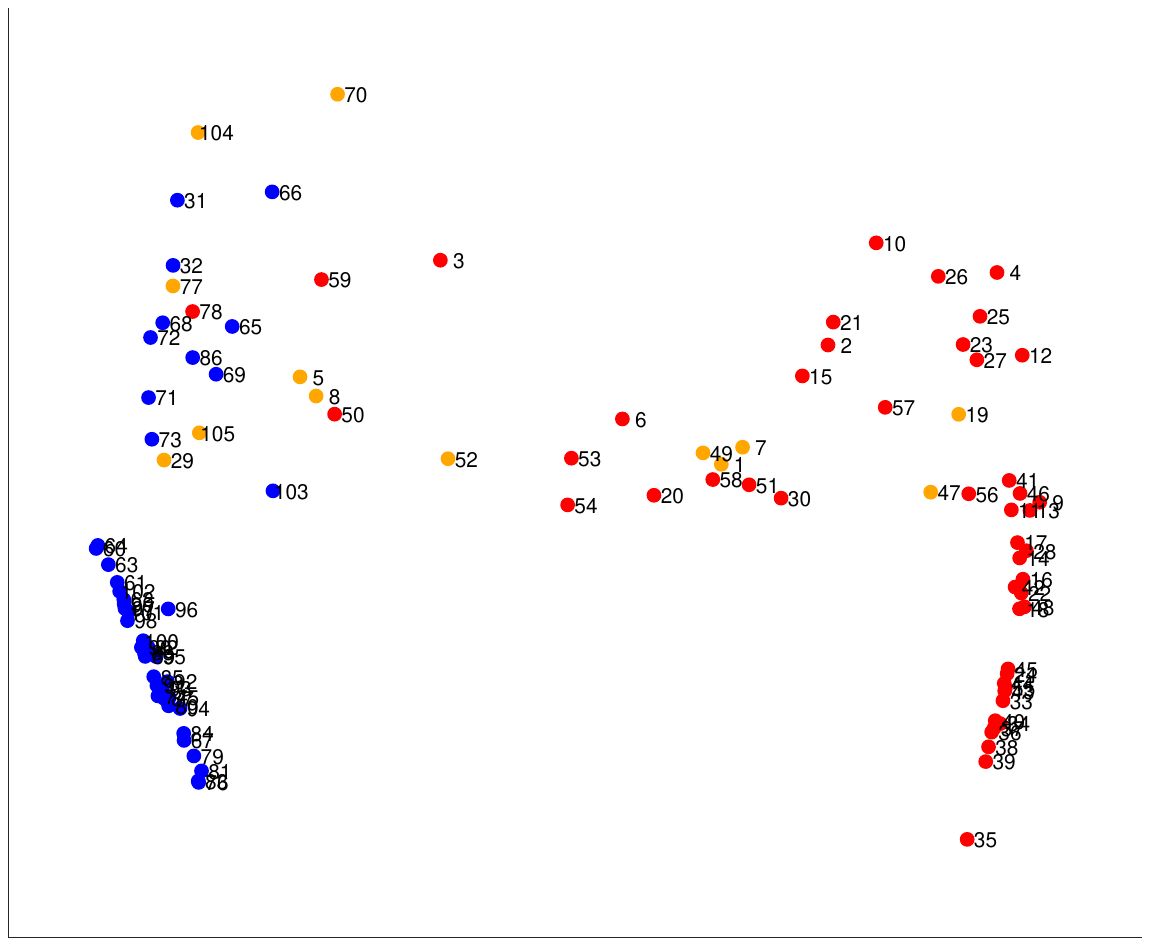}
	\end{subfigure}
			\begin{subfigure}[b]{0.46\textwidth}
					\includegraphics[scale=0.35,trim=55 160 30 120,clip]{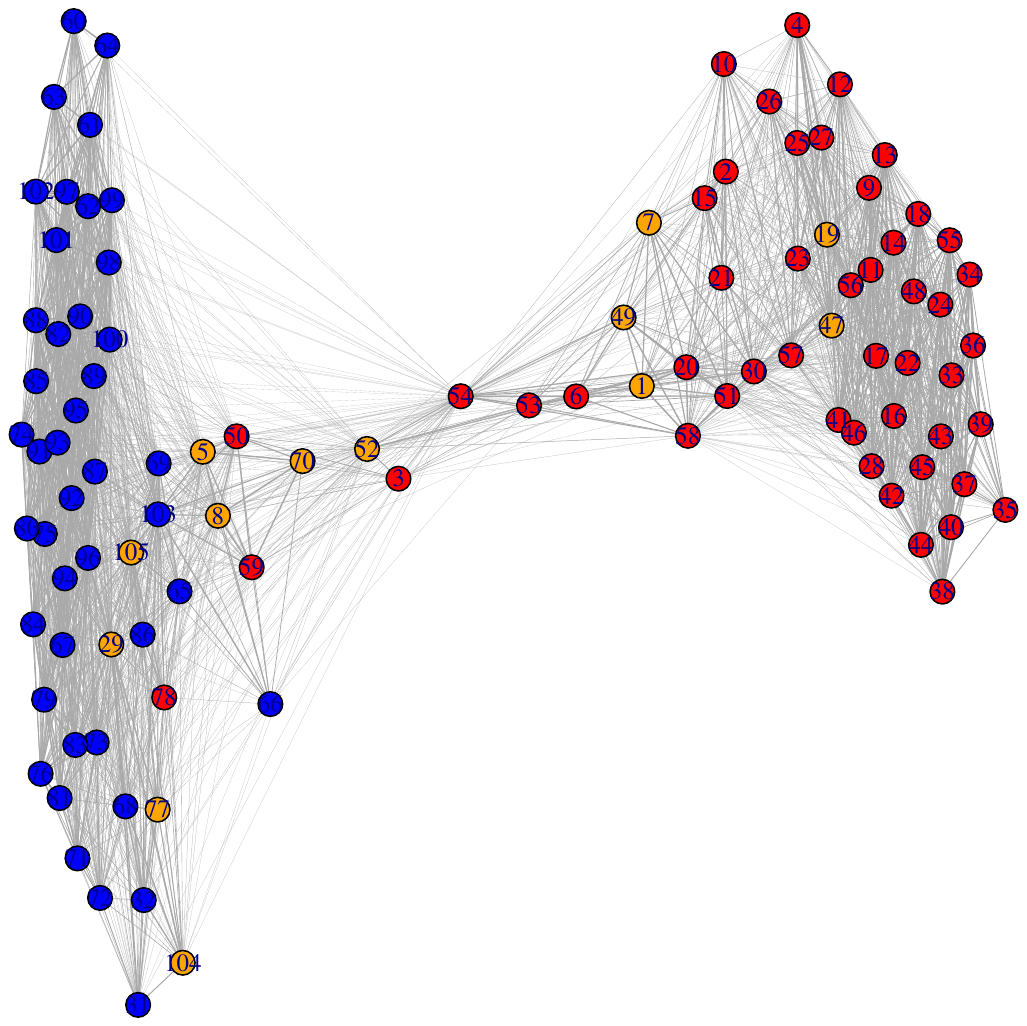}
			\end{subfigure}
	\caption{Left panel: the multidimensional scaling map of the nodes based on test statistics $\widehat G_{ij}$. Right panel: the connectivity graph generated from the thresholded p-valuate matrix based on $\widehat G_{ij}$. The nodes are color coded according to Newman's labels, with red representing ``conservative," blue representing ``liberal," and orange representing ``neutral."
	}\label{fig: Gplot}
\end{figure}

For a summary of our testing results, we also provide the multidimensional scaling map of the nodes based on test statistics $\widehat G_{ij}$ on the left panel of Figure \ref{fig: Gplot}. The graph on the right panel of Figure \ref{fig: Gplot} is defined by the pairwise p-value matrix calculated from $\widehat G_{ij}$. Specifically, we first apply the hard-thresholding to the p-value matrix by setting all entries below 0.05 to 0.   Denote by $\widetilde P$ the resulting matrix.  Then we plot the graph using the entries of $\widetilde P$ as edge weights so that zeros  correspond to unconnected pairs of nodes and larger entries mean more closely connected nodes with thicker edges. The nodes in both graphs are color coded according to Newman's labels, with red representing ``conservative," blue representing ``liberal," and orange representing ``neutral." It is seen that both graphs are mostly consistent with Newman's labels, with a few exceptions as partially discussed before. We also would like to mention that the hard-thresholding step in p-value graph is to make the graph less dense and easier to view. In fact, a small perturbation of the threshold does not change much of the overall layout of the graph.

\begin{table}[htbp]
	\centering
	\caption{Political books with labels}
	\begin{tabular}{l|l|c}
		\toprule
		\text{Title}   & \text{Label (by Newman)} &  \text{Node index} \\
		\hline
		\text{Empire}   & Neutral & 105 \\
		\text{The Future of Freedom}     & Neutral &104 \\
		\text{Rise of the Vulcans}     & Conservative &59  \\
		\text{All the Shah's Men}     & Neutral &29 \\
		\text{Bush at War}     & Conservative &78  \\
		\text{Plan of Attack}     & Neutral  &77 \\
		\text{Power Plays}     &Neutral  &47 \\
		\text{Meant To Be}     & Neutral  &19 \\
		\text{The Bushes}     & Conservative &50  \\
		\toprule
	\end{tabular}\label{tab10}%
\end{table}%
\begin{table}[htbp]
	\centering
	\caption{P-values based on test statistics $\widehat T_{ij}$. The labels provided by Newman are in the parentheses.}
	\begin{tabular}{rrrrrrrrrrr}
		\toprule
		\text{Node No.}	&105(N) &   104(N) & 59(C)  &29(N)   & 78(C) & 77(N) & 47(N) &19(N)&50(C)    \\
		\midrule
		105(N)    & 1.0000  & 0.6766  & 0.0298  & 0.3112  & 0.0248  & 0.0000  & 0.0574  & 0.1013  & 0.0449  \\
		104(N)     & 0.6766  & 1.0000  & 0.0261  & 0.2487  & 0.0204  & 0.0000  & 0.0643  & 0.1184  & 0.0407  \\
		59(C)     & 0.0298  & 0.0261  & 1.0000  & 0.1546  & 0.2129  & 0.0013  & 0.0326  & 0.0513  & 0.9249  \\
		29(N)     & 0.3112  & 0.2487  & 0.1546  & 1.0000  & 0.3206  & 0.0034  & 0.0236  & 0.0497  & 0.2121  \\
		78(C)     & 0.0248  & 0.0204  & 0.2129  & 0.3206  & 1.0000  & 0.0991  & 0.0042  & 0.0084  & 0.2574  \\
		77(N)     & 0.0000  & 0.0000  & 0.0013  & 0.0034  & 0.0991  & 1.0000  & 0.0000  & 0.0000  & 0.0035  \\
		47(N)     & 0.0574  & 0.0643  & 0.0326  & 0.0236  & 0.0042  & 0.0000  & 1.0000  & 0.9004  & 0.0834  \\
		19(N)     & 0.1013  & 0.1184  & 0.0513  & 0.0497  & 0.0084  & 0.0000  & 0.9004  & 1.0000  & 0.1113  \\
		50(C)     & 0.0449  & 0.0407  & 0.9249  & 0.2121  & 0.2574  & 0.0035  & 0.0834  & 0.1113  & 1.0000  \\

		\bottomrule
	\end{tabular}\label{tab8}%
\end{table}%

\begin{table}[htbp]
	\centering
	\caption{P-values based on test statistics $\widehat G_{ij}$. The labels provided by Newman are in the parentheses.}
	\begin{tabular}{rrrrrrrrrrr}
		\toprule
		\text{Node No.}	&105(N) &   104(N) & 59(C)  &29(N)   & 78(C) & 77(N) & 47(N) &19(N)&50(C)    \\	\midrule
		105(N)     & 1.0000  & 0.4403  & 0.1730  & 0.4563  & 0.8307  & 0.5361  & 0.0000  & 0.0000  & 0.1920  \\
		104(N)    & 0.4403  & 1.0000  & 0.0773  & 0.9721  & 0.3665  & 0.6972  & 0.0000  & 0.0000  & 0.1144  \\
		59(C)     & 0.1730  & 0.0773  & 1.0000  & 0.0792  & 0.1337  & 0.0885  & 0.0000  & 0.0000  & 0.8141  \\
		29(N)    & 0.4563  & 0.9721  & 0.0792  & 1.0000  & 0.4256  & 0.7624  & 0.0000  & 0.0000  & 0.1153  \\
		78(C)    & 0.8307  & 0.3665  & 0.1337  & 0.4256  & 1.0000  & 0.5402  & 0.0000  & 0.0000  & 0.1591  \\
		77(N)    & 0.5361  & 0.6972  & 0.0885  & 0.7624  & 0.5402  & 1.0000  & 0.0000  & 0.0000  & 0.1294  \\
		47(N)     & 0.0000  & 0.0000  & 0.0000  & 0.0000  & 0.0000  & 0.0000  & 1.0000  & 0.9778  & 0.0000  \\
		19(N)    & 0.0000  & 0.0000  & 0.0000  & 0.0000  & 0.0000  & 0.0000  & 0.9778  & 1.0000  & 0.0000  \\
		50(C)     & 0.1920  & 0.1144  & 0.8141  & 0.1153  & 0.1591  & 0.1294  & 0.0000  & 0.0000  & 1.0000  \\	
		\bottomrule
	\end{tabular}\label{tab9}
	\label{tab:addlabel}%
\end{table}%

\subsection{Stock data} \label{Secnewrealdata}
We consider a larger network of stocks in this section. Specifically, daily prices of stocks in the S\&P 500 from the period of January 2, 2009 to December 30, 2019 were collected and converted into log returns. After some pre-processing (e.g., removing stocks with missing values or very low node degrees), we ended up with 404 stocks. All data analyses in this section were conducted using those 404 stocks.   It is well known that much variation in stock excess returns can be captured by factors such as the Fama--French three factors. We first remove these common factors by fitting a factor model, and then the adjacency  matrix of stocks is constructed as the correlation matrix of idiosyncratic components from the factor model.  %{\bf Q: Not using 0-1?} {\color{blue} No. Since most correlation coefficients are close to 0.}

\begin{figure}[ht]
	\begin{center}
		\includegraphics[scale=0.8, trim=100 233 60 200,clip]{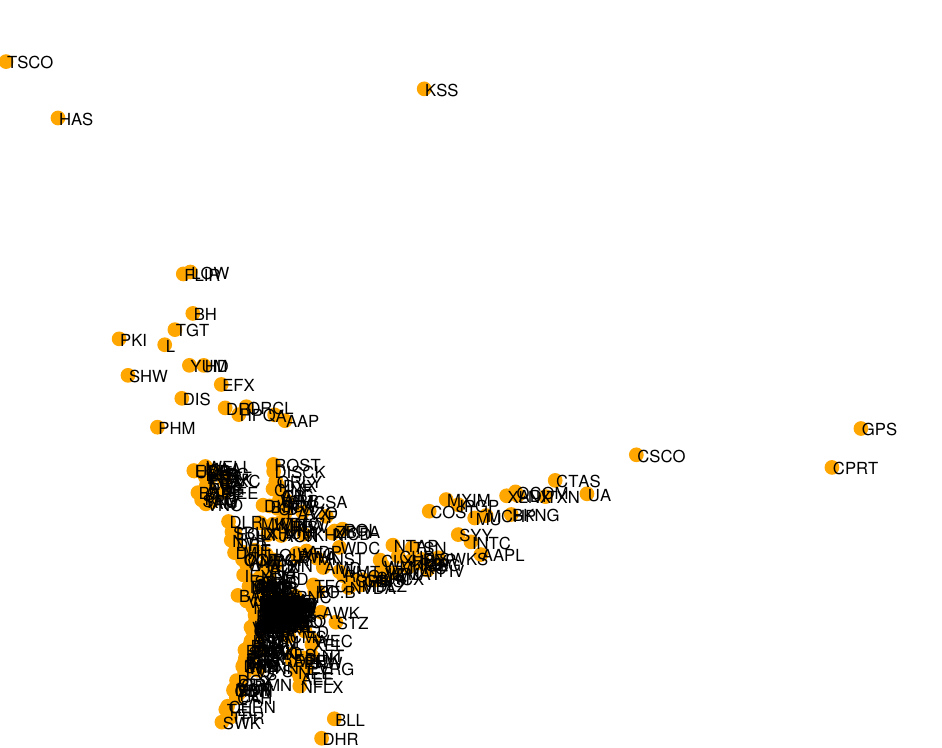}
		\end{center}
	\caption{Multiscale plot based on the distance matrix $\bf{1} - \bbA$, where $\mathbf 1$ is the matrix with all entries being 1 and $\bA$ is the p-value matrix based on $\widehat G_{ij}$.  It is seen that the scatter plot roughly has three legs and a central cluster. }\label{fig: Multi-stock}
\end{figure}

\begin{figure}[ht]
	\centering
	\includegraphics[scale=1, trim=96 220 0 200,clip]{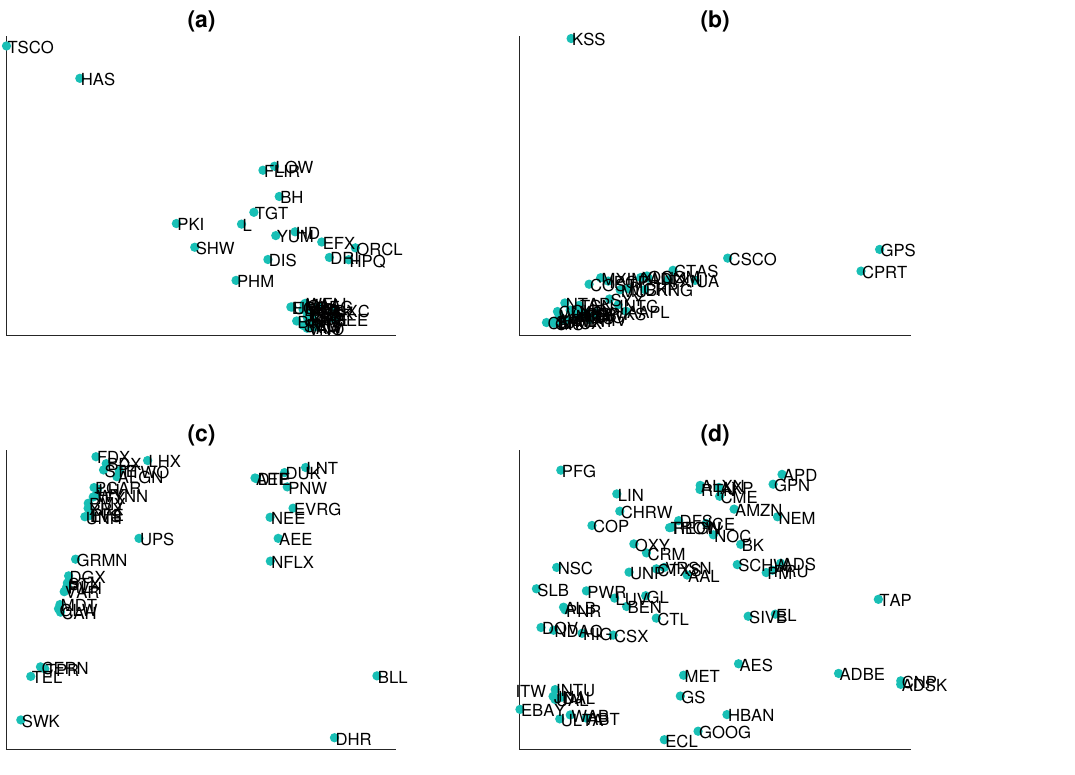}
	\caption{Zoomed multiscale plots based on the distance matrix $\bf{1} - \bbA$, where $\bA$ is the p-value matrix based on $\widehat G_{ij}$.}\label{fig: Multi-stock-zoom}
\end{figure}

Since stocks are commonly believed to have heterogeneous node degrees, we only apply $\widehat G_{ij}$ to the constructed adjacency matrix. The estimated number of communities is $\widehat K = 3$.   For each pair of stocks, we calculate its p-value using $\widehat G_{ij}$ and the asymptotic null distribution $\chi_2^2$.  This forms a p-value matrix, denoted as $\mathbf A$. To better visualize the results, we provide the multiscale plot of the distance matrix $\bf{1} - \bbA$ with $\bf{1}$ the matrix with all entries being 1, and present the results in Figure \ref{fig: Multi-stock}. It is seen that the scatter plot roughly has three legs and a central cluster. The three legs can be interpreted as the three communities with nodes having relatively more pure membership profiles, and the central cluster can be understood as for nodes with mixed membership profiles. For easier visualization, we provide zoomed plots for the three legs and the central cluster in Figure \ref{fig: Multi-stock-zoom}. The first three subplots a)--c) correspond to the three legs, and the last subplot d) corresponds to the central cluster.  We observe some interesting clustering effects.  Figure \ref{fig: Multi-stock-zoom}a) corresponds to the top leg in Figure \ref{fig: Multi-stock}.  When it is far away from the central cluster (i.e., top left  of this subplot), we have stocks mostly related to the retail and restaurant industry (e.g., TGT, HD, LOW, DRI),  and when it moves closer to the central cluster (i.e,  bottom right of this subplot), the companies are mostly in the real estate (e.g., EXR, VTR, PSA, AVB, PLD). Figure \ref{fig: Multi-stock-zoom}b) mostly consists of tech companies such as AAPL, MCHP, MU, INTC, XLNX, QCOM, ADI, among many others  in similar category. Figure \ref{fig: Multi-stock-zoom}c) roughly has two subclusters. The left cluster mostly consists of companies in or related to the health industry such as DGX, VAR, GLW, MDT, CERN, TEL, UNH, PFE, BMY, and many other similar ones. The right cluster has predominately companies in the energy industry such as AEE, NEE, EVRG, PNW, DUK, LNT, LNT, ES. %and etc.
Figure \ref{fig: Multi-stock-zoom}d) is a zoomed  plot that roughly shows the central cluster. It contains a wide range of companies including, but not limited to, risk management and investment companies (BEN, HIG, NDAQ),  transportation industry (AAL, NSC, UAL), and communication industry (CTL, VRSN, CTXS).

In Table \ref{tab: stock} below, we also present the p-value matrix for selected stocks.  The first three stocks (TGT, HD, L) are all in the retail industry, the next three stocks (APPL, INTC, MCHP) are all in the tech industry, stocks 7 to 9 (AEE, NEE, EVRG) are all in the energy industry, and the remaining one (ADBE) is taken from the central cluster. It is seen that the first three groups of stocks have high pairwise p-values within groups, but almost zero p-values with stocks from other groups. In particular, Adobe (ADBE) seems to be connected to most of these selected stocks, which is consistent with the common sense. %Ebay and Amazon have relatively p-value with each other, but almost zero p-values with others.
We would also like to point out that these results were obtained after removing the three common factors from the stock returns, and the clustering structure discovered here should be interpreted as complementary to the ones already captured by the factors.

\begin{table}[]
	\small
	\begin{tabular}{|l|lllllllll|}
		\hline
		& HD      & L       & AAPL    & INTC    & MCHP    & AEE     & NEE     & EVRG    & ADBE    \\ %& EBAY    & AMZN    \\
		\hline
		TGT  & 0.29643 & 0.71361 & 0.00000 & 0.00000 & 0.00000 & 0.00000 & 0.00000 & 0.00000 & 0.00033 \\
		HD   & 1.00000 & 0.14934 & 0.00000 & 0.00000 & 0.00000 & 0.00000 & 0.00000 & 0.00000 & 0.00025 \\
		L    & 0.14934 & 1.00000 & 0.00000 & 0.00000 & 0.00000 & 0.00000 & 0.00000 & 0.00000 & 0.00031 \\
		AAPL & 0.00000 & 0.00000 & 1.00000 & 0.00780 & 0.01933 & 0.00004 & 0.00003 & 0.00010 & 0.00395 \\
		INTC & 0.00000 & 0.00000 & 0.00780 & 1.00000 & 0.00024 & 0.00000 & 0.00000 & 0.00000 & 0.00148 \\
		MCHP & 0.00000 & 0.00000 & 0.01933 & 0.00024 & 1.00000 & 0.00000 & 0.00000 & 0.00000 & 0.00202 \\
		AEE  & 0.00000 & 0.00000 & 0.00004 & 0.00000 & 0.00000 & 1.00000 & 0.93719 & 0.46490 & 0.00467 \\
		NEE  & 0.00000 & 0.00000 & 0.00003 & 0.00000 & 0.00000 & 0.93719 & 1.00000 & 0.24407 & 0.00467 \\
		EVRG & 0.00000 & 0.00000 & 0.00010 & 0.00000 & 0.00000 & 0.46490 & 0.24407 & 1.00000 & 0.00465 \\
		ADBE & 0.00025 & 0.00031 & 0.00395 & 0.00148 & 0.00202 & 0.00467 & 0.00467 & 0.00465 & 1.00000 \\
		\hline
	\end{tabular}
\caption{The p-value matrix for selected stocks.}\label{tab: stock}
\end{table}

\section{Discussions} \label{Sec6}

In this paper, we have asked a simple yet practical question of how to determine whether any given pair of nodes in a network share the same profile of latent community memberships for large-scale social, economic, text, or health network data with precise statistical significance. Our work represents a first attempt to partially address such an important question. The suggested method of statistical inference on membership profiles in large networks (SIMPLE) provides theoretically justified network p-values in our context for both settings of mixed membership models and degree-corrected mixed membership models. We have formally shown that the two forms of SIMPLE test statistics can enjoy simple limiting distributions under the null hypothesis and appealing power under the contiguous alternative hypothesis. In particular, the tuning-free feature of SIMPLE makes it easy to use by practitioners. Our newly suggested method and established theory lay the foundation for practical policies or recommendations rooted on statistical inference for network data with quantifiable impacts.

To illustrate the key ideas of SIMPLE and simplify the technical analysis, we have focused our attention on the hypothesis testing problem for any preselected pair of nodes. It would be interesting to study the problem when one of or each of the nodes is replaced by a selected set of nodes. For example, in certain applications one may have some additional knowledge that all the nodes within the selected set indeed share the same membership profile information. It would also be interesting to quantify and control the statistical inference error rates when one is interested in performing a set of hypothesis tests simultaneously for network data. Moreover, it would be interesting to investigate the hypothesis testing problem for more general network models as well as for statistical models beyond network data such as for large collections of text documents.

In addition, it would be interesting to connect the growing literature on sparse covariance matrices and sparse precision matrices with that on network models. Such connections can be made via modeling the graph Laplacian through a precision matrix or covariance matrix \citep{BGG19}. A natural question is then how well the network profiles can be inferred from a panel of time series data. The same question also arises if the panel of time series data admits a factor structure \citep{FanFanLv2008,FLM13}. These problems and extensions are beyond the scope of the current paper and will be interesting topics for future research.

%\newpage

\bibliographystyle{chicago}
\bibliography{references}

\begin{thebibliography}{}

\bibitem[\protect\citeauthoryear{Abbe}{Abbe}{2017}]{Abbe2017CommunityDA}
Abbe, E. (2017).
\newblock Community detection and stochastic block models: recent developments.
\newblock {\em Journal of Machine Learning Research\/}~{\em 18}, 177:1--177:86.

\bibitem[\protect\citeauthoryear{{Abbe}, {Fan}, {Wang}, and {Zhong}}{{Abbe}
  et~al.}{2017}]{abbe2019entrywise}
{Abbe}, E., J.~{Fan}, K.~{Wang}, and Y.~{Zhong} (2017).
\newblock {Entrywise eigenvector analysis of random matrices with low expected
  rank}.
\newblock {\em arXiv preprint arXiv:1709.09565\/}.

\bibitem[\protect\citeauthoryear{Airoldi, Blei, Fienberg, and Xing}{Airoldi
  et~al.}{2008}]{Airoldi:2008}
Airoldi, E.~M., D.~M. Blei, S.~E. Fienberg, and E.~P. Xing (2008).
\newblock Mixed membership stochastic blockmodels.
\newblock {\em Journal of Machine Learning Research\/}~{\em 9}, 1981--2014.

\bibitem[\protect\citeauthoryear{Arias-Castro and Verzelen}{Arias-Castro and
  Verzelen}{2014}]{arias-castro2014}
Arias-Castro, E. and N.~Verzelen (2014, 06).
\newblock Community detection in dense random networks.
\newblock {\em Ann. Statist.\/}~{\em 42\/}(3), 940--969.

\bibitem[\protect\citeauthoryear{Bickel and Sarkar}{Bickel and
  Sarkar}{2016}]{Bickel2016}
Bickel, P.~J. and P.~Sarkar (2016).
\newblock Hypothesis testing for automated community detection in networks.
\newblock {\em Journal of the Royal Statistical Society Series B\/}~{\em 78},
  253--273.

\bibitem[\protect\citeauthoryear{Billingsley}{Billingsley}{1995}]{B1995}
Billingsley, P. (1995).
\newblock {\em Probability and Measure}.
\newblock Wiley.

\bibitem[\protect\citeauthoryear{Brownlees, Stefan, and Lugosi}{Brownlees
  et~al.}{2019}]{BGG19}
Brownlees, C., G.~G. Stefan, and G.~Lugosi (2019).
\newblock Community detection in partial correlation network models.
\newblock {\em Manuscript\/}.

\bibitem[\protect\citeauthoryear{Chen and Lei}{Chen and
  Lei}{2018}]{ChenLei2018}
Chen, K. and J.~Lei (2018).
\newblock Network cross-validation for determining the number of communities in
  network data.
\newblock {\em Journal of the American Statistical Association\/}~{\em 113},
  241--251.

\bibitem[\protect\citeauthoryear{Daudin, Picard, and Robin}{Daudin
  et~al.}{2008}]{Daudin2008}
Daudin, J.-J., F.~Picard, and S.~Robin (2008).
\newblock A mixture model for random graphs.
\newblock {\em Statistics and Computing\/}~{\em 18}, 173--183.

\bibitem[\protect\citeauthoryear{Fan, Fan, Han, and Lv}{Fan
  et~al.}{2020}]{FFHL19}
Fan, J., Y.~Fan, X.~Han, and J.~Lv (2020).
\newblock Asymptotic theory of eigenvectors for random matrices with diverging
  spikes.
\newblock {\em Journal of the American Statistical Association\/}, to appear.

\bibitem[\protect\citeauthoryear{Fan, Liao, and Mincheva}{Fan
  et~al.}{2013}]{FLM13}
Fan, J., Y.~Liao, and M.~Mincheva (2013).
\newblock Large covariance estimation by thresholding principal orthogonal
  complements (with discussion).
\newblock {\em Journal of the Royal Statistical Society Series B\/}~{\em 75},
  603--680.

\bibitem[\protect\citeauthoryear{Gao, Ma, Zhang, and Zhou}{Gao
  et~al.}{2018}]{gao2018}
Gao, C., Z.~Ma, A.~Y. Zhang, and H.~H. Zhou (2018, 10).
\newblock Community detection in degree-corrected block models.
\newblock {\em Ann. Statist.\/}~{\em 46\/}(5), 2153--2185.

\bibitem[\protect\citeauthoryear{Goldenberg, Zheng, Fienberg, and
  Airoldi}{Goldenberg et~al.}{2010}]{Goldenberg:2010}
Goldenberg, A., A.~X. Zheng, S.~E. Fienberg, and E.~M. Airoldi (2010).
\newblock A survey of statistical network models.
\newblock {\em Found. Trends Mach. Learn.\/}~{\em 2}, 129--233.

\bibitem[\protect\citeauthoryear{Holland, Laskey, and Leinhardt}{Holland
  et~al.}{1983}]{SBM1983}
Holland, P.~W., K.~B. Laskey, and S.~Leinhardt (1983).
\newblock Stochastic blockmodels: First steps.
\newblock {\em Social Networks\/}~{\em 5}, 109--137.

\bibitem[\protect\citeauthoryear{Jin}{Jin}{2015}]{jin2015}
Jin, J. (2015).
\newblock Fast community detection by {SCORE}.
\newblock {\em Ann. Statist.\/}~{\em 43}, 57--89.

\bibitem[\protect\citeauthoryear{Jin and Ke}{Jin and Ke}{2017}]{jin2017sharp}
Jin, J. and Z.~T. Ke (2017).
\newblock A sharp lower bound for mixed-membership estimation.
\newblock {\em arXiv preprint arXiv:1709.05603\/}.

\bibitem[\protect\citeauthoryear{Jin, Ke, and Luo}{Jin et~al.}{2017}]{JKL17}
Jin, J., Z.~T. Ke, and S.~Luo (2017).
\newblock Estimating network memberships by simplex vertex hunting.
\newblock {\em arXiv preprint arXiv:1708.07852\/}.

\bibitem[\protect\citeauthoryear{Karrer and Newman}{Karrer and
  Newman}{2011}]{DCSBM2011}
Karrer, B. and M.~E.~J. Newman (2011).
\newblock Stochastic blockmodels and community structure in networks.
\newblock {\em Phys. Rev. E\/}~{\em 83}, 016107.

\bibitem[\protect\citeauthoryear{Koutsourelakis and Eliassi-Rad}{Koutsourelakis
  and Eliassi-Rad}{2008}]{KF08}
Koutsourelakis, P.-S. and T.~Eliassi-Rad (2008).
\newblock Finding mixed-memberships in social networks.
\newblock {\em AAAI Spring Symposium: Social Information Processing\/}, 48--53.

\bibitem[\protect\citeauthoryear{Latouche, Birmel\'{e}, and Ambroise}{Latouche
  et~al.}{2012}]{Latouche-etal2012}
Latouche, P., E.~Birmel\'{e}, and C.~Ambroise (2012).
\newblock Variational bayesian inference and complexity control for stochastic
  block models.
\newblock {\em Statistical Modelling\/}~{\em 12}, 93--115.

\bibitem[\protect\citeauthoryear{Lei}{Lei}{2016}]{L16}
Lei, J. (2016).
\newblock A goodness-of-fit test for stochastic block models.
\newblock {\em Ann. Statist.\/}~{\em 44}, 401--424.

\bibitem[\protect\citeauthoryear{Lei and Rinaldo}{Lei and
  Rinaldo}{2015}]{lei2015}
Lei, J. and A.~Rinaldo (2015).
\newblock Consistency of spectral clustering in stochastic block models.
\newblock {\em Ann. Statist.\/}~{\em 43}, 215--237.

\bibitem[\protect\citeauthoryear{Liben-Nowell and Kleinberg}{Liben-Nowell and
  Kleinberg}{2007}]{Liben-Nowell:2007}
Liben-Nowell, D. and J.~Kleinberg (2007).
\newblock The link-prediction problem for social networks.
\newblock {\em J. Am. Soc. Inf. Sci. Technol.\/}~{\em 58}, 1019--1031.

\bibitem[\protect\citeauthoryear{Newman}{Newman}{2013a}]{Newman_2013}
Newman, M. E.~J. (2013a).
\newblock Community detection and graph partitioning.
\newblock {\em {EPL} (Europhysics Letters)\/}~{\em 103}, 28003.

\bibitem[\protect\citeauthoryear{Newman}{Newman}{2013b}]{PhysRevE.88.042822}
Newman, M. E.~J. (2013b).
\newblock Spectral methods for community detection and graph partitioning.
\newblock {\em Phys. Rev. E\/}~{\em 88}, 042822.

\bibitem[\protect\citeauthoryear{Newman and Peixoto}{Newman and
  Peixoto}{2015}]{Newman2015}
Newman, M. E.~J. and T.~P. Peixoto (2015).
\newblock Generalized communities in networks.
\newblock {\em Phys. Rev. Lett.\/}~{\em 115}, 088701.

\bibitem[\protect\citeauthoryear{Rai\u{c}}{Rai\u{c}}{2019}]{BEJ1072}
Rai\u{c}, M. (2019).
\newblock {A multivariate Berry-Esseen theorem with explicit constants}.
\newblock {\em Bernoulli\/}~{\em 25\/}(4A), 2824 -- 2853.

\bibitem[\protect\citeauthoryear{Rohe, Chatterjee, and Yu}{Rohe
  et~al.}{2011}]{rohe2011}
Rohe, K., S.~Chatterjee, and B.~Yu (2011).
\newblock Spectral clustering and the high-dimensional stochastic blockmodel.
\newblock {\em The Annals of Statistics\/}~{\em 39}, 1878--1915.

\bibitem[\protect\citeauthoryear{Saldana, Yu, and Feng}{Saldana
  et~al.}{2017}]{saldana2017many}
Saldana, D., Y.~Yu, and Y.~Feng (2017).
\newblock How many communities are there?
\newblock {\em Journal of Computational and Graphical Statistics\/}~{\em 26},
  171--181.

\bibitem[\protect\citeauthoryear{Tropp}{Tropp}{2012}]{T12}
Tropp, J. (2012).
\newblock User-friendly tail bounds for sums of random matrices.
\newblock {\em Found. Comput. Math.\/}~{\em 12}, 389--434.

\bibitem[\protect\citeauthoryear{Verzelen and Arias-Castro}{Verzelen and
  Arias-Castro}{2015}]{verzelen2015}
Verzelen, N. and E.~Arias-Castro (2015, 12).
\newblock Community detection in sparse random networks.
\newblock {\em Ann. Appl. Probab.\/}~{\em 25\/}(6), 3465--3510.

\bibitem[\protect\citeauthoryear{von Luxburg}{von Luxburg}{2007}]{von2007}
von Luxburg, U. (2007).
\newblock A tutorial on spectral clustering.
\newblock {\em Statistics and Computing\/}~{\em 17}, 395--416.

\bibitem[\protect\citeauthoryear{Wang and Wong}{Wang and
  Wong}{1987}]{WangWong1987}
Wang, Y.~J. and G.~Y. Wong (1987).
\newblock Stochastic blockmodels for directed graphs.
\newblock {\em Journal of the American Statistical Association\/}~{\em 82},
  8--19.

\bibitem[\protect\citeauthoryear{Wang and Bickel}{Wang and
  Bickel}{2017}]{wang2017}
Wang, Y. X.~R. and P.~J. Bickel (2017).
\newblock Likelihood-based model selection for stochastic block models.
\newblock {\em Ann. Statist.\/}~{\em 45}, 500--528.

\bibitem[\protect\citeauthoryear{{Wu}, {Levina}, and {Zhu}}{{Wu}
  et~al.}{2018}]{WuLevinaZhu2018}
{Wu}, Y.-J., E.~{Levina}, and J.~{Zhu} (2018).
\newblock {Link prediction for egocentrically sampled networks}.
\newblock {\em arXiv preprint arXiv:1803.040845\/}.

\bibitem[\protect\citeauthoryear{Zhang and Moore}{Zhang and
  Moore}{2014}]{Zhang18144}
Zhang, P. and C.~Moore (2014).
\newblock Scalable detection of statistically significant communities and
  hierarchies, using message passing for modularity.
\newblock {\em Proceedings of the National Academy of Sciences\/}~{\em 111},
  18144--18149.

\bibitem[\protect\citeauthoryear{Zhang, Levina, and Zhu}{Zhang
  et~al.}{2020}]{19M127}
Zhang, Y., E.~Levina, and J.~Zhu (2020).
\newblock Detecting overlapping communities in networks using spectral methods.
\newblock {\em SIAM Journal on Mathematics of Data Science\/}~{\em 2\/}(2),
  265--283.

\bibitem[\protect\citeauthoryear{Zhao, Levina, and Zhu}{Zhao
  et~al.}{2012}]{zhao12}
Zhao, Y., E.~Levina, and J.~Zhu (2012).
\newblock Consistency of community detection in networks under degree-corrected
  stochastic block models.
\newblock {\em Ann. Statist.\/}~{\em 40}, 2266--2292.

\end{thebibliography}

%%%%%%%%%%%%%%%%%%%%%%%%%%%%%%%%%%%%%%%%%%%%

\newpage
\appendix
\setcounter{page}{1}
\setcounter{section}{1}
\renewcommand{\theequation}{A.\arabic{equation}}
\renewcommand{\thesubsection}{A.\arabic{subsection}}
\setcounter{equation}{0}

\begin{center}{\bf \Large Supplementary Material to ``SIMPLE: Statistical Inference on Membership Profiles in Large Networks"}
	
\bigskip
	
Jianqing Fan, Yingying Fan, Xiao Han and Jinchi Lv
\end{center}

\noindent %This Supplementary Material contains the proofs of Lemmas \ref{prop: k-threshold}--\ref{0516-3}, some key lemmas and their proofs, and additional technical details. See the beginning of Section \ref{SecA} for additional notation and definitions that will be used throughout the technical presentation.
{\color{black} This Supplementary Material contains all the proofs and technical details. }
\appendix
\section{Proofs of main results} \label{SecA}
%We provide in this Appendix the proofs for the main results presented in Theorems \ref{0126-1}--\ref{0425-1}. The proofs of Lemmas \ref{prop: k-threshold}--\ref{0516-3}, some key lemmas with their proofs, and additional technical details are included in the Supplementary Material.

To facilitate the technical presentation, we list two definitions below, where $n$ represents the network size and dimensionality of eigenvectors.
\begin{deff} \label{0525.1h}
Let $\zeta$ and $\xi$ be a pair of random variables that may depend on $n$. We say that they satijsfy $\xi=O_{\prec}(\zeta)$ if for any pair of positive constants $(a,b)$, there exists  some positive integer $n_0(a,b)$ depending only on $a$ and $b$ such that $\mathbb{P}(|\xi|>n^{a}|\zeta|)\le n^{-b}$ for all $n\ge n_0(a,b)$.
\end{deff}
\begin{deff} \label{0525.1h.a}
We say that an event $\mathfrak{A}_n$ holds with high probability if for any positive constant $a$, there exists some positive integer  $n_0(a)$ depending only on $a$ such that $\mathbb{P}\left(\mathfrak{A}_n\right)\ge 1- n^{-a}$ for all $n\ge n_0(a)$.
\end{deff}
\noindent From Definitions \ref{0525.1h} and \ref{0525.1h.a} above, we can see that if $\xi=O_{\prec}(\zeta)$, then it holds that $\xi=O(n^{a}|\zeta|)$ with high probability for any positive constant $a$. The strong probabilistic bounds in the statements of Definitions \ref{0525.1h} and \ref{0525.1h.a} are in fact consequences of analyzing large binary random matrices given by networks.

Let us introduce some additional notation. Since the eigenvectors are always up to a sign change, for simplicity we fix the orientation of the empirical eigenvector $\widehat\bbv_k$ such that $\widehat\bbv_k^T \bbv_k \ge 0$ for each $1 \leq k \leq K$, where $\bbv_k$ is the $k$th population eigenvector of the low-rank mean matrix $\bH$ in our general network model (\ref{eq: model.general}). It is worth mentioning that all the variables are real-valued throughout the paper except that variable $z$ can be complex-valued. For any nonzero complex number $z$, deterministic matrices $\bbM_1$ and $\bbM_2$ of appropriate dimensions, $1 \leq k \leq K$, and $n$-dimensional unit vector $\bbu$, we define
\begin{align}\label{0619.1}
&\mathcal{P}(\bM_1, \bM_2, z)=z \mathcal{R}(\bM_1, \bM_2, z), \quad
\mathcal{\widetilde P}_{k,z}=\left[z^2 (A_{\bbv_k,k,z}/z)'\right]^{-1},\\
&
\bbb_{\bbu,k,z} = \bbu - \bbV_{-k} \left[(\bbD_{-k})^{-1}+\mathcal{R}(\bbV_{-k},\bbV_{-k},z)\right]^{-1} \mathcal{R}^T(\bbu,\bbV_{-k},z), \label{0619.3}
\end{align}
where $\mathcal{R}(\bM_1, \bM_2, z)$ is defined in (\ref{1119.1}),
\begin{equation}\label{0619.2}
A_{\bbu,k,z}=\mathcal{P}(\bbu,\bbv_k,z)-\mathcal{P}(\bbu,\bbV_{-k},z)\left[z(\bbD_{-k})^{-1}+\mathcal{P}(\bbV_{-k},\bbV_{-k},z)\right]^{-1}\mathcal{P}(\bbV_{-k},\bbv_k,z),
\end{equation}
$(A_{\bbv_k,k,z}/z)'$ denotes the derivative of $A_{\bbv_k,k,z}/z$ with respect to complex variable $z$, $\bbV_{-k}$ represents a submatrix of $\bV = (\bv_1, \cdots, \bv_K)$ by removing the $k$th column, and $\bD_{-k}$ stands for a principal submatrix of $\bD = \diag(d_1, \cdots, d_K)$ by removing the $k$th diagonal entry.

\subsection{Proof of Theorem \ref{0126-1}} \label{SecA.1}
We first prove the conclusion in the first part of Theorem \ref{0126-1} under the null hypothesis $H_0: \bpi_i=\bpi_j$, where $(i, j)$ with $1 \leq i < j \leq n$ represents a given pair of nodes in the network. In particular, Lemma \ref{prop-expansition} in Section \ref{SecB.8} of Supplementary Material plays a key role in the technical analysis. For the given pair $(i, j)$, let us define a new random matrix $\widetilde\bbX=(\widetilde\bbx_{lm})_{1 \leq l,m \leq n}$ based on the original random matrix $\bX = (\bx_{lm})_{1 \leq l,m \leq n}$ by swapping the roles of nodes $i$ and $j$, namely by setting
\begin{equation} \label{new.eq001}
\widetilde \bbx_{lm}=\begin{cases}
\bbx_{lm}, & l,m \in \{i,j\}^c \cr \bbx_{im}, &l=j,\, m\in \{i,j\}^c \cr \bbx_{jm}, &l=i, \,m\in \{i,j\}^c
\cr \bbx_{li}, &m=j, \,l\in \{i,j\}^c \cr \bbx_{lj}, &m=i, \,l\in \{i,j\}^c \end{cases} \ \text{ and } \ \widetilde \bbx_{lm}=\begin{cases}
\bbx_{ij}, & (l, m) = (i, j) \text{ or } (j, i)\cr \bbx_{ii}, &l=m=j \cr \bbx_{jj}, &l=m=i\end{cases},
\end{equation}
where $\{i,j\}^c$ stands for the complement of set $\{i,j\}$ in the node set $\{1, \cdots, n\}$. It is easy to see that the new symmetric random matrix $\widetilde\bbX$ defined in (\ref{new.eq001}) is simply the adjacency matrix of a network given by the mixed membership model \eqref{0115.1} by swapping the $i$th and $j$th rows, $\bpi_i$ and $\bpi_j$, of the community membership probability matrix $\bPi = (\bpi_1,\cdots, \bpi_n)^T$.

By the above definition of $\widetilde\bbX$, we can see that under the null hypothesis $H_0: \bpi_i=\bpi_j$, it holds that
\begin{eqnarray}\label{0119.2}
\widetilde\bbX \stackrel{d}{=} \bbX,
\end{eqnarray}
where $\stackrel{d}{=}$ denotes being equal in distribution. The representation in (\ref{0119.2}) entails that for each $1 \leq k \leq K$, the $i$th and $j$th components of the $k$th population eigenvector  $\bv_k$ are identical; that is,
$$\bbv_k(i)=\bbv_k(j). $$
This identity along with the asymptotic expansion of the empirical eigenvector $\widehat\bbv_k$ in (\ref{0426.1h}) given in Lemma \ref{prop-expansition} results in
\begin{equation}\label{eq: egv-diff}
\widehat\bbv_k(i)-\widehat\bbv_k(j)=\frac{(\bbe_i-\bbe_j)^T\bbW\bbv_k}{t_k}+O_{\prec}(\frac{\alpha_n^2}{\sqrt n |d_k|^2}+\frac{1}{\sqrt n|d_k|}).
\end{equation}

Note that although the expectation of $\bbe_i^T\bbW\bbv_k$ can be nonzero, the difference of expectations  $\mathbb{E}(\bbe_i-\bbe_j)^T\bbW\bbv_k=0$ under the null hypothesis by (\ref{0119.2}). It follows from Lemma \ref{lem: dk-alphan} in Section \ref{SecB.7} and Lemma \ref{lem: tk} in Section \ref{SecC.6} of Supplementary Material that
$$n^{1-c_2}\theta\lesssim d_k\sim t_k\lesssim n\theta \ \text{ and } \  \alpha_n=O(\sqrt{n\theta}),$$
where $\sim$ denotes the same asymptotic order. Condition \ref{cond2} ensures that  there exists some positive constant $\ep$ such that
 \begin{equation}\label{1117.1}
  \text{SD}\left((\bbe_i-\bbe_j)^T\bbW\bbv_k\right)\sim \sqrt{\theta}\gg n^{\ep}n^{c_2-1/2}\gtrsim  n^{\ep}\left(\frac{\alpha_n^2}{\sqrt n |d_k|}+\frac{1}{\sqrt n}\right),
   \end{equation}
 which guarantees that $O_{\prec}(\frac{\alpha_n^2}{\sqrt n d_k^2}+\frac{1}{\sqrt n|d_k|})$ in \eqref{eq: egv-diff} is negligible compared to the first term on the right hand side. Here $\text{SD}$ represents the standard deviation of a random variable.
  Moreover, by Lemma \ref{egv} in Section \ref{SecB.6} of Supplementary Material we have $\|\bbV\|_{\infty}=O(\frac{1}{\sqrt n})\ll \min_{1\leq k \leq K} \text{SD}((\bbe_i-\bbe_j)^T\bbW\bbv_k)\sim \sqrt{\theta}$,
and  hence $((\bbe_i-\bbe_j)^T\bbW\bbv_k)_{k=1}^K$ satisfies  the  conditions of Lemma \ref{jclt} in Section \ref{SecB.3} of Supplementary Material with $h_n=\theta$. Then it holds that
%{\color{blue}
	\begin{align}\label{1117.2}
 &\bold\Sigma_1^{-1/2}(\widehat\bbV(i)-\widehat\bbV(j))\non
 &=\bold\Sigma_1^{-1/2}\bbD^{-1}\left(\frac{(\bbe_i-\bbe_j)^T\bbW\bbv_1}{t_1/d_1},\cdots,\frac{(\bbe_i-\bbe_j)^T\bbW\bbv_K}{t_K/d_K}\right)^T+o_p(1) \toD N(\bf{0},\bbI),
   \end{align}
   %}
which proves (\ref{0126.1h}).

We next establish (\ref{0120.1}) under the condition of  $\sqrt{n\theta}\|\bpi_i-\bpi_j\|\rightarrow \infty$.
By (\ref{0426.1h}) in Lemma \ref{prop-expansition}, we have
\begin{align}\label{1117.3}
&\bbD(\widehat\bbV(i)-\widehat\bbV(j))\non
&=\bbD(\bbV(i)-\bbV(j))
+\left(\frac{(\bbe_i-\bbe_j)^T\bbW\bbv_1}{t_1/d_1},\cdots,\frac{(\bbe_i-\bbe_j)^T\bbW\bbv_K}{t_K/d_K}\right)^T+O_{\prec}(\frac{\alpha_n^2}{\sqrt n |d_K|}).
  \end{align}
In view of (\ref{1117.1}), it holds that
$$\left(\frac{(\bbe_i-\bbe_j)^T\bbW\bbv_1}{t_1/d_1},\cdots,\frac{(\bbe_i-\bbe_j)^T\bbW\bbv_K}{t_K/d_K}\right)=O_p(\sqrt\theta).$$
Thus it suffices to show that
$$\|\bbD(\bbV(i)-\bbV(j))\|\gg \sqrt{\theta}.$$
In fact, it follows from (\ref{0119.5}) that
$$\bbD(\bbV(i)-\bbV(j))=\bbD\bbB(\bpi_i-\bpi_j).$$
This along with (\ref{0119.4}) and Condition \ref{cond1} leads to
\begin{align*}
\|\bbD(\bbV(i)-\bbV(j))\|&=\|\bbD(\bpi_i-\bpi_j)^T\bbB\|\ge |d_K|\sqrt{(\bpi_i-\bpi_j)^T(\bPi^T\bPi)^{-1}(\bpi_i-\bpi_j)}\\&\gtrsim \|\pi_1-\pi_2\|n^{1/2-c_2}\theta\gg \sqrt{\theta},
\end{align*}
which concludes the proof of (\ref{0120.1}).

Finally, we prove (\ref{0126.1}). The conclusion follows immediately from (\ref{1117.3}) and $(\bbV(i)-\bbV(j))^T\bSig_1^{-1}(\bbV(i)-\bbV(j))\rightarrow \mu$ as $n \rightarrow \infty$.
This completes the proof of Theorem \ref{0126-1}.

\subsection{Proof of Theorem \ref{0519-1}} \label{SecA.2}
As guaranteed by Slutsky's lemma, the asymptotic distributions of test statistics after replacing $\bSig_1$ with $\widehat\bS_1$ stay the same. Thus we need only to prove that the asymptotic distributions are the same  after replacing $K$ with its estimate $\widehat K$ in the test statistics.

To ease the presentation, we write $T_{ij}=T_{ij}(K)$ and $\widehat T_{ij}=T_{ij}(\widehat K)$ to emphasize their dependency on $K$  and $\widehat K$, respectively. By (\ref{0126.1h}) of Theorem \ref{0126-1}, we have for any $t>0$,
\begin{eqnarray}\label{0519.1}
	\lim_{n\rightarrow \infty}\mathbb{P}(T_{ij}(K)<t)=\mathbb{P}(\chi_K^2<t).
\end{eqnarray}
By the condition on $\widehat K$, it holds that
\begin{equation}\label{1217.11h}
	\mathbb{P}(\widehat K=K)=1-o(1).
\end{equation}
Then by the properties of conditional probability, we deduce
\begin{eqnarray}\label{0519.2}
	&&\mathbb{P}(T_{ij}(\widehat K)<t)=\mathbb{P}(T_{ij}(\widehat K)<t|\widehat K=K)\mathbb{P}(\widehat K=K)+\mathbb{P}(T_{ij}(\widehat K)<t|\widehat K\neq K)\mathbb{P}(\widehat K\neq K)\non
	&&=\mathbb{P}(T_{ij}(K)<t|\widehat K=K)\mathbb{P}(\widehat K=K)+o(1)\non
	&&=\mathbb{P}(T_{ij}(K)<t|\widehat K=K)\mathbb{P}(\widehat K=K)+\mathbb{P}(T_{ij}(K)<t|\widehat K\neq K)\mathbb{P}(\widehat K\neq K)+o(1)\non
	&&=\mathbb{P}(T_{ij}(K)<t)+o(1).
\end{eqnarray}
Observe that the $o(1)$ term comes from $(\ref{1217.11h})$ and thus it holds uniformly for any $t$. Combining \eqref{0519.2} with (\ref{0519.1}), we can show that
\begin{eqnarray}\label{0519.3}
	\lim_{n\rightarrow \infty}\mathbb{P}(T_{ij}(\widehat K)<t)=\mathbb{P}(\chi_K^2<t).
\end{eqnarray}
Therefore, the same conclusion as in (\ref{0126.1h}) of Theorem \ref{0126-1} is proved. Results in (\ref{0120.1}) and (\ref{0126.1}) can be shown using similar arguments and are omitted here for simplicity. This concludes the proof of Theorem \ref{0519-1}.

\subsection{Proof of Corollary \ref{coro2}}
Recall that in the proof of Theorem \ref{0519-1}, we denote by  $T_{ij}=T_{ij}(K)$ and  $\widehat T_{ij}=T_{ij}(\widehat K)$ to emphasize their dependency on $K$ and $\widehat K$.  It suffices to prove that the impact of the use of $\widehat K$ in place of $K$ is asymptotically negligible. In fact, we can deduce that
	\begin{align}\label{hatK}
	\mathbb{P} (&T_{ij}(\widehat K)>\chi^2_{\widehat K,1-\alpha})  =\mathbb{P}(T_{ij}(\widehat K)>\chi_{\widehat K, 1-\alpha}^2|\widehat K=K)\mathbb{P}(\widehat K=K)\nonumber\\
	&\quad+\mathbb{P}(T_{ij}(\widehat K)>\chi_{\hat K, 1-\alpha}^2|\widehat K\neq K)\mathbb{P}(\widehat K\neq K)\nonumber\\
	&=\mathbb{P}(T_{ij}(K)>\chi_{K, 1-\alpha}^2|\widehat K=K)\mathbb{P}(\widehat K=K)+o(1)\nonumber\\
	&=\mathbb{P}(T_{ij}(K)>\chi_{K, 1-\alpha}^2|\widehat K=K)\mathbb{P}(\widehat K=K)\nonumber\\
	&\quad+\mathbb{P}(T_{ij}(K)>\chi_{K, 1-\alpha}^2|\widehat K\neq K)\mathbb{P}(\widehat K\neq K)+o(1)\nonumber\\
	&=\mathbb{P}(T_{ij}(K)>\chi_{K, 1-\alpha}^2)+o(1).
	\end{align}

	By \eqref{hatK}, under the null hypothesis we have
	\begin{align}\label{0901.1h}
	\lim_{n\rightarrow \infty}\mathbb{P} (&\widehat T_{ij}> \chi_{\widehat K, 1-\alpha}^2)=\lim_{n\rightarrow \infty}\mathbb{P}(T_{ij}>\chi_{K, 1-\alpha}^2)=\alpha
	\end{align}
	for any constant $\alpha \in (0,1)$. Moreover, by \eqref{0519.2}, under the alternative hypothesis, for any arbitrarily large constant $C>0$ it holds that
	\begin{align}\label{0901.2h}
	\lim_{n\rightarrow \infty}\mathbb{P} (&\widehat T_{ij}> C)=\lim_{n\rightarrow \infty}\mathbb{P}(T_{ij}>C)=1.
	\end{align}
	Therefore, combining \eqref{0901.1h} and \eqref{0901.2h}  completes the proof of Corollary \ref{coro2}.

\subsection{Proof of Theorem \ref{0104-1}} \label{SecA.3}
We begin with listing some basic properties of $\bbv_k$ and $d_k$:
\begin{itemize}
\item[1).] We can choose a direction such that all components of $\bbv_1$ are nonnegative. Moreover, $\min_{1 \leq l \leq n}\{\bbv_1(l)\}\sim \frac{1}{\sqrt n}$.
\item[2).] $\max_{1\le k\le K}\|\bbv_k\|_{\infty}\le \frac{C}{\sqrt n}$ for some positive constant $C$.
\item[3).] $\alpha_n\le \sqrt n \theta_{\max}$.
\item[4).] $|d_K|\ge cn^{1-2c_2}\theta^2_{\min}$ and $|d_1|\le c^{-1}n\theta^2_{\max}$ for some positive constant $c$.
\end{itemize}
Here the second  statement is ensured by Lemma \ref{egv}. The third and fourth statements are guaranteed by Lemma \ref{lem: dk-alphan}, and the remaining properties are entailed by Lemma B.2 of \cite{JKL17}. One should notice that the proof of Lemma B.2 of \cite{JKL17} does not require $\{d_k\}_{k=1}^K$ have the same order. %Notice that our  Conditions \ref{as3}, \ref{cond5}--\ref{cond7} satisfy the conditions of  Lemmas 3.1-3.3 of \cite{JKL17}. Thus we have the following statements:

%Similarly, we have
%$$\widehat \bbv_k(j)=\bbv_k(j)+\frac{\bbe_j^T\bbW\bbv_k}{t_k}+O_{\prec}(\frac{1}{n^{3/2}\theta^{2}_{\max}}),$$
%and
%$$\widehat \bbv_1(i)=\bbv_1(i)+\frac{\bbe_i^T\bbW\bbv_1}{t_1}+O_{\prec}(\frac{1}{n^{3/2}\theta^{2}_{\max}}).$$
By Condition \ref{cond6}  and Statement 4 above, we have %\textcolor{red}{Condition 5 is only for node $i$ and $j$ so the bext inequality only holds for $i$ and $j$, not for all $1\leq i\leq n$} {\color{blue} Yes, it is
	$$\frac{1}{n^{1/2-c_2}|t_k|}\ll \frac{\min_{1\le k\le K,\, t=i,j}\sqrt{\var(\be_t^T\bbW\bbv_k)}}{|t_k|}.$$
By (\ref{1229.1h}), there exists some $K\times K$ matrix $\bbB$ such that
\begin{equation}\label{0119.5h}
\bbV=\bTheta\bPi\bbB.
\end{equation}
Recall that $\bTheta$ is a diagonal matrix. Then it follows from (\ref{0119.5h}) that under the null hypothesis, we have
\begin{equation}\label{1231.6h}
\frac{\bbv_k(i)}{\bbv_1(i)}=\frac{\bbv_k(j)}{\bbv_1(j)}, \quad k=1,\cdots,K.
\end{equation}
Here we use the exchangeability between rows $i$ and $j$ of matrix $\bPi \bB$ under the null hypothesis as argued under the mixed membership model (see the beginning of the proof of Theorem~\ref {0126-1}).
%Therefore $\frac{\bbv_k-\frac{\bbv_k(i)}{\bbv_1(i)}\bbv_1}{\bbv_1(i)}=\frac{\bbv_k-\frac{\bbv_k(j)}{\bbv_1(j)}\bbv_1}{\bbv_1(j)}$.
In light of the asymptotic expansion in Lemma \ref{prop-expansition}, we deduce
\begin{equation}\label{1231.7h}
\widehat \bbv_k(i)=\bbv_k(i)+\frac{\bbe_i^T\bbW\bbv_k}{t_k}+O_{\prec}(\frac{1}{n^{1/2-c_2}|t_k|}).
\end{equation}
Moreover, it follows from Corollary \ref{cor} in Section \ref{SecC.2} of Supplementary Material, Condition \ref{cond5}, and the statements at the beginning of this proof that
\begin{equation}\label{aa1}
\frac{\bbe_s^T\bbW\bbv_k}{t_k}=O_{\prec}(\frac{\theta_{\max}}{|t_k|}), \quad s=i,j, \ k=1,\cdots,K.
\end{equation}
Thus, by (\ref{1231.6h})--(\ref{aa1}) and Statement 1 above we have under the null hypothesis that
\begin{align}\label{0104.5}
	& \bbY(i,k) -\bbY(j,k)=\frac{\widehat\bv_{k}(i)}{\widehat\bv_{1}(i)}-\frac{\widehat\bv_{k}(j)}{\widehat\bv_{1}(j)}\nonumber \\
&=	\frac{\bbv_k(i)+\frac{\bbe_i^T\bbW\bbv_k}{t_k}+O_{\prec}(\frac{1}{n^{1/2-c_2}|t_k|})}{\bbv_1(i)+\frac{\bbe_i^T\bbW\bbv_k}{t_1}+O_{\prec}(\frac{1}{n^{1/2-c_2}|t_1|})}-\frac{\bbv_k(j)+\frac{\bbe_j^T\bbW\bbv_k}{t_k}+O_{\prec}(\frac{1}{n^{1/2-c_2}|t_k|})}{\bbv_1(j)+\frac{\bbe_j^T\bbW\bbv_k}{t_1}+O_{\prec}(\frac{1}{n^{1/2-c_2}|t_1|})} \nonumber\\ &=\frac{\be_i^T\bW\bbv_k}{t_k\bbv_1(i)}-\frac{\be_j^T\bbW\bbv_k}{t_k\bbv_1(j)}-\frac{\bbv_k(i)\be_i^T\bbW\bbv_1}{t_1\bbv^2_1(i)}+\frac{\bbv_k(j)\be_j^T\bbW\bbv_1}{t_1\bbv^2_1(j)}+O_{\prec}(\frac{n^{c_2}}{|t_k|})\nonumber\\
	&=\frac{\bbe_i^T\bbW[\bbv_k-\frac{t_k\bbv_k(i)}{t_1\bbv_1(i)}\bbv_1]}{t_k\bbv_1(i)}-\frac{\bbe_j^T\bbW[\bbv_k-\frac{t_k\bbv_k(j)}{t_1\bbv_1(j)}\bbv_1]}{t_k\bbv_1(j)}+O_{\prec}(\frac{n^{c_2}}{|t_k|}).
	\end{align}
	
Denote by  $\bby_k=\frac{\bbv_k-\frac{t_k\bbv_k(i)}{t_1\bbv_1(i)}\bbv_1}{t_k\bbv_1(i)}$ and $\bbz_k=\frac{\bbv_k-\frac{t_k\bbv_k(j)}{t_1\bbv_1(j)}\bbv_1}{t_k\bbv_1(j)}$. Then we have $f_k = \bbe_i^T\bW\bby_k-\bbe_j^T\bW\bbz_k$ with $f_k$ defined in (\ref{new.eq002}),   and
\begin{align}\label{1231.10h}
\bbY(i,k)-\bbY(j,k) = f_k + O_{\prec}(\frac{n^{c_2}}{|t_k|}).
\end{align}
To establish the central limit theorem, we need to compare the ord4er of the variance of $f_k$ with that of the residual term $O_{\prec}(\frac{n^{2c_2}}{t_k^2})$. The variance of $f_k$ is 	\begin{equation}\label{1231.8h}
	\var(f_k)=\sum_{l=1}^n\var(w_{il})\bby_k^2(l)+\sum_{l=1}^n\var(w_{jl})\bbz_k^2(l)-\var(w_{ij})\left[\bby_k(i)\bbz_k(j)+\bby_k(j)\bbz_k(i)\right].
	\end{equation}
By Statements 1 and 2 at the beginning of this proof and (\ref{1231.6h}), we can conclude that $\max_{1\leq l \leq n}\{|\bby_k(l)|,|\bbz_k(l)|\}=O(\frac{1}{|t_k|})$ and $\bby_k(l)\sim \bbz_k(l)$, $l=1,\cdots,n$. Consequently, we obtain % \textcolor{red}{should it be $1/t_k^2$ on the right hand side?}{\color{blue} Yes it is.}
	\begin{align}\label{1231.9hh}
	\var(w_{ij})\left[\bby_k(i)\bbz_k(j)+\bby_k(j)\bbz_k(i)\right]
	=O(\frac{1}{t_k^2}).
	\end{align}
 By Condition \ref{cond7}, it holds that  $(n\theta^2_{\max})^{-1}d_k^2\var(f_k)=(n\theta^2_{\max})^{-1}d_k^2\var(\bbe_i^T\bW\bby_k-\bbe_j^T\bW\bbz_k)\sim 1$. Combining the previous two results and by Statement 4, the last term on the left hand side of \eqref{1231.8h} is asymptotically negligible compared to the right hand side.

Note that under the null hypothesis $\bpi_i=\bpi_j$ and model (\ref{DCMM}), we have
	$\frac{\bbH_{il}}{\theta_{i}}=\frac{\bbH_{jl}}{\theta_j}$. Since $\bbX = \bbH + \bbW$ with $\bbW$ a generalized Wigner matrix, it follows from the properties of Bernoulli random variables that $\var(w_{il})\sim \var(w_{jl})$. Thus the first two terms on the left hand side of (\ref{1231.8h}) are comparable and satisfy that
	\begin{align}\label{1231.9h}
	\nonumber &(n\theta^2_{\max})^{-1}d_k^2\var\left(\bbe_i^T\bW\bby_k\right)=(n\theta^2_{\max})^{-1}d_k^2\sum_{l=1}^n2\var(w_{il})\bby_k^2(l)\\
	&\sim (n\theta^2_{\max})^{-1}d_k^2\sum_{l=1}^n2\var(w_{jl})\bbz_k^2(l)= (n\theta^2_{\max})^{-1}d_k^2\var\left(\bbe_j^T\bW\bbz_k\right) \nonumber \\
	&\sim (n\theta^2_{\max})^{-1}d_k^2\var(f_k)\sim 1.
	\end{align}
%where $t_k^2\sim n^2\theta_{\min}^4$ by Lemmas \ref{lem: dk-alphan}  and \ref{lem: tk}.
Consequently, $\var(\bbe_i^T\bW\bby_k)\sim \var(\bbe_i^T\bW\bbz_k)\sim \var(f_k)$.

Now we are ready  to check the conditions of Lemma \ref{jclt}.   By  $\max_l\{|\bby_k(l)|,|\bbz_k(l)|\}=O(\frac{1}{|t_k|})$ (see (\ref{1231.9hh}) above) and noticing that the expectations of the off-diagonal entries of $\bbW$ are zero, we have
	$|\mathbb{E}(f_k)|=|\mathbb{E}(\bbe_i^T\bW\bby_k-\bbe_j^T\bW\bbz_k)|=|\mathbb{E}(w_{ii}\bby_k(i)-w_{jj}\bbz_k(j))|\le |\bby_k(i)|+|\bbz_k(j)|=O(\frac{1}{|t_k|})$, which means that the expectation of $\bbe_i^T\bW\bby_k-\bbe_j^T\bW\bbz_k$ is asymptotically negligible compared to its standard deviation. Moreover, by \eqref{1231.9h} it holds that  $\max_l\{|\bby_k(l)|,|\bbz_k(l)|\}\ll \min_{1\le k\le K}\min\{\text{SD}(\bbe_i^T\bW\bby_k), \text{SD}(\bbe_j^T\bW\bbz_k)\}$ and hence they satisfy the conditions of Lemma \ref{jclt} with $h_n=n\theta_{\max}^2$. Thus we arrive at
	\begin{equation}\label{1117.4}
	\cov(\bbe_i^T\bW\bby_2,\bbe_j^T\bW\bbz_2,\cdots,\bbe_j^T\bW\bbz_K)^{-1/2}(\bbe_i^T\bW\bby_2,\bbe_j^T\bW\bbz_2,\cdots,\bbe_j^T\bW\bbz_K)^T \toD N(\bf{0},\bbI).
	\end{equation}
Using the compact notation, (\ref{1117.4}) can be rewritten as
\begin{equation}\label{1117.4h}
\bSig_2^{-1/2}(f_2,\cdots,f_K)^T \toD N(\mathbf{0},\bbI).
\end{equation}
Furthermore, there exists some positive constant $\ep$ such that $\text{SD}(f_k)\sim \frac{\sqrt{n}\theta_{\max}}{|t_k|}\gg n^{\ep}\frac{n^{c_2}}{|t_k|}$ by Condition \ref{cond5}. Hence $O_{\prec}(\frac{n^{c_2}}{|t_k|})$ involved in (\ref{1231.10h})  is negligible compared to $f_k$. Finally, we can obtain from \eqref{1231.10h} and (\ref{1117.4h}) that
$$\bSig_2^{-1/2}(\bbY_i-\bbY_j) \toD N(\bf{0},\bbI), $$
which completes the proof for part i) of Theorem \ref{0104-1}.

It remains to prove part ii) of Theorem \ref{0104-1}. Under the alternative hypothesis that $\bpi_i\neq \bpi_j$, we have the generalized asymptotic expansion
 \begin{eqnarray}\label{1217.12h}
	\bbY(i,k)-\bbY(j,k)=\frac{\bbv_k(i)}{\bbv_1(i)}-\frac{\bbv_k(j)}{\bbv_1(j)}+\bbe_i^T\bbW\bby_k-\bbe_j^T\bbW\bbz_k+O_{\prec}(\frac{n^{c_2}}{|t_k|}).
	\end{eqnarray}
In view of (\ref{1117.4}), to complete the proof it suffices to show that
\begin{eqnarray}\label{1217.13h}
\left\|\frac{\bbV(i)}{\bbv_1(i)}-\frac{\bbV(j)}{\bbv_1(j)}\right\|\gg \frac{1}{n^{1/2-c_2}\theta_{\min}}.
\end{eqnarray}
Denote by $\bbB(i)$ the $i$th column of matrix $\bbB$ in (\ref{0119.5h}).  It follows from (\ref{0119.5h}) that
$$\frac{\bbV(i)}{\bbv_1(i)}=\frac{\bpi_i^T\bbB}{\bpi_i^T\bbB(1)} \  \text{ and } \  \frac{\bbV(j)}{\bbv_1(j)}=\frac{\bpi_j^T\bbB}{\bpi_j^T\bbB(1)}.$$
Let $a_i=\bpi_i^T\bbB(1)$ and $a_j=\bpi_j^T\bbB(1)$. Note that by Statements 1 and 2 at the beginning of this proof, we have  $\bbv_1(i)\sim \bbv_1(j)\sim \frac{1}{
	\sqrt n}$. In light of (\ref{0119.5h}), it holds that  $\bbv_1(i)=\theta_ia_i$ and  $\bbv_1(j)=\theta_ja_j$. Combining these two results yields
$$a_i\sim a_j\sim \frac{1}{\sqrt n \theta_{\min}}.$$
Moreover, it holds that
$$\frac{\bpi_i^T\bbB}{\bpi_i^T\bbB(1)}-\frac{\bpi_j^T\bbB}{\bpi_j^T\bbB(1)}=(a_i^{-1},-a_j^{-1})(\bpi_i,\bpi_j)^T\bbB,$$
which entails that
$$\left\|\frac{\bbV(i)}{\bbv_1(i)}-\frac{\bbV(j)}{\bbv_1(j)}\right\|^2\ge \|(a_i^{-1},-a_j^{-1})\|^2\lambda_{\min}((\bpi_i,\bpi_j)^T(\bpi_i,\bpi_j))\lambda_{\min}(\bbB\bbB^T). $$
Here $\lambda_{\min}(\cdot)$ stands for the smallest eigenvalue. By (\ref{0119.5h}), similar to (\ref{0119.4}) we can show that
$$\bbB\bbB^T=(\bPi^T\bTheta^2\bPi)^{-1}.$$
Thus $\lambda_{\min}(\bbB\bbB^T)\sim \frac{1}{n\theta_{\min}^2}$. By the condition that
$\lambda_2(\bpi_i\bpi_i^T+\bpi_j\bpi_j^T)\gg \frac{1}{n^{1-2c_2}\theta^2_{\min}}$ in Theorem \ref{0104-1}, it holds that
$$\lambda_{\min}((\bpi_i,\bpi_j)^T(\bpi_i,\bpi_j))=\lambda_2(\bpi_i\bpi_i^T+\bpi_j\bpi_j^T)\gg \frac{1}{n^{1-2c_2}\theta^2_{\min}}.$$
Therefore, combining the above arguments results in
$$\left\|\frac{\bbV(i)}{\bbv_1(i)}-\frac{\bbV(j)}{\bbv_1(j)}\right\|^2\gg \frac{1}{n^{1-2c_2}\theta^2_{\min}},$$
which concludes the proof of Theorem \ref{0104-1}.

\subsection{Proof of Theorem \ref{0524-1}} \label{SecA.4}
The arguments for the proof of Theorem \ref{0524-1} are similar to those for the proof of Theorem \ref{0519-1} in Section \ref{SecA.2}.

\subsection{Proof of Theorem \ref{0425-1}} \label{SecA.5}
By Lemma \ref{prop: k-threshold}, \eqref{eq: khat} holds. Since $\widehat K$ is bounded with probability tending to 1,  it suffices to show the entrywise convergence of {\color{black}$\widehat\bSig_1=\theta^{-1}\bbD\widehat\bbS_1\bbD$} and $\widehat\bSig_2=(n\theta_{\max})^{-1}\bbD\widehat\bbS_2\bbD$. As will be made clear later, the proof relies heavily on the asymptotic expansions of $(\widehat\bSig_1)_{11}$, $(\widehat\bSig_1)_{12}$, $(\widehat\bSig_2)_{11}$, and $(\widehat\bSig_2)_{12}$. We will provide only the full details on the convergence of  $(\widehat\bSig_1)_{11}$. For the other cases, the asymptotic expansions will be provided and the technical details will be mostly omitted since the arguments of the proof are similar. Throughout the proof, we will use repeatedly the results in Lemma \ref{prop-expansition}, and the node indices $i$ and $j$ are fixed.

%This section highly depends on Lemma \ref{prop-expansition}. Here $\|\bbV\|_{\infty}=O(\frac{1}{\sqrt n})$ in Lemma \ref{prop-expansition} holds by Lemma \ref{egv}.
%Since $w_{ij}$ follows bernoulli distribution, by Chebyshev inequality it is straight forward to get that
%\begin{equation}\label{1119.2}
% w_{ij}=O_{\prec}()
%\end{equation}

We start with considering  $(\widehat{\bSig}_1)_{11}$. First, by definitions of $\widehat\bW$ we have the following expansions
%{\color{blue}
\begin{eqnarray}\label{0425.4}
(\theta^{-1}\bbD\bold\Sigma_1\bbD)_{11}=\theta^{-1}\sum_{t=i,j,\ 1\le l\le n }\Big[\sigma^2_{tl}\bbv^2_1(l)-2\sigma^2_{ij}\bbv_1(j)\bbv_1(i)\Big]
\end{eqnarray}
and
\begin{eqnarray}\label{0425.4h}
(\widehat\bSig_1)_{11}=(\theta^{-1}\bbD\widehat\bbS_1\bbD)_{11}=\theta^{-1}\sum_{t=i,j,\ 1\le l\le n}\Big[\widehat w^2_{tl}\bbv^2_1(l)-2\widehat w^2_{ij}\widehat \bbv_1(l)\widehat  \bbv_1(i)\Big].
\end{eqnarray}
%}
%Recall that we have proved in  (\ref{1117.10}) in Section \ref{SecB.9} of Supplementary Material that $\widehat d_k=t_k+O_{\prec}(\sqrt{\theta})$. This together with $|t_k|\sim n\theta$ entails
%\begin{eqnarray}\label{0425.5}
%\frac{n^2\theta}{t_k^2}=\frac{n^2\theta}{\widehat d_k^2}+O_{\prec}(\frac{1}{n\theta^{3/2}}).
%\end{eqnarray}
It follows from  Lemma \ref{0429-11} in Section \ref{SecB.9} of Supplementary Material that $\widehat w^2_{ij}\widehat\bbv_1(j)\widehat\bbv_1(i)=O_{\prec}(\frac{1}{n})$.
In addition,  by Lemmas \ref{egv} and \ref{lem: dk-alphan} it holds that
	\begin{eqnarray}\label{0425.6h}
	\var\Big[\sum_{1\le l\le n}(w^2_{il}-\sigma^2_{il})\bbv^2_1(l)\Big]\le \sum_{1\le l \le n}\bbv^4_1(l)\mathbb{E}w^2_{il}=O(\frac{1}{n^2})(\alpha_n^2+1)=O(\frac{\theta}{n}).
	\end{eqnarray}
The same inequality also holds for $\var[\sum_{1\le l\le n}(w^2_{jl}-\sigma^2_{jl})\bbv^2_1(l)]$. %\textcolor{red}{this is the same as in (77) above, did you mean  $j$ instead of $i$?}
	Thus we have
	\begin{eqnarray}\label{0425.6h.1}
	\sum_{t=i,j,\, 1\le l\le n}(w^2_{tl}-\sigma^2_{tl})\bbv^2_1(l)%=\sum_{t=i,j}\mathbb{E}(w^2_{tl}-\sigma^2_{tl})^2\bbv^4_1(l)
	=O_{p}(\frac{\sqrt \theta}{\sqrt n}),
	\end{eqnarray}
which implies that
\begin{eqnarray}\label{0425.7}
\sum_{t=i,j,\,1\le l\le n}w^2_{tl}\bbv^2_1(l)=\sum_{t=i,j,\,1\le l\le n}\sigma^2_{tl}\bbv^2_1(l)+O_{p}(\frac{\sqrt \theta}{\sqrt n}).
\end{eqnarray}

By Lemmas \ref{lem: dk-alphan} and \ref{prop-expansition}, we have
$$\widehat\bbv_k(j)=\bbv_k(j)+\frac{\bbe_j^T\bbW\bbv_k}{t_k}+O_{\prec}(\frac{1}{n^{3/2-2c_2}\theta}).$$
It follows from Corollary \ref{cor} in Section \ref{SecC.2} and Lemma \ref{precl} in Section \ref{SecC.4} of Supplementary Material that
\begin{eqnarray}\label{0425.8}
&&\sum_{t=i,j,1\le l\le n}w_{tl}^2[\bbv^2_1(l)-\widehat\bbv^2_1(l)]=2\sum_{t=i,j,1\le l\le n}w_{tl}^2\bbv_1(j)[\bbv_1(l)-\widehat\bbv_1(l)]+O_{\prec}(n^{2c_2-1})\non
&&=-\frac{2}{t_1}\sum_{t=i,j,1\le l\le n}w_{tl}^2\bbv_1(l)\bbe_l^T\bbW\bbv_1+O_{\prec}(\frac{1}{n^{1-2c_2}})\non
&&=O_{\prec}(\frac{\sqrt{\theta}}{n^{1/2-c_2}}).
\end{eqnarray}
Similarly, by Lemma \ref{0429-11} we have
\begin{eqnarray}\label{0425.9}
\sum_{t=i,j,1\le l\le n}^nw^2_{tl}\widehat\bbv^2_1(l)=\sum_{t=i,j,1\le l\le n}^n\widehat w^2_{tl}\widehat\bbv^2_k(l)+O_{\prec}(\frac{\sqrt\theta}{n^{1/2-c_2}}).
\end{eqnarray}
Combining the equalities (\ref{0425.4})--(\ref{0425.9}) yields
\begin{equation}\label{0425.11}
(\widehat\bSig_1)_{11}=\theta^{-1}(\bbD\bSig_{1}\bbD)_{11}+O_{\prec}(\frac{1}{n^{1/2-c_2}\sqrt{\theta} })+O_{p}(\frac{1}{\sqrt{n\theta} })=\theta^{-1}(\bSig_{1})_{11}+o_p(1),
\end{equation}
where we have used $O_{\prec}(\frac{1}{n^{1/2-c_2}\sqrt{\theta} })=o_p(1)$ by Condition \ref{cond1}.
This has proved the convergence of $(\widehat\bSig_1)_{11}$ to $(\bSig_1)_{11}$.

We next consider $(\widehat{\bSig}_{1})_{12}$. By definitions, we have the following expansions
\begin{eqnarray}\label{0425.10}
(\theta^{-1}\bbD\bold\bSig_1\bbD)_{12}=\theta^{-1}\Big\{\sum_{t=i,j}\sigma^2_{tl}\bbv_1(l)\bbv_2(l)-\sigma^2_{ij}[\bbv_1(j)\bbv_2(i)+\bbv_1(i)\bbv_2(j)]\Big\}
\end{eqnarray}
and
\begin{eqnarray}\label{0425.10h}
(\widehat\bSig_1)_{12}=\theta^{-1}\Big\{\sum_{t=i,j}\widehat w^2_{tl}\widehat\bbv_1(l)\widehat\bbv_2(l)-\widehat w^2_{ij}[\widehat\bbv_1(j)\widehat\bbv_2(i)+\widehat\bbv_1(i)\widehat\bbv_2(j)]\Big\}.
\end{eqnarray}
Based on the above two expansions, using similar arguments to those for proving (\ref{0425.11}) we can show that
\begin{equation}\label{0425.12}
(\widehat\bSig_1)_{12}=\theta^{-1}(\bbD\bold\Sigma_1\bbD)_{12}+o_p(1).
\end{equation}

Now let us consider $\widehat{\bSig}_2$. Similar as above, we will provide only the asymptotic expansions for  $(\widehat{\bold\Sigma}_2)_{11}$ and $(\widehat{\bold\Sigma}_2)_{12}$, and the remaining arguments are similar. By definitions, we can deduce that
%{\color{blue}
	\begin{align*}
&((n\theta^2_{\max})^{-1}\bbD\bold\Sigma_2\bbD)_{11}=(n\theta^2_{\max})^{-1}d_2^2\var(f_2)\\
&=\frac{d_2^2}{t^2_2n\theta_{\max}^2}\Big\{\sum_{l\neq j}\sigma^2_{il}\Big[\frac{\bbv_2(l)}{\bbv_1(i)}-\frac{t_2\bbv_2(i)\bbv_1(l)}{t_1\bbv_1(i)^2}\Big]^2+\sum_{l\neq i}\sigma^2_{jl}\Big[\frac{\bbv_2(l)}{\bbv_1(j)}-\frac{t_2\bbv_2(j)\bbv_1(l)}{t_1\bbv_1(j)^2}\Big]^2\\
&\quad+\sigma^2_{ij}\Big[\frac{\bbv_2(j)}{\bbv_1(i)}-\frac{t_2\bbv_2(i)\bbv_1(j)}{t_1\bbv_1(i)^2}-\frac{\bbv_2(i)}{\bbv_1(j)}+\frac{t_2\bbv_2(j)\bbv_1(i)}{t_1\bbv_1(j)^2}\Big]^2\Big\}
\end{align*}
%}
and
$$(\widehat\bSig_2)_{11}=\frac{d_2^2}{\widehat d^2_2n\theta_{\max}^2}\Big\{\sum_{l\neq j}\widehat w^2_{il}\Big[\frac{\widehat\bbv_2(l)}{ \widehat\bbv_1(i)}-\frac{\widehat d_2\widehat\bbv_2(i)\widehat\bbv_1(l)}{\widehat d_1\widehat\bbv_1(i)^2}\Big]^2+\sum_{l\neq i} \widehat w^2_{jl}\Big[\frac{\widehat\bbv_2(l)}{\widehat\bbv_1(j)}-\frac{\widehat d_2\widehat\bbv_2(j)\widehat\bbv_1(l)}{\widehat d_1\widehat\bbv_1(j)^2}\Big]^2$$
$$+\widehat w^2_{ij}\Big[\frac{\widehat\bbv_2(j)}{\widehat\bbv_1(i)}-\frac{\widehat d_2\widehat\bbv_2(i)\widehat\bbv_1(j)}{\widehat d_1\widehat\bbv_1(i)^2}-\frac{\widehat\bbv_2(i)}{\widehat\bbv_1(j)}+\frac{\widehat d_2\widehat\bbv_2(j)\widehat\bbv_1(i)}{\widehat d_1\widehat\bbv_1(j)^2}\Big]^2\Big\}.$$
Note that the expression of $(n\theta^2_{\max}\bbD\bold\Sigma_2\bbD)_{11}$ is essentially the same as (\ref{0425.4}) up to a normalization factor involving $\bbv_1(i)$ and $\bbv_1(j)$. Thus applying the similar arguments to those for proving (\ref{0425.1}), we can establish the desired result.

Finally, the consistency of $(\widehat\bSig_2)_{12}$ can also be shown similarly using the following expansions
%{\color{blue}
\begin{align*}
&((n\theta^2_{\max})^{-1}\bbD\bold\Sigma_2\bbD)_{12}\\
&=\frac{d_2d_3}{t_2t_3n\theta_{\max}^2}\Big\{\sum_{l\neq j}\sigma^2_{il}\Big[\frac{\bbv_2(l)}{\bbv_1(i)}-\frac{t_2\bbv_2(i)\bbv_1(l)}{t_1\bbv_1(i)^2}\Big]\Big[\frac{\bbv_3(l)}{\bbv_1(i)}-\frac{t_3\bbv_3(i)\bbv_1(l)}{t_1\bbv_1(i)^2}\Big]\non
&\quad+\sum_{l\neq i}\sigma^2_{jl}\Big[\frac{\bbv_2(l)}{\bbv_1(j)}-\frac{t_2\bbv_2(j)\bbv_1(l)}{t_1\bbv_1(j)^2}\Big]\Big[\frac{\bbv_2(l)}{\bbv_1(j)}-\frac{t_3\bbv_3(j)\bbv_1(l)}{t_1\bbv_1(j)^2}\Big]\\
&\quad+\sigma^2_{ij}\Big[\frac{\bbv_2(j)}{\bbv_1(i)}-\frac{t_2\bbv_2(i)\bbv_1(j)}{t_1\bbv_1(i)^2}-\frac{\bbv_2(i)}{\bbv_1(j)}+\frac{t_2\bbv_2(j)\bbv_1(i)}{t_1\bbv_1(j)^2}\Big]\\
&\quad \times \Big[\frac{\bbv_3(j)}{\bbv_1(i)}-\frac{t_3\bbv_3(i)\bbv_1(j)}{t_1\bbv_1(i)^2}-\frac{\bbv_3(i)}{\bbv_1(j)}+\frac{t_3\bbv_3(j)\bbv_1(i)}{t_1\bbv_1(j)^2}\Big]\Big\}
\end{align*}
%}
and
\begin{align*}
&(\widehat\Sigma_2)_{12}=\frac{d_2d_3}{\widehat d_2\widehat d_3n\theta_{\max}^2}\Big\{\sum_{l\neq j}\widehat w^2_{il}\Big[\frac{\widehat\bbv_2(l)}{\widehat\bbv_1(i)}-\frac{\widehat d_2\widehat\bbv_2(i)\widehat\bbv_1(l)}{\widehat d_1\widehat\bbv_1(i)^2}\Big]\Big[\frac{\widehat\bbv_3(l)}{\widehat\bbv_1(i)}-\frac{\widehat d_3\widehat\bbv_3(i)\widehat\bbv_1(l)}{\widehat d_1\widehat\bbv_1(i)^2}\Big]\non
&\quad+\sum_{l\neq i} \widehat w^2_{jl}\Big[\frac{\widehat\bbv_2(l)}{\widehat\bbv_1(j)}-\frac{\widehat d_2\widehat\bbv_2(j)\widehat\bbv_1(l)}{\widehat d_1\widehat\bbv_1(j)^2}\Big]\Big[\frac{\widehat\bbv_3(l)}{\widehat\bbv_1(j)}-\frac{\widehat d_3\widehat\bbv_3(j)\widehat\bbv_1(l)}{\widehat d_1\widehat\bbv_1(j)^2}\Big]\non
&\quad+\widehat w^2_{ij}\Big[\frac{\widehat\bbv_2(j)}{\widehat\bbv_1(i)}-\frac{\widehat d_2\widehat\bbv_2(i)\widehat\bbv_1(j)}{\widehat d_1\widehat\bbv_1(i)^2}-\frac{\widehat\bbv_2(i)}{\widehat\bbv_1(j)}+\frac{\widehat d_2\widehat\bbv_2(j)\widehat\bbv_1(i)}{\widehat d_1\widehat\bbv_1(j)^2}\Big]\\
&\quad\times\Big[\frac{\widehat\bbv_3(j)}{\widehat\bbv_1(i)}-\frac{\widehat d_3\widehat\bbv_3(i)\widehat\bbv_1(j)}{\widehat d_1\widehat\bbv_1(i)^2}-\frac{\widehat\bbv_3(i)}{\widehat\bbv_1(j)}+\frac{\widehat d_3\widehat\bbv_3(j)\widehat\bbv_1(i)}{\widehat d_1\widehat\bbv_1(j)^2}\Big]\Big\}.
\end{align*}
This completes the proof of Theorem \ref{0425-1}.

\renewcommand{\theequation}{B.\arabic{equation}}
\renewcommand{\thesubsection}{B.\arabic{subsection}}
\setcounter{equation}{0}

\section{Some key lemmas and their proofs} \label{SecB}

\subsection{Proof of Lemma \ref{prop: k-threshold}} \label{SecB.1}
For each pair $(i,j)$ with $i\neq j$, let us define a matrix $\bbW(i,j)=w_{ij}(\bbe_i\bbe_j^T+\bbe_j\bbe_i^T)$.  For $i=j$, we define a matrix $\bbW(i,j)=(w_{ii}-\mathbb{E}w_{ii})\bbe_i\bbe_j^T$.
Then it is easy to see that
\begin{equation}\label{1217.15h}
	\|\sum_{1\le i\le j\le n}\bbW(i,j)-\bbW\|=\|\diag(\bbW-\mathbb{E}\bbW)\|\le  1.
\end{equation}
It is straightforward to show that
$$\|\sum_{1\le i\le j\le n}\mathbb{E}\bbW(i,j)^2\|=\alpha_n^2.$$
By Theorem 6.2 of \cite{T12}, for any constant $c>\sqrt 2$ we have
\begin{eqnarray}\label{0531.1h}
	\mathbb{P}(\|\sum_{1\le i\le j\le n}\bbW(i,j)\|\ge c\sqrt{\log n}\alpha_n-1)\le n\exp\Big[\frac{-(c\sqrt{\log n}\alpha_n-1)^2}{2\alpha_n^2+2(c\sqrt{\log n}\alpha_n-1)}\Big]=o(1).
\end{eqnarray}
This together with (\ref{1217.15h}) entails that
\begin{eqnarray}\label{0531.1}
	\mathbb{P}(\|\bbW\|\le c\sqrt{\log n}\alpha_n)\ge 1-o(1).
\end{eqnarray}
Note that this result is weaker than Lemma \ref{0505-1} in Section \ref{SecC.5}.

By (\ref{0531.1})  and $|\widehat d_K-d_K|\le \|\bbW\|$, and using the assumption of $|d_K|\gg \sqrt{\log n}\alpha_n$, it holds that
\begin{equation}\label{1120.1}
	|\widehat d_K|\gg \sqrt{\log n}\alpha_n
\end{equation}
with probability tending to one.
Finally, by Weyl's inequality we
have
$$\lambda_n(\bbW)=\lambda_n(\bbW)-\lambda_{K+1}(\bbH)\le \lambda_{K+1}(\bbX)= \lambda_{K+1}(\bbH+\bbW)\le \lambda_1(\bbW)+\lambda_{K+1}(\bbH)=\lambda_1(\bbW),$$
which leads to
\begin{equation}\label{1120.2}
	|\widehat d_{K+1}|=|\lambda_{K+1}(\bbX)|\le \|\bbW\|.
\end{equation}
Let us choose $c=\sqrt{2.01}$ and define
\begin{equation}\label{0531.2}
	\widetilde K=\#\left\{|\widehat d_i|>\sqrt{2.01\log n}\alpha_n, i=1,\cdots,n\right\}.
\end{equation}
Then by (\ref{1120.1})--(\ref{1120.2}), we can show that
\begin{equation}\label{0531.3}\mathbb{P}(\widetilde K=K)=1-o(1).
\end{equation}

Recall that $X_{ij}$ follows the Bernoulli distribution. %and $\max\{\mathbb{E}X_{ij}\}<1$.
 Thus it holds that
$$\sum_{j=1}\mathbb{E}w^2_{ij}\le \sum_{j=1}\mathbb{E}X_{ij}.$$% \ \text{ and } \ \sum_{j=1}\mathbb{E}w^2_{ij}\sim \sum_{j=1}\mathbb{E}X_{ij}.$$
By Lemma \ref{0426-1h} in Section \ref{SecC.1}, choosing $l=1$, $\bbx=\bbe_i$, and $\bby=\frac{1}{\sqrt n}\mathbf{1}$ yields
$$\sum_{j=1}\mathbb{E}X_{ij}=\sum_{j=1}X_{ij}+O_{\prec}(\alpha_n),$$
where we have used $X_{ij}-\mathbb{E}X_{ij}=w_{ij}$.
Thus it holds that
$$\max_{i}\sum_{j=1}X_{ij}\ge \max_{i}\sum_{j=1}\mathbb{E}w^2_{ij}+O_{\prec}(\alpha_n)= \alpha_n^2+O_{\prec}(\alpha_n).$$
This together with \eqref{0531.2} and (\ref{0531.3}) results in
\begin{equation}\label{0531.4}
	\mathbb{P}(\widehat K=K)=1-o(1),
\end{equation}
which completes the proof of Lemma \ref{prop: k-threshold}.

\subsection{Proofs of Lemmas \ref{0516-2} and \ref{0516-3}} \label{SecB.2}
The proofs of Lemmas \ref{0516-2} and \ref{0516-3} involve standard calculations and thus are omitted for brevity.

\subsection{Lemma \ref{jclt} and its proof} \label{SecB.3}
\begin{lem} \label{jclt}
Let $m$ be a fixed positive integer, $\bbx_i$ and $\bby_i$ be  $n$-dimensional unit vectors for $1 \leq i \leq m$, and $\bSig = (\Sigma_{ij})$ the covariance matrix with
$\Sigma_{ij}=\cov(\bbx_i^T\bbW\bby_i,\bbx_j^T\bbW\bby_j)$. Assume that there exists some positive sequence $(h_n)$
%(may tend to $0$)
such that  $\|\Sig^{-1}\|\sim\|\Sig\|\sim h_n$ and $\max_{k} \{\|\bbx_k\|_\infty \|\bby_k\|_\infty\}
%\|\bbx_k\bby^T_k\|_{\infty}
\ll\|\Sig^{1/2}\|$. Then it holds that
	\begin{eqnarray}\label{0625.1}
		\Sig^{-1/2}\left(\bbx_1^T(\bbW-\mathbb{E}\bbW)\bby_1,\cdots, \bbx_m^T(\bbW-\mathbb{E}\bbW)\bby_m\right)^T\toD N(\bf{0},\bbI).
	\end{eqnarray}
\end{lem}

\noindent \textit{Proof}. Note that it suffices to show that for any unit vector $\bbc=(c_1,\cdots,c_m)^T$, we have
\begin{equation}\label{neq1}
\bbc^T\Sig^{-1/2}\big(\bbx_1^T(\bbW-\mathbb{E}\bbW)\bby_1,\cdots, \bbx_m^T(\bbW-\mathbb{E}\bbW)\bby_m\big)^T\toD N(0,1).
\end{equation}
Let $\bx_i = (x_{1i}, \cdots, x_{ni})^T$ and $\by_i = (y_{1i}, \cdots, y_{ni})^T$, $i=1,\cdots,m$.
Since $\bbW$ is a symmetric random matrix of independent entries on and above the diagonal, we can deduce
\begin{equation} \label{neweq020}
\bbx^T_i\bbW\bby_i-\bbx_i^T\mathbb{E}\bbW\bby_i=\sum_{1 \leq s, t \leq n,\, s < t}w_{st}(x_{si}y_{ti}+x_{ti}y_{si})+\sum_{1 \le s \leq n}(w_{ss}-\mathbb{E}w_{ss}) x_{si}y_{si}
\end{equation}
and
\begin{eqnarray} \label{neweq021}
&&s^2_n\coloneqq\var\left[\bbc^T\Sig^{-1/2}(\bbx_1^T(\bbW-\mathbb{E}\bbW)\bby_1,\cdots, \bbx_m^T(\bbW-\mathbb{E}\bbW)\bby_m)^T\right]\non
&&=\bbc^T\Sig^{-1/2}\cov\left[(\bbx_1^T\bbW\bby_1,\cdots, \bbx_m^T\bbW\bby_m)^T\right]\Sig^{-1/2}\bbc=\bbc^T\bbc=1.
\end{eqnarray}
Denote by $\widetilde\bbc=\Sig^{-1/2}\bbc=(\widetilde c_1,\cdots,\widetilde c_m)^T$. Then it holds that
$$\bbc^T\Sig^{-1/2}\big(\bbx_1^T(\bbW-\mathbb{E}\bbW)\bby_1,\cdots, \bbx_m^T(\bbW-\mathbb{E}\bbW)\bby_m\big)^T=\tr\Big[(\bbW-\mathbb{E}\bbW)\sum_{s=1}^m\widetilde c_s\bby_s\bbx_s^T\Big].$$

Let $\bbM=(M_{ij})=\sum_{s=1}^m\widetilde c_s\bby_s\bbx_s^T$. By assumption, we have  $\max_{k}\|\bbx_k\bby^T_k\|_{\infty}\ll\|\Sig^{1/2}\|\sim \|\Sig^{-1/2}\|$, which entails that
\begin{equation}\label{neq2}
\|\bbM\|_{\infty}\ll 1.
\end{equation}
Then it follows from the assumption of $\max_{1\le i, j\le n}|w_{ij}|\le 1$ and (\ref{neq2}) that
\begin{align} \label{neweq022}
\frac{1}{|s_n|^3} & \Big(\sum_{1 \leq i, j \leq n,\, i < j}\mathbb{E}|w_{ij}|^3|M_{ij}+M_{ji}|^3+\sum_{1 \le i \leq n}\mathbb{E}|w_{ii}-\mathbb{E}w_{ii}|^3|M_{ii}|^3\Big)\non
& \quad \le \frac{2}{|s_n|^3}\Big(\sum_{1 \leq i, j \leq n,\, i < j}\mathbb{E}|w_{ij}|^2|M_{ij}+M_{ji}|^3+\sum_{1 \le i \leq n}\mathbb{E}|w_{ii}-\mathbb{E}w_{ii}|^2|M_{ii}|^3\Big) \nonumber \\
& \quad
\ll \frac{2}{|s_n|^3}\Big(\sum_{1 \leq i, j \leq n,\, i < j}\mathbb{E}|w_{ij}|^2|M_{ij}+M_{ji}|^2+\sum_{1 \le i \leq n}\mathbb{E}|w_{ii}-\mathbb{E}w_{ii}|^2|M_{ii}|^2\Big) \nonumber \\
& \quad \leq 2.
\end{align}
Since $w_{ij}$ with $1\le i<j\le n$ and $w_{ii}-\mathbb{E}w_{ii}$ with $1\le i\le n$ are independent random variables with zero mean, by the Lyapunov condition (see, for example, Theorem 27.3 of \cite{B1995}) we can conclude that (\ref{neq1}) holds.
This concludes the proof of Lemma \ref{jclt}.

\subsection{Lemma \ref{lem2} and its proof} \label{SecB.4}
\begin{lem} \label{lem2}
	Under either model \eqref{0115.1} and  Conditions \ref{as3}--\ref{cond1}, or model (\ref{DCMM})  and Conditions \ref{as3} and  \ref{cond5},	it holds that
	\begin{equation}\label{1217.3h}
		\  \|(\bbD_{-k})^{-1}+\mathcal{R}(\bbV_{-k},\bbV_{-k},z)\|=O(|z|) \text{ for any } z\in [a_k,b_k],
	\end{equation}
	where $a_k$ and $b_k$ are defined in \eqref{csl}.
	%	\textcolor{red}{first result is a special case of the second so we do not need the 1st one, right?}{\color{blue} Yes it is. I delete the first result}
\end{lem}

\noindent \textit{Proof}. The conclusion of Lemma \ref{lem2} has been proved in (A.16) of \cite{FFHL19}.

\subsection{Lemma \ref{egv} and its proof} \label{SecB.6}
\begin{lem} \label{egv}
	Under model \eqref{0115.1} and Conditions \ref{as3}--\ref{cond1}, we have
	\begin{equation}\label{0119.3}
		\max_{1\le k\le K}\|\bbv_k\|_{\infty}=O(\frac{1}{\sqrt n}).
	\end{equation}
	The same conclusion also holds under model \eqref{DCMM} and  Conditions  \ref{as3} and \ref{cond5}.
\end{lem}

\noindent \textit{Proof}. We first consider model \eqref{0115.1}  and prove (\ref{0119.3}) under Conditions \ref{as3} and  \ref{cond1}.
	In light of
	$\theta\bPi\bbP\bPi^T=\bbV\bbD\bbV^T$,
	we have
	$\theta\bPi(\bbP\bPi^T\bbV\bbD^{-1})=\bbV$.
	This shows that $\bbV$ belongs to the space expanded by $\bPi$. Thus there exists some $K\times K$ matrix $\bbB$ such that
	\begin{equation}\label{0119.5}
	\bbV=\bPi\bbB.
	\end{equation}
	Since $\bbV^T\bbV=\bbI$, it holds that
	$\bbB^T\bPi^T\bPi\bbB=\bbI$,
	which entails that
	$\bbB\bbB^T\bPi^T\bPi\bbB\bbB^T=\bbB\bbB^T$ and
	\begin{equation}\label{0119.4}
	\bbB\bbB^T=(\bPi^T\bPi)^{-1}.
	\end{equation}
	By Condition \ref{cond1}, we can conclude that $\|(\bPi^T\bPi)^{-1}\|=O(n^{-1})$ and thus each entry of matrix $\bbB$ is of order $O(\frac{1}{\sqrt n})$. Hence in view of \eqref{0119.5}, the desired result can be established.
	%$$\max_{1\le k\le K}\|\bbv_k\|_{\infty}=O(\frac{1}{\sqrt n}).$$
	
	Now let us consider model (\ref{DCMM}) under Conditions \ref{as3} and \ref{cond5}. For this model, we also have
	$\bTheta\bPi\bbP\bPi^T\bTheta=\bbV\bbD\bbV^T$ and thus
	\begin{equation}\label{1229.1h}
	\bTheta\bPi(\bbP\bPi^T\bTheta\bbV\bbD^{-1})=\bbV.
	\end{equation}
	Since $\bTheta$ is a diagonal matrix, we can see that $\bbV$ belongs to the space expanded by $\bPi$. Let $\widetilde\bPi=(\widetilde\bpi_1,\cdots,\widetilde\bpi_n)^T$ be the submatrix of $\bPi$ such that
	$$\widetilde \bpi_i=\begin{cases}
	\bpi_i & \text{if there exists some $1\le k\le K$ such that}\ \bpi_i(k)=1, \cr 0 &\text{otherwise}.\end{cases}$$
	By Condition \ref{cond5}, it holds that $c_2n^{c_2}\bbI\le \widetilde\bPi^T\widetilde\bPi=\sum_{i=1}^n\widetilde \bpi_i\widetilde\bpi_i^T\le \sum_{i=1}^n \bpi_i\bpi_i^T=\bPi^T\bPi$, which leads to $\|(\bPi^T\bPi)^{-1}\|=O(n^{-1})$. % having the same order as model \eqref{0115.1}.
	Therefore, an application of similar arguments to those for (\ref{0119.5})--(\ref{0119.4}) concludes the proof of Lemma \ref{egv}.

\subsection{Lemma \ref{lem: dk-alphan} and its proof} \label{SecB.7}
\begin{lem} \label{lem: dk-alphan}
	Under  model \eqref{0115.1} and  Condition \ref{cond1} , it holds that
	\begin{equation}\label{1231.2h}
		\alpha_n^2 \leq n\theta, \quad d_k\gtrsim n^{1-c_2}\theta,\quad d_1=O(n\theta), \quad k=1,\cdots,K.
	\end{equation}
	Under  model (\ref{DCMM})  and Condition  \ref{cond5}, similarly we have
	\begin{equation}\label{1231.3h}
		\alpha_n^2 \leq n\theta^2_{\max}, \quad d_k\gtrsim n^{1-c_2}\theta_{\min}^2,\quad d_1=O(n\theta_{\max}^2), \quad k=1,\cdots,K.
	\end{equation}
\end{lem}

\noindent \textit{Proof}. We show (\ref{1231.2h}) first. It follows from  $\sum_{k=1}^K\bpi_i(k)=1$ that $\|\bPi\|_F^2=\sum_{i=1}^n\sum_{k=1}^K\bpi_i^2(k)\le n$ and  $\lambda_1(\bPi^T\bPi)=O(n)$.  By Condition \ref{cond1}, we have
	$$d_K=\theta\lambda_K(\bbP\bPi^T\bPi)\ge \theta\lambda_K(\bPi^T\bPi)\lambda_K(\bbP)\ge c_0^2\theta n^{1-c_2}$$
	and $$\ d_1\le \theta\lambda_1(\bPi^T\bPi)\lambda_1(\bbP)=O (\theta n).$$
	Thus the second result in \eqref{1231.2h} is proved. Next by model (\ref{0115.1}), the $(i,j)$th entry $h_{ij}$ of matrix $\bbH$ satisfies that
	\begin{equation}\label{1231.4h}
	h_{ij}=\theta\sum_{s,t=1}^K\bpi_i(s)\bpi_j(t)p_{st}\le \theta.
	\end{equation}
	Since the entries of $\bbX$ follow the Bernoulli distributions, it follows from (\ref{1231.4h}) that
	$\var(w_{ij})\le \theta.$
	Therefore, in view of the definition of $\alpha_n$, we have
	$$\alpha_n^2=\max_{j}\sum_{i=1}^n\var(w_{ij})\le n\theta.$$
	
	The results in (\ref{1231.3h}) can also be proved using similar arguments. This completes the proof of Lemma \ref{lem: dk-alphan}.
	%By Condition \ref{cond5}, we have
	%$$d_K=\lambda_K(\bbP\bPi^T\Theta^2\bPi)\ge \theta_{\min}^2\lambda_K(\bPi^T\bPi)\lambda_K(\bbP)\ge c_2^2\theta n, \ d_1\le \theta_{\max}^2\lambda_1(\bPi^T\bPi)\lambda_1(\bbP)=O (\theta_{\max}^2 n).$$
	%Together with the relation that $\theta_{\min}\sim \theta_{\max}$ by Condition \ref{cond5}, we conclude that
	%$$d_k\sim n\theta_{\max}^2, k=1,\cdots,K.$$
	%The remaining inequality $\alpha_n^2\le n\theta_{\max}^2$ can be proved by similar strategy as (\ref{1231.2h}) from the inequality for model \ref{DCMM} that
	%\begin{equation}\label{1231.5h}
	%\bbH_{ij}\le\theta_{\max}^2\sum_{s,t=1}^K\bpi_i(s)\bpi_j(t)p_{st}\le \theta_{\max}.
	% \end{equation}
\subsection{Lemma \ref{lem3}} \label{SecB.5}
The following 3 Lemmas follow from Lemma \ref{1212-1} and exactly the same proof as \cite{FFHL19}
\begin{lem} \label{lem3}
	Under either model \eqref{0115.1} and  Conditions \ref{as3}--\ref{cond1}, or model (\ref{DCMM})  and Conditions \ref{as3} and \ref{cond5}, for $\bbu=\bbe_i$ or $\bbv_k$ we have the following asymptotic expansions
	%For $\bu=\be_i$ or $\bbv_k$, we will prove the following improved version of  (\ref{0515.5}) in Lemma \ref{lem3}
	\begin{align} \label{0515.5h}
		\bbu^T  \widehat \bbv_k\widehat \bbv^T_k\bbv_k &%=\frac{\widehat t^2_k\bu^T\left[\bbG(\widehat t_k)-\bbF_k(\widehat t_k)\right]\bbv_k\bbv_k^T\left[\bbG(\widehat t_k)-\bbF_k(\widehat t_k)\right]\bbv_k}{\widehat t^2_k\bbv_k^T\left[\bbG'(\widehat t_k)-\bbF'_k(\widehat t_k)\right]\bbv_k}\non
		= \Big[ \mathcal{\widetilde P}_{k,t_k}-2t_k^{-1}\mathcal{\widetilde P}^2_{k,t_k}\bbv_k^T\bbW\bbv_k+O_{\prec}(\frac{\alpha_n^2 }{\sqrt nt^{2}_k})\Big] \Big[A_{\bbu,k,t_k}-t_k^{-1}\bbb^T_{\bbu,k,t_k}\bbW\bbv_k+O_{\prec}(\frac{\alpha_n^2 }{\sqrt n t^{2}_k})\Big] \nonumber\\
		&\quad\times \Big[A_{\bbv_k,k,t_k}-t_k^{-1}\bbb^T_{\bbv_k,k,t_k}\bbW\bbv_k+O_{\prec}(\frac{\alpha_n^2 }{\sqrt nt^{2}_k})\Big],\\
		\widehat d_k&=t_k+\bbv_k^T\bbW\bbv_k+O_{\prec}(\frac{\alpha_n^2}{\sqrt n |d_k|}).\label{1117.7}
	\end{align}
\end{lem}

\subsection{Lemma \ref{prop-expansition}} \label{SecB.8}
\begin{lem} \label{prop-expansition}
	Under model \eqref{0115.1} and Conditions \ref{as3}--\ref{cond1}, we have
	\begin{equation}\label{0426.1h}
		{t_k\left[\bbe_i^T\widehat\bbv_k-\bbv_k(i)\right]=\bbe_i^T\bbW\bbv_k}+O_{\prec}(\frac{\alpha_n^2}{\sqrt n|t_k| }+\frac{1}{\sqrt n}).
	\end{equation}
	The same conclusion also holds under model \eqref{DCMM} and  Conditions  \ref{as3} and \ref{cond5}.
\end{lem}

\subsection{Lemma \ref{0429-11}} \label{SecB.9}
\begin{lem} \label{0429-11}
Assume that $\widehat K=K$. Then under the mixed membership model (\ref{0115.1}) and Conditions \ref{as3}--\ref{cond1}, it holds  uniformly over all $i,j$ that
	\begin{equation}\label{1129.1}
		\widehat w_{ij}=w_{ij}+O_{\prec}(\frac{\sqrt{\theta}}{\sqrt{n}}).
	\end{equation}
Under the degree-corrected mixed membership model (\ref{DCMM}),  if Conditions \ref{as3} and \ref{cond5}--\ref{cond6} are satisfied, then it holds uniformly over all $i, j$ that
	\begin{equation}\label{1129.1h}
		\widehat w_{ij}=w_{ij}+O_{\prec}(\frac{\theta_{\max}}{\sqrt{n}}).
	\end{equation}
	%	uniformly for all $i,j$. The definition of $O_{\prec}$ is given in Definition \ref{0525.1h}.
\end{lem}

\renewcommand{\thesubsection}{C.\arabic{subsection}}
\renewcommand{\theequation}{C.\arabic{equation}}
\setcounter{equation}{0}

\section{Additional technical details} \label{SecC}

\subsection{Lemma \ref{0426-1h} and its proof} \label{SecC.1}
\begin{lem}\label{0426-1h}
	For any $n$-dimensional unit vectors $\bbx$,  $\bby$ and any positive integer $r$, we have
	\begin{equation}\label{0506.1h}
	\mathbb{E}\left[\bbx^T(\bbW^l-\mathbb{E}\bbW^l)\bby\right]^{2r}
	\leq C_r(\min\{\alpha_n^{l-1}, d_{\sbx}\alpha_n^l, d_{\sby}\alpha_n^l\})^{2r},
	\end{equation}
	where $ l$ is any positive integer and $C_r$ is some positive constant determined only by $r$.
\end{lem}

%The bounds for the higher moments of $\bbx^T(\bbW^l-\mathbb{E}\bbW^l)\bby$ is crucial for this paper.

\noindent \textit{Proof}. The main idea of the proof is similar to that for Lemma 4 in \cite{FFHL19}, which is to count the number of nonzero terms in the expansion of $\mathbb{E}[\bbx^T(\bbW^l-\mathbb{E}\bbW^l)\bby]^{2r}$. It will be made clear that the nonzero terms in the expansion consist of terms such as $w_{ij}^s$ with $s\ge 2$.  In counting the nonzero terms, we will fix one index, say $i$, and vary the other index $j$ which ranges from $1$ to $n$. Note that for any $i=1,\cdots, n$ and $s\ge 2$, we have $\sum_{j=1}^n\mathbb{E}|w_{ij}|^s\le \alpha_n^2$  since $|w_{ij}|\le 1$. Thus roughly speaking, counting the maximal moment of $\alpha_n$ is the crucial step in our proof. % In the proof of Lemma 4 in \cite{FFHL19}, we consider two cases separately: 1. Special case: The diagonal entries of $\bbW$ are zero. 2. General case: The diagonal entries of $\bbW$ can be non zero.   In this proof we directly consider the general case and then simplify the proof.
	
Let $\bx = (x_1, \cdots, x_n)^T$, $\by = (y_1, \cdots, y_n)^T$, and  $C_r$ be a positive constant depending only on $r$ and whose value may change from line to line. Recall that  $l,r \geq 1$ are two integers. We can  expand $\mathbb{E}(\bbx^T\bbW^l\bby-\mathbb{E}\bbx^T\bbW^l\bby)^{2r}$ to obtain the following expression
	\begin{align} \label{1212.1h}
	\nonumber & \mathbb{E} (\bbx^T\bbW^l\bby-\mathbb{E}\bbx^T\bbW^l\bby)^{2r}\\
	& =\sum_{1\le i_1,\cdots,i_{l+1},i_{l+2},\cdots,i_{2l+2}, \cdots,\atop i_{(2r-1)(l+1)+1},\cdots,i_{2r(l+1)}\le n
	}\mathbb{E} \Big[\left(x_{i_1}w_{i_1i_2}w_{i_2i_3}\cdots w_{i_li_{l+1}}y_{i_{l+1}}-\mathbb{E}x_{i_1}w_{i_1i_2}w_{i_2i_3}\cdots w_{i_li_{l+1}}y_{i_{l+1}}\right)\times \cdots \nonumber \\
	& \quad \times \big(x_{i_{(2r-1)(l+1)+1}}w_{i_{(2r-1)(l+1)+1}i_{(2r-1)(l+1)+2}}w_{i_{(2r-1)(l+1)+2}i_{(2r-1)(l+1)+3}}\cdots w_{i_{2r(l+1)-1}i_{2r(l+1)}}y_{i_{2r(l+1)}}\nonumber \\
	&-\mathbb{E}x_{i_{(2r-1)(l+1)+1}}w_{i_{(2r-1)(l+1)+1}i_{(2r-1)(l+1)+2}}w_{i_{(2r-1)(l+1)+2}i_{(2r-1)(l+1)+3}}\cdots w_{i_{2r(l+1)-1}i_{2r(l+1)}}y_{i_{2r(l+1)}}\big)\Big].
	\end{align}
	Let $\bbi^{(j)}=(i_{(j-1)(l+1)+1},\cdots,i_{j(l+1)})$, $j=1,\cdots,2r$, be $2r$ vectors taking values in $\{1,\cdots,n\}^{l+1}$. Then for each $\bbi^{(j)}$, we define a graph $\mathcal{G}^{(j)}$  whose vertices represent distinct values of the components of $\bbi^{(j)}$. Each adjacent component of $\bbi^{(j)}$ is  connected by an undirected edge in $\mathcal G^{(j)}$.  It can be seen that for each  $j$, $\mathcal{G}^{(j)}$ is a connected graph, which means that there exists some path connecting any two nodes in $\mathcal{G}^{(j)}$. For each fixed $i_1,\cdots,i_{l+1}, \cdots,i_{(2r-1)(l+1)+1},\cdots,i_{2r(l+1)}$, consider the following term
	\begin{align}\label{0928.3h}
	&\mathbb{E} \Big[\left(x_{i_1}w_{i_1i_2}w_{i_2i_3}\cdots w_{i_li_{l+1}}y_{i_{l+1}}-\mathbb{E}x_{i_1}w_{i_1i_2}w_{i_2i_3}\cdots w_{i_li_{l+1}}y_{i_{l+1}}\right)\times \cdots \\
	& \quad \times \big(x_{i_{(2r-1)(l+1)+1}}w_{i_{(2r-1)(l+1)+1}i_{(2r-1)(l+1)+2}}w_{i_{(2r-1)(l+1)+2}i_{(2r-1)(l+1)+3}}\cdots w_{i_{2r(l+1)-1}i_{2r(l+1)}}y_{i_{2r(l+1)}}\nonumber \\
	&-\mathbb{E}x_{i_{(2r-1)(l+1)+1}}w_{i_{(2r-1)(l+1)+1}i_{(2r-1)(l+1)+2}}w_{i_{(2r-1)(l+1)+2}i_{(2r-1)(l+1)+3}}\cdots w_{i_{2r(l+1)-1}i_{2r(l+1)}}y_{i_{2r(l+1)}}\big)\Big], \nonumber  \end{align}
	which corresponds to graph $\mathcal{G}^{(1)}\cup\cdots\cup\mathcal{G}^{(2r)}$. If there exists one graph $\mathcal{G}^{(s)}$ that is unconnected to the remaining graphs $\mathcal{G}^{(j)}$, $j\neq s$ , then the corresponding expectation in (\ref{0928.3h}) is equal to zero. This  shows that for any graph $\mathcal{G}^{(s)}$, there exists at least one connected $\mathcal{G}^{(s')}$ to ensure the nonzero expectation in (\ref{0928.3h}). To analyze each nonzero (\ref{0928.3h}), we next calculate how many distinct vertices are contained in the  graph $\mathcal{G}^{(1)}\cup\cdots\cup\mathcal{G}^{(2r)}$.
	
	Denote by $\mathfrak{S}(2r)$ the set of  partitions of the integers $\{1,2,\cdots,2r\}$ and $\mathfrak{S}_{\ge 2}(2r)$ the subset of $\mathfrak{S}(2r)$  whose block sizes are at least two. To simplify the notation, define
	$$\mathfrak{h}_j=x_{i_{(j-1)(l+1)+1}}w_{i_{(j-1)(l+1)+1}i_{(j-1)(l+1)+2}}w_{i_{(j-1)(l+1)+2}i_{(j-1)(l+1)+3}}\cdots w_{i_{j(l+1)-1}i_{j(l+1)}}y_{i_{j(l+1)}}.$$ Let $\mathcal{A}\in \mathfrak{S}_{\ge 2}(2r)$ be a partition of $\{1,2,\cdots,2r\}$ and $|\mathcal{A}|$ the number of groups in $\mathcal{A}$. We can further define $A_j\in \mathcal{A}$ as the $j$th group in $\mathcal{A}$ and $|A_j|$ as the number of integers in $A_j$. For example, let us consider $\mathcal{A}=\{\{1,2,3\}, \{4,5,\cdots,2r\}\}$. Then we have $|\mathcal{A}|=2$, set $A_1=\{1,2,3\}\in \mathcal A$, and $|A_1|=3$. It is easy to see that there is a one-to-one correspondence between the partitions of $\{1,2,\cdots,2r\}$ and the graphs $\mathcal{G}^{(1)},\cdots,\mathcal{G}^{(2r)}$ such that $\mathcal{G}^{(s)}$ and $\mathcal{G}^{(s')}$ are connected if and only if $s$ and $s'$ belong to one group  in the partition.  For any $A_j\in \mathcal{A} \in \mathfrak{S}_{\ge 2}(2r)$, there are $|A_j|l$  edges in the graph $\bigcup_{w\in A_j}\mathcal{G}^{(j)}$  since for each integer $w\in A_j$, there is a chain containing $l$ edges by $\mathfrak{h}_w$. Since $\mathbb{E}w_{ss'}=0$ for $s\neq s'$, in order to obtain a nonzero value of (\ref{0928.3h}) each edge in $\bigcup_{w\in A_j}\mathcal{G}^{(j)}$ should have at least one additional copy. Thus for each nonzero (\ref{0928.3h}), we have $[\frac{|A_j|l}{2}]$ distinct edges without self loops in  $\bigcup_{w\in A_j}\mathcal{G}^{(j)}$. Since the  graph  $\bigcup_{w\in A_j}\mathcal{G}^{(j)}$ is connected, we can  conclude that there are at most $[\frac{|A_j|l}{2}]+1$ distinct vertices in  $\bigcup_{w\in A_j}\mathcal{G}^{(j)}$.  Let $\mathcal{S}(\mathcal{A})$ be the collection of all choices of $\bigcup_{s=1}^{2r}\bbi^{(s)}$ such that
	
	1). $\bigcup_{s=1}^{2r}\mathcal{G}^{(s)}$ has the same partition as $\mathcal{A}$ such that they are connected within the same group and unconnected between groups;
	
	2). Within each group $A_j$, there are at most $[\frac{|A_j|l}{2}]$ distinct edges without self loops and $[\frac{|A_j|l}{2}]+1$ distinct vertices.
	
	Similarly we can define $\mathcal{S}(A_j)$ since $A_j$ can be regarded as a special partition of $A_j$ with only one group.
	Summarizing the arguments above, (\ref{1212.1h}) can be rewritten as %Denote by $\mathcal{S}$ the set of all such pairs $(\bbi, \bbj)$. Combining the above arguments, we can conclude that
	\begin{align}\label{0928.4h}
	(\ref{1212.1h})&=\sum_{ \mathcal{A}\in \mathfrak{S}_{\ge 2}(2r)}\sum_{ \bigcup_{s=1}^{2r}\bbi^{(s)}\in \mathcal{S}(\mathcal{A})}\prod_{j=1}^{|\mathcal{A}|}\Big[\mathbb{E}\prod_{\gamma\in A_j}(\mathfrak{h}_{\gamma}-\mathbb{E}\mathfrak{h}_{\gamma})\Big].
	\end{align}
	Let us further simplify $\mathbb{E}\prod_{\gamma\in A_j}(\mathfrak{h}_{\gamma}-\mathbb{E}\mathfrak{h}_{\gamma})$. Let $\mathcal{B}_j$ be the set of partitions of  $A_j$ such that each partition contains exactly two groups.  Without loss of generality, let $\mathcal{B}_j=\{b_{j_1},b_{j_2}\}$, where for any $w\in A_j$, we have $w\in b_{j_1}$ or $w\in b_{j_2}$. Then it holds that
	\begin{align}\label{0928.5h}
	& |\mathbb{E}\prod_{\gamma\in A_j}(\mathfrak{h}_{\gamma}-\mathbb{E}\mathfrak{h}_{\gamma})|\le \sum_{\gamma\in \mathcal{B}_j}\mathbb{E}\Big| \prod_{\gamma\in b_{j_1}}\mathfrak{h}_{\gamma}\Big|\prod_{\gamma\in b_{j_2}}\Big|\mathbb{E} \mathfrak{h}_{\gamma}\Big|.
	\end{align}
	
	Observe that by definition, $\mathfrak{h}_{\gamma}$ is  the product of some independent random variables, and $\mathfrak{h}_{\gamma_1}$ and $\mathfrak{h}_{\gamma_2}$ may share some dependency through factors $w_{ab}^{m_1}$ and $w_{ab}^{m_2}$, respectively, for some $w_{ab}$ and nonnegative integers $m_1$ and $m_2$. Thus in light of the inequality
	\[ \mathbb{E}|w_{ab}|^{m_1}\mathbb{E}|w_{ab}|^{m_2}\le \mathbb{E}|w_{ab}|^{m_1+m_2}, \]
	(\ref{0928.5h}) can be bounded as
	\begin{align}\label{0928.6h}
	(\ref{0928.5h}) & \le 2^{|A_j|}\mathbb{E}\Big| \prod_{\gamma\in A_{j}}\mathfrak{h}_{\gamma}\Big|.
	\end{align}
	By (\ref{0928.6h}), we can deduce
	\begin{align}\label{1116.1}
	(\ref{0928.4h})  &\le 2^{2r}\sum_{ \mathcal{A}\in \mathfrak{S}_{\ge 2}(2r)}\sum_{ \bigcup_{s=1}^{2r}\bbi^{(s)}\in \mathcal{S}(\mathcal{A})}\prod_{j=1}^{|\mathcal{A}|}\mathbb{E}\Big| \prod_{\gamma\in A_{j}}\mathfrak{h}_{\gamma}\Big|\non
	&\le 2^{2r}\sum_{ \mathcal{A}\in \mathfrak{S}_{\ge 2}(2r)}\prod_{j=1}^{|\mathcal{A}|}(\sum_{ \bbi^{(s)}\in \mathcal{S}(A_j)}\mathbb{E}\Big| \prod_{\gamma\in A_{j}}\mathfrak{h}_{\gamma}\Big|).
	\end{align}
	Thus it suffices to show that
	$$\sum_{ \bbi^{(s)}\in \mathcal{S}(A_j)}\mathbb{E}\Big| \prod_{\gamma\in A_{j}}\mathfrak{h}_{\gamma}\Big|=C_{|A_j|}(\min\{\alpha_n^{l-1},d_{\sbx}\alpha_n^l, d_{\sby}\alpha_n^l\})^{|A_j|},$$
	using the fact that  $\sum_{j=1}^{|\mathcal{A}|}|A_j|=2r$.  Without loss of generality, we prove the most difficult case of $|\mathcal{A}|=1$, that is, there is only one connected chain which is $A=\{1,2,\cdots,2r\}$. It has the most components in the chain $ \prod_{\gamma\in A}\mathfrak{h}_{\gamma}$. Other cases with smaller $|A|$ can be shown in the same way. Using the same arguments as those for (\ref{0928.4h}), we have the basic property for this chain that there are at most $[\frac{|A|l}{2}]+1=rl+1$ distinct vertices and $rl$ distinct edges without self loops.
	{% Dear Prof. Lv, the following proof in red is almost the same as lemma 4 in \cite{FFHL19} since the chain is essentially almost the same as Lema 4 in \cite{FFHL19}. I am not sure how to deal with it.
		
		%The following arguments until (\ref{0928.11hh}) are almost the same as Lemma 4 in \cite{FFHL19}.
		%In order to make the proof self consistent, we provide them here.
		To facilitate our technical presentation, let us introduce some additional notation. Denote by $\psi(r,l)$  the set of partitions of the edges $\{(i_s,i_{s+1}),1\le s\le 2rl, i_s\neq i_{s+1}\}$  and $\psi_{\ge 2}(r,l)$ the subset of $\psi(r,l)$ whose blocks have size at least two. Let $\widetilde \bbi=\bigcup_{s=1}^{2r}\bbi^{(s)}$ and $P(\widetilde \bbi)\in\psi_{\ge 2}(2l+2)$ be the partition of $\{(i_s,i_{s+1}),1\le s\le 2rl, i_s\neq i_{s+1}\}$ that is associated with the equivalence relation $(i_{s_1},i_{s_1+1})\sim (i_{s_2},i_{s_2+1})$,  which is defined as if and only if $(i_{s_1},i_{s_1+1})=(i_{s_2},i_{s_2+1})$ or $(i_{s_1},i_{s_1+1})=(i_{s_2+1},i_{s_2})$. Denote by $|P(\widetilde \bbi)|=m$ the number of groups in the partition $P(\widetilde \bbi)$ such that the edges are equivalent within each group. We further denote the distinct edges in the partition $P(\widetilde \bbi)$ as $(s_1,s_2), (s_3,s_4), \cdots, (s_{2m-1},s_{2m})$ and the corresponding counts in each group as  $r_1, \cdots, r_m$, and define $\widetilde \bbs=(s_1,s_2,\cdots,s_{2m})$. For the vertices, let $\phi(2m)$ be the set of partitions of $\{1,2,\cdots,2m\}$ and $Q(\widetilde \bbs)\in\phi(2m)$ the partition that is associated with the equivalence relation $a\sim b$, which is defined as if and only if $s_a=s_b$. Note that $s_{2j-1}\neq s_{2j}$ by the definition of the partition. By $|w_{aa}|\le 1$,  we can deduce
		\begin{align}\label{0928.7h}
		&\sum_{ \bbi^{(s)}\in \mathcal{S}(A)}\mathbb{E}\Big| \prod_{\gamma\in A}\mathfrak{h}_{\gamma}\Big|=\sum_{ \bbi^{(s)}\in \mathcal{S}(A)}\mathbb{E}\Big| \prod_{\gamma=1}^{2r}\mathfrak{h}_{\gamma}\Big|\non
		&\le \sum_{1\le |P(\widetilde \bbi)|=m\le rl\atop
			P(\widetilde \bbi)\in \psi_{\ge 2}(2l+2)}\sum_{\widetilde \bbi \text{ with partition } P(\widetilde \bbi)\atop
			r_1,\cdots,r_m\ge 2}\sum_{Q(\widetilde \bbs)\in\phi(2m)}\sum_{\widetilde \bbs \text{ with partition } Q(\widetilde \bbs)\atop 1\le s_1,\cdots,s_{2m}\le n}\prod_{j=1}^{2r}(|x_{i_{(j-1)(l+1)+1}}||y_{i_{j(l+1)}}|)\nonumber\\
		&\quad \times \prod_{j=1}^m\mathbb{E}\big|w_{s_{2j-1}s_{2j}}|^{r_j}.
		\end{align}
		
		Denote by $\mathcal{F}_{\widetilde \bbs}$ the graph constructed by the edges of $\widetilde \bbs$. Since the edges in $\widetilde \bbs$ are the same as those of the edges in $\bigcup_{s=1}^{2r}\mathcal{G}^{(s)}$ with the structure $\mathcal{S}(A)$, we can see that $\mathcal{F}_{\widetilde \bbs}$ is also a connected graph. In view of  (\ref{0928.7h}), putting term $|x_{i_1}y_{i_{l+1}}x_{i_{l+2}}y_{i_{2l+2}}|$ aside we need to analyze the summation
		\[ \sum_{\widetilde \bbs \text{ with partition } Q(\widetilde \bbs)\atop 1\le s_1,\cdots, s_{2m}\le n}\prod_{j=1}^m\mathbb{E}\big|w_{s_{2j-1}s_{2j}}|^{r_j}. \]
		If index $s_{2k-1}$ satisfies that $s_{2k-1}\neq s$ for all $s\in \{s_1,\cdots,s_{2m}\}\setminus \{s_{2k-1}\}$, that is, index $s_{2k-1}$ appears only in one $w_{s_{2j-1}s_{2j}}$, we call $s_{2k-1}$ a single index (or single vertex). If there exists some single index $s_{2k-1}$, then it holds that
		\begin{align}\label{0928.10h}
		&\sum_{\widetilde \bbs \text{ with partition } Q(\widetilde \bbs)\atop 1\le s_1,\cdots,s_{2m}\le n}\prod_{j=1}^m\mathbb{E}\big|w_{s_{2j-1}s_{2j}}|^{r_j}  \nonumber \\
		&\quad \le \sum_{\tiny\substack{\widetilde \bbs\setminus \{s_{2k-1}\} \text{ with partition } Q(\widetilde \bbs\setminus \{s_{2k-1}\})\\ 1\le s_1,\cdots,s_{2k-2},s_{2k+2},s_{2m}\le n\\
				s_{2k}=s_j \text{ for some } 1\le j\le 2m}} \prod_{j=1}^m\mathbb{E}\big|w_{s_{2j-1}s_{2j}}|^{r_j} \sum_{s_{2k-1}=1}^n\mathbb{E}\big|w_{s_{2k-1}s_{2k}}|^{r_k}.
		\end{align}
		Note that since graph $\mathcal{F}_{\widetilde \bbs}$ is connected and  index $s_{2k-1}$ is single, there exists some $j$ such that $s_j=s_{2k}$, which means that in the summation $\sum_{s_{2k-1}=1}^n\mathbb{E}\big|w_{s_{2k-1}s_{2k}}|^{r_k}$, index $s_{2k}$ is fixed. Then it follows from the definition of $\alpha_n$, $|w_{ij}|\le 1$, and $r_k\ge 2$ that
		\[ \sum_{s_{2k-1}=1}^n\mathbb{E}\big|w_{s_{2k-1}s_{2k}}|^{r_k}\le \alpha_n^2. \]
		
		After taking the summation over index $s_{2k-1}$, we can see that there is one less edge in $\mathcal{F}(\widetilde \bbs)$. That is, by taking the summation above we will have one additional $\alpha_n^2$ in the upper bound while removing one edge from graph $\mathcal{F}(\widetilde \bbs)$. For the single index $s_{2k}$, we also have the same bound.  If $s_{2k_1-1}$ is not a single index, without loss of generality we assume that $s_{2k_1-1}=s_{2k-1}$. Then this vertex $s_{2k-1}$ needs some delicate analysis. By the assumption of $|w_{ij}|\le 1$, we have
		\[ \mathbb{E}|w_{2k-1,2k}|^{r_k}|w_{2k_1-1,2k_1}|^{r_{k_1}}\le \frac{\mathbb{E}|w_{2k-1,2k}|^{r_k}+\mathbb{E}|w_{2k_1-1,2k_1}|^{r_{k_1}}}{2}.
		\]
		Then it holds that
		\begin{align}\label{0928.9h}
		&\sum_{\widetilde \bbs \text{ with partition } Q(\widetilde \bbs)\atop 1\le s_1,\cdots,s_{2m}\le n}\prod_{j=1}^m\mathbb{E}\big|w_{s_{2j-1}s_{2j}}|^{r_j}  \nonumber\\
		&\le \frac{1}{2}\sum_{\widetilde \bbs\setminus (s_{2k-1},s_{2k_1-1}) \text{ with partition } Q(\widetilde \bbs\setminus(s_{2k-1},s_{2k_1-1}))\atop 1\le s_1,\cdots,s_{2m}\le n}\prod_{j=1,\,j\neq k}^m\mathbb{E}\big|w_{s_{2j-1}s_{2j}}|^{r_j} \nonumber\\
		&\quad+\frac{1}{2}\sum_{\widetilde \bbs\setminus(s_{2k-1},s_{2k_1-1}) \text{ with partition } Q(\widetilde \bbs\setminus (s_{2k-1},s_{2k_1-1}))\atop 1\le s_1,\cdots,s_{2m}\le n}\prod_{j=1,\,j\neq k_1}^m\mathbb{E}\big|w_{s_{2j-1}s_{2j}}|^{r_j}.
		\end{align}
		Note that since $\mathcal{F}_{\widetilde \bbs}$ is a connected graph, if we delete either edge $(s_{2k-1},s_{2k})$ or edge $(s_{2k_1-1},s_{2k_1})$ from graph $\mathcal{F}_{\widetilde \bbs}$, the resulting graph is also connected. Then the two summations on the right hand side of (\ref{0928.9h}) can be reduced to the case in (\ref{0928.10h}) for the graph with edge $(s_{2k-1},s_{2k})$ or $(s_{2k_1-1},s_{2k_1})$ removed, since $s_{2k-1}$ or $s_{2k_1-1}$ is a single index in the subgraph. Similar to (\ref{0928.10h}), after taking the summation over index $s_{2k-1}$ or $s_{2k_1-1}$ there are two less edges in graph $\mathcal{F}_{\widetilde \bbs}$ and thus we now obtain $2\alpha_n^2$ in the upper bound.
		
		For the general case when there are $m_1$ vertices belonging to the same group, without loss of generality we denote them as $w_{ab_1},\cdots,w_{ab_{m_1}}$. If for any $k$ graph $\mathcal{F}_{\widetilde \bbs}$ is still connected after deleting edges $(a,b_1),\cdots,(a,b_{k-1}),(a,b_{k+1}),\cdots,(a,b_{m_1})$, then we repeat the process in (\ref{0928.9h}) to obtain a new connected graph by deleting $k-1$ edges in $w_{ab_1},\cdots,w_{ab_{m_1}}$ and thus obtain $k\alpha_n^2$ in the upper bound. Motivated by the key observations above, we carry out an iterative process in calculating the upper bound as follows.
		\begin{itemize}
			\item[(1)] If there exists some single index in $\widetilde \bbs$, using  (\ref{0928.10h}) we can calculate the summation over such an index and then delete the edge associated with this vertex in $\mathcal{F}_{\widetilde \bbs}$. The corresponding vertices associated with this edge are also deleted. For simplicity, we also denote the new graph as $\mathcal{F}_{\widetilde \bbs}$. In this step, we obtain $\alpha_n^2$ in the upper bound.
			
			\item[(2)] Repeat (1) until there is no single index in graph $\mathcal{F}_{\widetilde \bbs}$.
			
			\item[(3)] Suppose there exists some index associated with $k$ edges such that graph $\mathcal{F}_{\widetilde \bbs}$ is still connected after deleting any $k-1$ edges. Without loss of generality, let us consider the case of $k=2$. Then we can apply (\ref{0928.10h}) to obtain $\alpha_n^2$ in the upper bound. Moreover, we delete $k$ edges associated with this vertex in $\mathcal{F}_{\widetilde \bbs}$.
			
			\item[(4)] Repeat (3) until there is no such index.
			
			\item[(5)] If there still exists some single index, go back to (1).  Otherwise stop the iteration.
		\end{itemize}
		
		Completing the graph modification process mentioned above, we can obtain a final graph $\bbQ$ that enjoys the following properties:
		\begin{itemize}
			\item[i)] Each edge does not contain any single index;
			
			%	\item[ii)] The degree of each vertex is at least two.
			
			\item[ii)] Deleting any vertex makes the graph disconnected.
		\end{itemize}
		Let $\bbS_{\bQ}$ be the spanning tree of graph $\bbQ$, which is defined as the subgraph of $\bbQ$ with the minimum possible number of edges. Since $\bbS_{\bQ}$ is a subgraph of $\bQ$, it also satisfies property ii) above. Assume that $\bbS_{\bQ}$ contains $p$ edges. Then the number of vertices in $\bbS_{\bQ}$ is $p+1$. Denote by $q_1,\cdots,q_{p+1}$ the vertices of $\bbS_{\bQ}$ and $\text{deg}(q_i)$ the degree of vertex $q_i$. Then by the degree sum formula, we have $\sum_{i=1}^{p+1}\text{deg}(q_i)=2p$. As a result, the spanning tree has at least two vertices with degree one and thus there exists a subgraph of $\bbS_{\bQ}$ without either of the vertices that is connected. This will result in a contradiction with property ii) above unless the number of vertices in graph $\bbQ$ is exactly one. Since $l$ is a bounded constant, the numbers of partitions $P(\widetilde \bbi)$ and $Q(\widetilde \bbs)$ are also bounded. It follows that
		\begin{equation}\label{0928.11h}
		(\ref{0928.7h}) \le C_rd_{\sbx}^{2r}d_{\sby}^{2r}\sum_{\widetilde \bbs \text{ with partition } Q(\widetilde \bbs)\atop 1\le s_1,\cdots,s_{2m}\le n}\prod_{j=1}^m\mathbb{E}\big|w_{s_{2j-1}s_{2j}}|^{r_j},
		\end{equation}
		where $d_{\sbx}=\|\bbx\|_{\infty}$, $d_{\sby}=\|\bbx\|_{\infty}$, and $C_r$ is some positive constant determined by $l$. Combining these arguments above and noticing that there are at most $l$ distinct edges in graph $\mathcal{F}_{\widetilde \bbs}$, we can obtain
		\begin{align}\label{0928.11hh}
		(\ref{0928.11h}) & \le C_rd_{\sbx}^{2r}d_{\sby}^{2r}\alpha_n^{2rl-2}\sum_{1\le s_{2k_0-1}, s_{2k_0}\le n,\,  (s_{2k_0-1},s_{2k_0})=\bbQ}\mathbb{E}\big|w_{s_{2k_0-1}s_{2k_0}}|^{r_{k_0}}\non
		&\le C_rd_{\sbx}^{2r}d_{\sby}^{2r}\alpha_n^{2rl}n.
		\end{align}
		Therefore, we have established a simple upper bound of $C_rd_{\sbx}^{2r}d_{\sby}^{2r}\alpha_n^{2rl}n$.
	}
	
	In fact, we can improve the aforementioned upper bound to $C_r\alpha_n^{r(l-1)}$. Note that the process mentioned above did not utilize the condition that both $\bx$ and $\by$ are unit vectors, that is, $\|\bbx\|=\|\bby\|=1$. Since term $\prod_{j=1}^{2r}(|x_{i_{(j-1)(l+1)+1}}||y_{i_{j(l+1)}}|)$ is involved in (\ref{0928.7h}), we can analyze them together with random variables $w_{ij}$.  First, we need to deal with some distinct lower indices with low moments in $\prod_{j=1}^{2r}(|x_{i_{(j-1)(l+1)+1}}||y_{i_{j(l+1)}}|)$. If there are two distinct lower indices, without loss of generality denoted them as $i_s$ and $i_{s'}$ and then the corresponding entries are $x_{i_s}$ (or $y_{i_s}$) and $y_{i_{s'}}$ (or $x_{i_{s'}}$). Moreover, there are only one $x_{i_s}$ and $y_{i_{s'}}$ involved in $\prod_{j=1}^{2r}(|x_{i_{(j-1)(l+1)+1}}||y_{i_{j(l+1)}}|)$. Without loss of generality, let us assume that $s=1$ and $s'=l+1$. Then it holds that
	\begin{align}\label{1116.2}
	&\prod_{j=1}^{2r}(|x_{i_{(j-1)(l+1)+1}}||y_{i_{j(l+1)}}|)=|x_{i_1}||y_{i_{l+1}}|\prod_{j=2}^{2r}(|x_{i_{(j-1)(l+1)+1}}||y_{i_{j(l+1)}}|)\non
	& \le \frac{x^2_{i_1}}{2}\prod_{j=2}^{2r}(|x_{i_{(j-1)(l+1)+1}}||y_{i_{j(l+1)}}|)+\frac{y^2_{i_{l+1}}}{2}\prod_{j=2}^{2r}(|x_{i_{(j-1)(l+1)+1}}||y_{i_{j(l+1)}}|).
	\end{align}
	That is, if we have two lower indices and each index  appears only once in the product above, we can use  (\ref{1116.2}) to increase the moment of $x_{i_s}$( or $y_{i_{s'}}$) and  delete the other one.  For (\ref{1116.2}), it is equivalent for us to consider the case when the lower index $i_1=i_{l+1}$. Repeating the procedure (\ref{1116.2}), finally we can obtain a product $\prod_{j=1}^{2r}(|x_{i_{(j-1)(l+1)+1}}||y_{i_{j(l+1)}}|)$ with the following properties:
	
	1). Except for one vertex $i_{s_0}$, for each $i_s$ with $s\neq s_0$ there exists some $i_{s'}$ such that $i_{s}=i_{s'}$ with $s\neq s'$.
	
	2). Except for one vertex $i_{s_0}$, for each $i_s$ with $s\neq s_0$ the term  $x^{m_1}_{i_s}y^{m_2}_{i_s}$ involved in $\prod_{j=1}^{2r}(|x_{i_{(j-1)(l+1)+1}}| |y_{i_{j(l+1)}}|)$ satisfies the condition that $m_1+m_2\ge 2$. Moreover, at least one of $m_1$ and $m_2$ is larger than one.
	
	By the properties above, let us denote by $\Upsilon(2r)$ the  set of partitions of the vertices $\{i_{(j-1)(l+1)+1}, i_{j(l+1)}, j=1,\cdots,2r\}$ such that except for one group, the remaining groups in $\Upsilon$ with $\Upsilon\in \Upsilon(2r)$ have blocks with  size at least two. %One should notice that we do not need to distinguish $\bbx$ with $\bby$ since $|x^{m_1}_{i_s}y^{m_2}_{i_s}|\le |x|^{m_1+m_2}_{i_s}+ |y|^{m_1+m_2}_{i_s}$. Hence it suffices for us to consider the product of $\prod_{j=1}^{2r}(|x_{i_{(j-1)(l+1)+1}}||y_{i_{j(l+1)}}|)$.
	There are three different cases to consider.
	
	\textit{Case 1)}.  All the groups in $\Upsilon$ have block size two. Then it follows that
	\begin{equation}\label{1116.5}
	|\prod_{j=1}^{2r}(|x_{i_{(j-1)(l+1)+1}}||y_{i_{j(l+1)}}|)|=\prod_{k=1}^{|\Upsilon|}|x|^{m_{1k}}_{i_s}|y|^{m_{2k}}_{i_k},
	\end{equation}
	where $m_{1k}+m_{2k}=2$. In fact, by the second property of $\Upsilon$ above, $m_{1k}=0$ or $m_{2k}=0$. Without loss of generality, we assume that $m_{2k}=0$. Then we need only to consider the equation
	$$|\prod_{j=1}^{2r}(|x_{i_{(j-1)(l+1)+1}}||y_{i_{j(l+1)}}|)|=\prod_{k=1}^{|\Upsilon|}|x|^{2}_{i_k}.$$
	Then by (\ref{0928.7h}), it remains to bound
	\begin{equation}\label{0928.12h}
	\sum_{\widetilde \bbs \text{ with partition } Q(\widetilde \bbs)\atop 1\le s_1,\cdots,s_{2m}\le n}\prod_{k=1}^{|\Upsilon|}|x|^{2}_{i_k}\prod_{j=1}^m\mathbb{E}\big|w_{s_{2j-1}s_{2j}}|^{r_j}.
	\end{equation}
	
	To simplify the presentation, assume without loss of generality that $i_k=s_k$, $k=1,\cdots,|\Upsilon|$. Then the summation in (\ref{0928.12h}) becomes
	\[ \sum_{\widetilde \bbs \text{ with partition } Q(\widetilde \bbs)\atop 1\le s_1,\cdots,s_{2m}\le n}\prod_{j=1}^{|\Upsilon|}|x|^{2}_{s_j}\prod_{j=1}^m\mathbb{E}\big|w_{s_{2j-1}s_{2j}}|^{r_j}. \]
	By repeating the iterative process (1)--(5) mentioned before, we can bound the summation for fixed $s_2,\cdots,s_{|\Upsilon|}$ and obtain an alternative upper bound
	\[ \sum_{s_1=1}^nx_{s_1}^2\mathbb{E}\big|w_{s_{2j-1}s_{2j}}|^{r_j}\le \sum_{s_1=1}^nx_{s_1}^2=1 \]
	since $\bx$ is a unit vector.  Thus for this step of the iteration, we obtain term one instead of $\alpha_n^2$ in the upper bound. Repeat this step until there is only $x^2_{s_{|\Upsilon|}}$ left. Since the graph is always connected during the iteration process, there exists another vertex $b$ such that $w_{s_{|\Upsilon|}b}$ is involved in (\ref{0928.12h}). For index $s_{|\Upsilon|}$, we do not delete the edges containing $s_{|\Upsilon|}$ in the graph during the iterative process (1)--(5). Then after the iteration stops, the final graph $\bQ$ satisfies properties i) and ii) defined earlier except for vertex $s_{|\Upsilon|}$. Since there are at least two vertices with degree one in  $\bbS_{\bQ}$, we will also reach a contradiction unless the number of vertices in graph $\bbQ$ is exactly one. By (\ref{1116.5}), it holds that $2|\Upsilon|=4r$. As a result, we can obtain the upper bound
	\begin{align}\label{0928.13h}
	(\ref{0928.7h})\le C_r\alpha_n^{2rl-2|\Upsilon|}\sum_{1\le s_{2},b\le n, \,  (s_2,b)=\bbQ}\mathbb{E}x_{s_{|\Upsilon|}}^2\big|w_{s_{|\Upsilon|}b}|^{r}\le C_r\alpha_n^{2rl-2r}
	\end{align}
	with $C_r$ some positive constant. Therefore, the improved bound  $C_r\alpha_n^{2r(l-1)}$ is shown for this case.
	
	\textit{Case 2)}. All the groups in $\Upsilon$ have block size at least two and there is at least one block with size larger than two.  Then it follows that
	$$|\prod_{j=1}^{2r}(|x_{i_{(j-1)(l+1)+1}}||y_{i_{j(l+1)}}|)|=\prod_{k=1}^{|\Upsilon|}|x|^{m_{1k}}_{i_s}|y|^{m_{2k}}_{i_k}.$$
	Since $m_{1k}+m_{2k}\ge 2$ by the second property of $\Upsilon$ above, define the nonnegative integer $r_1=\sum_{k=1}^{|\Upsilon|}(m_{1k}+m_{2k}-2)$. There are at most $[\frac{2rl+2-r_1}{2}]$ distinct vertices in the graph $\mathcal{F}_{\widetilde \bbs}$ and  at most $[\frac{2rl+2-r_1}{2}]-1$ distinct edges. Similar to Case  1 with less distinct edges, we have
	\begin{align}\label{0928.14h}
	(\ref{0928.7h})\le C\alpha_n^{2[\frac{2rl+2-r_1}{2}]-2|\Upsilon|-2}\sum_{1\le s_{1},b\le n,\,  (s_1,b)=\bbQ}\mathbb{E}x_{s_1}^2\big|w_{s_1b}|^{r}\le C\alpha_n^{2[\frac{2rl+2-r_1}{2}]-2|\Upsilon|}.
	\end{align}
	By the definition of $r_1$ and $\sum_{k=1}^{|\Upsilon|}(m_{1k}+m_{2k})=4r$, it holds that
	$$r_1+2|\Upsilon|=4r.$$
	Thus $r_1$ is an even number and $2[\frac{2rl+2-r_1}{2}]-2|\Upsilon|=2rl-r_1-2|\Upsilon|+2\le 2rl-2r$. The improved bound $C_r\alpha_n^{2r(l-1)}$ is also shown for this case.
	
	\textit{Case 3)}. Except for one index $i_{k_0}$, the other groups in $\Upsilon$ have block size at least two. Let us define $r'_1=\sum_{k=1, k\neq k_0}^{|\Upsilon|}(m_{1k}+m_{2k}-2)$. There are at most $[\frac{2rl+2-r'_1}{2}]$ distinct vertices and at most $[\frac{2rl+2-r'_1}{2}]-1$ distinct edges. For the parameter $|x_{i_{k_0}}|$ (or $|y_{i_{k_0}}|$), we can bound it by one since $\bbx$ and $\bby$ are unit vectors. Then similar to Case 2, we can deduce
	\begin{align}\label{0916.3}
	(\ref{0928.7h})\le C\alpha_n^{2[\frac{2rl+2-r'_1}{2}]-2|\Upsilon|}\sum_{1\le s_{1},b\le n,\,  (s_1,b)=\bbQ}\mathbb{E}x_{s_1}^2\big|w_{s_1b}|^{r}\le C\alpha_n^{2[\frac{2rl+2-r'_1}{2}]-2|\Upsilon|+2}.
	\end{align}
	By the definition of $r'_1$ in this case, it holds that
	$$r'_1+2|\Upsilon|=4r+1.$$
	Then $r'_1$ is an odd number and thus
	$$2[\frac{2rl+2-r'_1}{2}]-2|\Upsilon|+2=2rl-r_1-2|\Upsilon|+3\le 2rl-2r.$$
	Summarizing the arguments above, for this case we can also obtain the desired bound $C_r\alpha_n^{2r(l-1)}$.
	
	In addition, we can also improve the upper bound to $C_r(\min\{d_{\sbx}^{2r}\alpha_n^{2rl}, d_{\sby}^{2r}\alpha_n^{2rl}\})$. The technical arguments for this refinement are similar to those for the improvement to order $C_r\alpha_n^{2r(l-1)}$ above. As an example, we can bound the components of $\bby$ by $d_{\sby} = \|\by\|_\infty$, which leads to
	$|\prod_{j=1}^{2r}(|x_{i_{(j-1)(l+1)+1}}||y_{i_{j(l+1)}}|)|\le d_{\sby}^{2r}|\prod_{j=1}^{2r}|x_{i_{(j-1)(l+1)+1}}|$. Then the analysis becomes similar to the three cases above. The only difference is that $\sum_{k=1}^{|\Upsilon|}m_{1k}=2r$ instead of $\sum_{k=1}^{|\Upsilon|}(m_{1k}+m_{2k})=4r$.  For this case, we have
	\begin{align}\label{0928.15h.new}
	(\ref{0928.7h})\le Cd_{\sby}^{2r}\alpha_n^{2rl-2|\Upsilon|}\sum_{1\le s_{2},b\le n,\,  (s_2,b)=\bbQ}\mathbb{E}x_{s_1}^2\big|w_{s_1b}|^{r}\le C_rd_{\sby}^{2r}\alpha_n^{2rl}.
	\end{align}
	Thus we can obtain the claimed upper bound $C_r(\min\{d_{\sbx}^{2r}\alpha_n^{2rl}, d_{\sby}^{2r}\alpha_n^{2rl}\})$. Therefore, combining the two aforementioned improved bounds yields the desired upper bound of $$C_r(\min\{\alpha_n^{2r(l-1)}, d_{\sbx}^{2r}\alpha_n^{2rl}, d_{\sby}^{2r}\alpha_n^{2rl}\}),$$
	which completes the proof of Lemma  \ref{0426-1h}.

\subsection{Corollary \ref{cor} and its proof} \label{SecC.2}
Lemma \ref{0426-1h} ensures the following corollary immediately.
\begin{coro}\label{cor}
	Under the conditions of Lemma \ref{0426-1h}, it holds that for any positive constants $a$ and $b$,  there exists some $n_0(a,b)>0$ such that
	\begin{equation}\label{1119.1hh}
	\sup_{\|\bbx\|=\|\bby\|=1}\mathbb{P}\left(\bbx^T(\bbW^l-\mathbb{E}\bbW^l)\bby\ge n^{a}\min\{\alpha_n^{l-1}, d_{\sbx}\alpha_n^l,  d_{\sby}\alpha_n^l\}\right)\le n^{-b}
	\end{equation}
	for any $n\ge n_0(a,b)$ and $l\geq 1$.
	Moreover,  we have
	\begin{equation}\label{1119.1h}
	\bbx^T(\bbW^l-\mathbb{E}\bbW^l)\bby=O_{\prec}(\min\{\alpha_n^{l-1}, d_{\sbx}\alpha_n^l, d_{\sby}\alpha_n^l\}).
	\end{equation}
\end{coro}

\noindent \textit{Proof}. It suffices to show (\ref{1119.1hh}) because then \eqref{1119.1h} follows from the definition. For any positive constants $a$ and $b$, there exists some integer $r$ such that $2ar\ge b+1$. By the Chebyshev inequality, it holds that
	$$\sup_{\|\bbx\|=\|\bby\|=1}\mathbb{P}(|\bbx^T(\bbW^l-\mathbb{E}\bbW^l)\bby|\ge n^{a}\min\{\alpha_n^{l-1},d_{\sbx}\alpha_n^l, d_{\sby}\alpha_n^l\})$$
	$$\le \sup_{\|\bbx\|=\|\bby\|=1}\frac{\mathbb{E}(\bbx^T(\bbW^l-\mathbb{E}\bbW^l)\bby)^{2r}}{n^{2ar}(\min\{\alpha_n^{l-1},d_{\sbx}\alpha_n^l, d_{\sby}\alpha_n^l\})^{2r}}\le \frac{C_r}{n^{b+1}},$$
	which can be further bounded by $n^{-b}$ as long as $n\ge C_r$. It is seen that $C_r$ is determined completely by $a$ and $b$. This concludes the proof of Corollary \ref{cor}.

\subsection{Lemma \ref{1212-1} and its proof} \label{SecC.3}
%By similar proof to Lemma \ref{0426-1h}, we conclude the following theorem and omit the detailed proof.
\begin{lem}\label{1212-1}
	For any $n$-dimensional unit vectors $\bbx$ and $\bby$, we have
	\begin{equation}\label{1212.3a}
	\mathbb{E}\bbx^T\bbW^l\bby=O(\alpha_n^{l}),
	\end{equation}
	where $ l\ge 2$ is a positive integer. Furthermore, if the number of nonzero components of $\bbx$ is bounded, then it holds that
	\begin{equation}\label{1212.3}
	\mathbb{E}\bbx^T\bbW^l\bby=O(\alpha_n^{l}d_{\sby}),
	\end{equation}
	where $d_{\sby} = \|\by\|_\infty$.
\end{lem}

\noindent \textit{Proof}. The result in \eqref{1212.3a} follows directly from Lemma 5 of \cite{FFHL19}. Thus it remains to show  (\ref{1212.3}).
	The main idea of the proof is similar to that for the proof of Lemma \ref{0426-1h}. Denote by $\mathfrak{C}$ the set of positions of the nonzero components of $\bbx$. Then we have
	\begin{align}\label{0928.19h}
	\mathbb{E}\bbx^T\bbW^l\bby&=\sum_{i_1\in\mathfrak{C}, 1\le i_2,\cdots,i_{l+1}\le n\atop i_s\neq i_{s+1}}\mathbb{E} \left(x_{i_1}w_{i_1i_2}w_{i_2i_3}\cdots w_{i_li_{l+1}}y_{i_{l+1}}\right).
	\end{align}
	Note that the cardinality of set $\mathfrak{C}$ is bounded. Thus it suffices to show that for fixed $i_1$, we have
	\begin{align}\label{0928.19hh}
	\sum_{ 1\le i_2,\cdots,i_{l+1}\le n\atop i_s\neq i_{s+1}}\mathbb{E} \left(x_{i_1}w_{i_1i_2}w_{i_2i_3}\cdots w_{i_li_{l+1}}y_{i_{l+1}}\right)=O(d_{\sby}\alpha_n^l).
	\end{align}
	By the definition of graph $\mathcal{G}^{(1)}$ in the proof of Lemma \ref{0426-1h}, we can also get a similar expression as (\ref{0928.6h}) that
	\begin{align}
	\nonumber&|(\ref{0928.19h})| \\
	& \le d_{\sby}\sum_{\mathcal{G}^{(1)} \text{ with at most $[l/2]$ distinct edges without self loops and $[l/2]+1$ distinct vertices, $i_1$ is fixed }}\mathbb{E}\big|w_{i_1i_2}w_{i_2i_3} \nonumber\\
	&\quad \cdots w_{i_li_{l+1}}\big|.
	\end{align}

Using similar arguments for bounding the order of the summation through the iterative process as those for (\ref{0928.11h})--(\ref{0928.11hh}) in the proof of Lemma \ref{0426-1h}, we can obtain a similar bound
	\begin{equation}\label{0928.21h}
	\mathbb{E}\bbx^T\bbW^l\bby\le Cd_{\sby}\alpha_n^{l-2}\sum_{i_{k_0}=1}^n\mathbb{E}\big|w_{i_{1}i_{k_0}}|^{r_0}\le Cd_{\sby}\alpha_n^{l}
	\end{equation}
	with $r_0\ge 2$. Here we do not remove the lower index $i_1$ during the iteration procedure. The additional factor $n$ on the right hand side of (\ref{0928.11hh}) can be eliminated since $i_1$ is fixed. This completes the proof of Lemma \ref{1212-1}.

\subsection{Lemma \ref{precl} and its proof} \label{SecC.4}
\begin{lem}\label{precl}
	Assume that $\xi_1=O_{\prec}(\zeta), \cdots,\xi_{m}=O_{\prec}(\zeta)$ with $m=\lfloor n^{c}\rfloor$ and $c$ some positive constant.
	If \begin{equation}\label{0101.1h}
	\mathbb{P}\left[|\xi_i|>n^{a}|\zeta|\right]\le n^{-b}
	\end{equation}
	uniformly for $\xi_i$, $i=1,\cdots,m$, and any positive constants $a$,$b$ with $n\ge n_0(a,b)$, then for any positive random variables $X_1, \cdots, X_m$, we have
	$$\sum_{i=1}^mX_i\xi_i=O_{\prec}\Big(\sum_{i=1}^mX_i\zeta\Big).$$
	%	That is to say $\xi_i$ can be considered as additive parameters and the order can be calculated by simple summation.
\end{lem}

\noindent \textit{Proof}. For any positive constants $a$ and $b$, let $b_1=c+b$. By (\ref{0101.1h}), it holds that
	$$\mathbb{P}\left[|\xi_i|>n^{a}|\zeta|\right]\le n^{-b_1} $$
	for all $n\ge n_0(a,b_1)$, where $n_0(a,b_1)$ is  determined completely by $a$ and $b_1$. Then we have
	$$\mathbb{P}\left[|\sum_{i=1}^mX_i\xi_i|>n^{a}|\zeta|\sum_{i=1}^mX_i\right]\le \sum_{i=1}^m\mathbb{P}\left[|\xi_i|>n^{a}|\zeta|\right]\le n^{-b} $$
	for large enough $n\ge n_0(a,b_1)$. Since $b_1=c+b$ and $c$ is fixed, the constant $n_0(a,b_1)$ is determined essentially by $a$ and $b$. This concludes the proof of Lemma \ref{precl}.

\subsection{Lemma \ref{0505-1} and its proof} \label{SecC.5}
\begin{lem} \label{0505-1}
	For any positive constant $\mathfrak{L}$, it holds that
	$$\mathbb{P}(\|\bbW\|\ge \alpha_n\log n)\le n^{-\mathfrak{L}}$$
	for all sufficiently large $n$.
\end{lem}

\noindent \textit{Proof}. The conclusion of Lemma \ref{0505-1} follows directly from Theorem 6.2 of \cite{T12}. We can also prove it by (\ref{1217.15h}) and the inequality with $c\sqrt{\log n}\alpha_n-1$ replaced by $\alpha_n\log n$ in (\ref{0531.1h}).

\subsection{Lemma \ref{lem: tk}} \label{SecC.6}
\begin{lem}[\cite{FFHL19}] \label{lem: tk}
There exists a unique solution $z = t_k$ to equation \eqref{0515.3.1} on the interval $[a_k, b_k ]$, and thus $t_k$'s are well defined. In addition, for each $k=1,\cdots, K$, we have $t_k/d_k \to 1$ as $n\rightarrow \infty$.
\end{lem}

{
\renewcommand{\thesubsection}{D.\arabic{subsection}}
\renewcommand{\theequation}{D.\arabic{equation}}
\setcounter{equation}{0}
\section{Sufficient conditions for Condition \ref{cond2}}\label{SecCCond2}

\subsection{Lemma \ref{lemcond2} and its proof}

\begin{lem}\label{lemcond2}
	Under Conditions \ref{as3}--\ref{cond1}, if $\theta<1$ and $\min_{1\le i,j\le K}\bbP_{ij}\ge c$ for some positive constant $c$, then Condition \ref{cond2} holds.
\end{lem}

\noindent \textit{Proof}. The key step of the proof is to calculate $\cov[(\be_i-\be_j)^T\bW\bV]$. Without loss of generality, let us assume that $(i,j) = (1,2)$.
Note that the main difference between the null and alternative hypotheses is that the mean value of $(\be_1-\be_2)^T\mathbb{E}\bW$ is 0 under the former and is $(\mathbb Ew_{1,1}, -\mathbb Ew_{2,2}, 0, \cdots, 0)^T$, which may be nonzero, under the latter. However, since the main idea of the proof applies to both cases, we will provide only the technical details under the null hypothesis.

First, some direct calculations show that
\begin{align}\label{le.1h}
\theta^{-1}\bD\bSig_1 \bD& =  \theta^{-1}\cov[(\be_i-\be_j)^T\bW\bV]\non
&=\theta^{-1}\bV^T\mathbb{E}(\bbW(\be_i-\be_j)(\be_i-\be_j)^T\bbW)\bV\non
&=\theta^{-1}\bV^T\bbQ\bV,
\end{align}
where $\bbQ=\diag(\mathbb{E}(\bbw_{i1}-\bbw_{j1})^2,\cdots,\mathbb{E}(\bbw_{in}-\bbw_{jn})^2)+\mathbb{E}\bbw^2_{ij}\bbe_i\bbe_j^T+\mathbb{E}\bbw^2_{ij}\bbe_j\bbe_i^T$.
By the assumptions that  $\theta<1$ and $\min_{1\le i,j\le K}\bbP_{ij}\ge c$, we see that the entries of the mean matrix $\bbH = (h_{ij})$ are bounded from below by $c\theta$ and from above by $\theta$. Since $\mathbb Ew_{ij}^2 \sim h_{ij}$ and $w_{ik}$ and $w_{jk}$ are independent for $i\neq j$, it holds that
\begin{equation}\label{le.2h}
\theta\bbI\lesssim  \diag(\mathbb{E}(w_{i1}-w_{j1})^2,\cdots,\mathbb{E}(w_{in}-w_{jn})^2)\lesssim  \theta\bbI.
\end{equation}
%By Conditions 1--2, we have $ n\theta\bbI\lesssim  \bbD\lesssim   n\theta\bbI$(One can also check the equation above {\color{blue}(36)} \yfnote{the equation number doesn't seem to be correct}). This together with
%{\color{blue}Dear Prof. Fan, we modify condition 3 and therefore this proof should be modified correspondingly. I also copy this proof into Lemma 16 of our main paper.}

Then it follows from \eqref{le.2h} that
\begin{equation}\label{le.3h}
\bbI\lesssim \theta^{-1}\bV^T\diag(\mathbb{E}(w_{i1}-w_{j1})^2,\cdots,\mathbb{E}(w_{in}-w_{jn})^2)\bV\lesssim \bbI.
\end{equation}
Since $\bSig_1 \in \mathbb R^{K\times K}$ with $K$ a finite integer, we can deduce that
\begin{equation}\label{le.4h}
\|\theta^{-1}\bV^T(\mathbb{E}w^2_{ij}\bbe_i\bbe_j^T+\mathbb{E}w^2_{ij}\bbe_j\bbe_i^T)\bV\|\lesssim \frac{1}{n}.
\end{equation}
Therefore, combining \eqref{le.1h}--\eqref{le.4h}, we can obtain the desired conclusion under the null hypothesis. This completes the proof of Lemma \ref{lemcond2}.

\renewcommand{\thesubsection}{E.\arabic{subsection}}
\renewcommand{\theequation}{E.\arabic{equation}}
\setcounter{equation}{0}

\section{Uniform convergence} \label{sec: uniform}
%In this section, we consider the uniform convergence of the statistics for any set $\bbA\subseteq \{1,\ldots,n\}^2$.% with  $\log(|\bbA|)=O(\log\log n)$.
\begin{thm}
Assume that the null hypotheses $H_{0,ij}: \bpi_i=\bpi_j$ hold for all $1\leq i \neq j\leq n$. Then
\begin{itemize}
\item[1)] Under Conditions \ref{as3}--\ref{cond2} and the mixed membership model \eqref{0115.1}, we have for any  $x\in \mathbb{R}$,
\begin{equation}\label{0310.4h2}
\lim_{n\rightarrow \infty}\sup_{1\leq i \neq j\leq n}|\mathbb{P}(T_{ij}\le x)-\mathbb{P}(\chi_K^2\le x)|=0.
\end{equation}
\item[2)] Under Conditions \ref{as3} and \ref{cond5}--\ref{cond8} and DCMM \eqref{DCMM}, we have for any  $x\in \mathbb{R}$,
\begin{equation}\label{0310.5h2}
\lim_{n\rightarrow \infty}\sup_{1\leq i \neq j\leq n}|\mathbb{P}(G_{ij}\le x)-\mathbb{P}(\chi_{K-1}^2\le x)|=0.
\end{equation}
\end{itemize}
\end{thm}

\noindent \textit{Proof}. We provide the detailed proof only for \eqref{0310.4h2} since the proof of \eqref{0310.5h2} is almost identical. Recall that $T_{ij}=\|\bSig_1^{-1/2}(\widehat\bbV(i)-\widehat\bbV(j))\|^2$. Let us investigate the asymptotic behavior of random vector $\bSig_1^{-1/2}(\widehat\bbV(i)-\widehat\bbV(j))$.
Checking the proof of Theorem \ref{0126-1} in Section \ref{SecA.1} carefully, we can see that there exists some positive constant $\ep$ such that
	\begin{align}\label{1117.2h2}
 &\bold\Sigma_1^{-1/2}(\widehat\bbV(i)-\widehat\bbV(j))\non
 &=\bold\Sigma_1^{-1/2}\bbD^{-1}\left(\frac{(\bbe_i-\bbe_j)^T\bbW\bbv_1}{t_1/d_1},\cdots,\frac{(\bbe_i-\bbe_j)^T\bbW\bbv_K}{t_K/d_K}\right)^T+O_{\prec}(n^{-\ep}),
   \end{align}
 where the $o_p(1)$ term in \eqref{1117.2} is replaced by $O_{\prec}(n^{-\ep})$.   By %the condition $\log(|\bbA|)=O(\log\log n)$,
 \eqref{1117.2h2} and the continuity of the standard multivariate Gaussian distribution, it suffices to show that for any convex set $\bbS \subset \mathbb{R}^K$, we have
 \begin{align}\label{0310.1h2}
& \lim_{n\rightarrow \infty}\sup_{i\neq j}\left|\mathbb{P}\left(\bold\Sigma_1^{-1/2}\bbD^{-1}\left(\frac{(\bbe_i-\bbe_j)^T\bbW\bbv_1}{t_1/d_1},\cdots,\frac{(\bbe_i-\bbe_j)^T\bbW\bbv_K}{t_K/d_K}\right)^T\in \bbS\right)-\mathbb{P}\left(\bbx_K\in \bbS\right)\right| \nonumber\\
&\quad =0,
\end{align}
where $\bbx_K\sim N({\bf 0},\bbI_K)$.

For an application of Theorem 1.1 in \cite{BEJ1072}, we need to rewrite $$\bold\Sigma_1^{-1/2}\bbD^{-1}\left(\frac{(\bbe_i-\bbe_j)^T\bbW\bbv_1}{t_1/d_1},\cdots,\frac{(\bbe_i-\bbe_j)^T\bbW\bbv_K}{t_K/d_K}\right)^T$$ as the sum of independent random vectors.  Indeed, some direct calculations yield
\begin{align}\label{0310.2h2}
&\bold\Sigma_1^{-1/2}\bbD^{-1}\left(\frac{(\bbe_i-\bbe_j)^T\bbW\bbv_1}{t_1/d_1},\cdots,\frac{(\bbe_i-\bbe_j)^T\bbW\bbv_K}{t_K/d_K}\right)^T\non
&=\sum_{l=1}^n\bold\Sigma_1^{-1/2}\bbD^{-1}\left(\frac{(w_{il}-w_{jl})\bbv_{1l}}{t_1/d_1},\cdots,\frac{(w_{il}-w_{jl})\bbv_{Kl}}{t_K/d_K}\right)^T\non
&=\sum_{l\neq i,j}\bold\Sigma_1^{-1/2}\bbD^{-1}(w_{il}-w_{jl})\left(\frac{\bbv_{1l}}{t_1/d_1},\cdots,\frac{\bbv_{Kl}}{t_K/d_K}\right)^T\non
&\quad+\sum_{l\in \{i, j\}}\bold\Sigma_1^{-1/2}\bbD^{-1}(w_{il}-w_{jl})\left(\frac{\bbv_{1l}}{t_1/d_1},\cdots,\frac{\bbv_{Kl}}{t_K/d_K}\right)^T,
\end{align}
where the first term in the last step is the sum of independent random vectors. Then it follows from Lemma \ref{egv} and Condition \ref{cond2} that
$$\sum_{l\in \{i,  j\}}\bold\Sigma_1^{-1/2}\bbD^{-1}(w_{il}-w_{jl})\left(\frac{\bbv_{1l}}{t_1/d_1},\cdots,\frac{\bbv_{Kl}}{t_K/d_K}\right)^T=O(\frac{1}{\sqrt{n\theta}}).$$
Combining this with \eqref{0310.1h2} and \eqref{0310.2h2}, we see that it remains to show that
 \begin{align}\label{0310.3h2}
&\lim_{n\rightarrow \infty}\sup_{i\neq j}\left|\mathbb{P}\left(\sum_{l\neq i,j}\bold\Sigma_1^{-1/2}\bbD^{-1}(w_{il}-w_{jl})\left(\frac{\bbv_{1l}}{t_1/d_1},\cdots,\frac{\bbv_{Kl}}{t_K/d_K}\right)^T\in \bbS\right)-\mathbb{P}(\bbx_K\in \bbS)\right|\nonumber\\
&\quad=0.
\end{align}

From Theorem 1.1 in \cite{BEJ1072}, Condition \ref{cond2}, and Lemma \ref{egv}, we can deduce that for any fixed $i,j$, there exists some positive constant $C$ (independent of $i,j$) such that
\begin{align*}
&\left|\mathbb{P}\left(\sum_{l\neq i,j}\bold\Sigma_1^{-1/2}\bbD^{-1}(w_{il}-w_{jl})\left(\frac{\bbv_{1l}}{t_1/d_1},\cdots,\frac{\bbv_{Kl}}{t_K/d_K}\right)^T\in \bbS\right)-\mathbb{P}(\bbx_K\in \bbS)\right| \\
&\le C\sum_{l\neq i,j}\mathbb E\|\bold\Sigma_1^{-1/2}\bbD^{-1}(w_{il}-w_{jl})\left(\frac{\bbv_{1l}}{t_1/d_1},\cdots,\frac{\bbv_{Kl}}{t_K/d_K}\right)^T\|_2^3\non
&=C\sum_{l\neq i,j}\left(\|\bold\Sigma_1^{-1/2}\bbD^{-1}\left(\frac{\bbv_{1l}}{t_1/d_1},\cdots,\frac{\bbv_{Kl}}{t_K/d_K}\right)^T\|_2^3\times\mathbb E|w_{il}-w_{jl}|^3\right)\non
& \le \frac{C^2}{\sqrt n}\max_{l\neq i, j} \mathbb{E}|\frac{w_{il}-w_{jl}}{\sqrt \theta}|^3\\
&=O(\frac{1}{\sqrt{n\theta}}),
\end{align*}
which entails \eqref{0310.3h2}.  Therefore, the desired conclusions of the theorem follow immediately, which concludes the proof of Theorem \ref{sec: uniform}.

\end{document}